\documentclass[longauth]{aa}
\usepackage{graphicx}
\usepackage{siunitx}
\usepackage{subcaption}
\usepackage{lscape}
\usepackage{comment}
\usepackage{txfonts}
\usepackage{hyperref}
\hypersetup{
    colorlinks = true,
    linkcolor = {blue},
    citecolor = {blue},
    urlcolor = {blue}
}

\begin{document}

   \title{The Jet Paths of Radio AGN and their Cluster Weather}

   \author{E. Vardoulaki\inst{\ref{inst1},\ref{inst2}\thanks{email: elenivard@gmail.com}\texorpdfstring{\href{https://orcid.org/0000-0002-4437-1773}{\protect\includegraphics{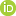}}}{0000-0002-4437-1773}}
            \and 
            V. Backöfer
          \inst{\ref{inst2}}
          \and
          A. Finoguenov\inst{\ref{inst3}\texorpdfstring{\href{https://orcid.org/0000-0002-4606-5403}{\protect\includegraphics{ORCID-iD_icon-16x16.png}}}{0000-0002-4606-5403}}
          \and
          F. Vazza\inst{\ref{inst4},\ref{inst5},\ref{inst6}}
          \and 
          J. Comparat\inst{\ref{inst7}}
          \and
          G. Gozaliasl\inst{\ref{inst3},\ref{inst8}\texorpdfstring{\href{https://orcid.org/0000-0002-0236-919X}{\protect\includegraphics{ORCID-iD_icon-16x16.png}}}{0000-0002-0236-919X}}
          \and
          I. H. Whittam\inst{\ref{inst9}}
          \and 
          C. L. Hale\inst{\ref{inst9}}
          \and
          J. R. Weaver\inst{\ref{inst10},\ref{inst11},\ref{inst12}}
          \and
          A. M. Koekemoer\inst{\ref{inst13}}
          \and
           J. D. Collier\inst{\ref{inst14},\ref{inst15},\ref{inst16}}
           \and
           B. Frank\inst{\ref{inst14},\ref{inst17},\ref{inst18}}
           \and
           I. Heywood\inst{\ref{inst9},\ref{inst19}}
           \and
           S. Sekhar\inst{\ref{inst14},\ref{inst21}}
           \and
           A. R. Taylor\inst{\ref{inst20},\ref{inst14},\ref{inst16}}
           \and
          S. Pinjarkar\inst{\ref{inst22}}
          \and 
          M. J. Hardcastle\inst{\ref{inst22}}
          \and
          T. Shimwell\inst{\ref{inst23},\ref{inst24}}
           \and
          M. Hoeft\inst{\ref{inst2}}
          \and
          S. V. White\inst{\ref{inst30}} 
          \and 
          F. An\inst{\ref{inst25},\ref{inst26}}
          \and
          F. Tabatabaei\inst{\ref{inst27},\ref{inst28},\ref{inst29}} 
          \and
          Z. Randriamanakoto\inst{\ref{inst30},\ref{inst31}}
          \and
          M. D. Filipovic\inst{\ref{inst15}}
        }

 \institute{IAASARS, National Observatory Athens, Lofos Nymfon, 11852 Athens, Greece\\
        \label{inst1}
        \and
 Thüringer Landessternwarte, Sternwarte 5, 07778 Tautenburg, Germany\\
        \label{inst2}
         \and
             Department of Physics, University of Helsinki, P.O. Box 64, FI-00014 Helsinki, Finland\\
        \label{inst3}
        \and
            Dipartimento di Fisica e Astronomia, Universita di Bologna, Via Gobetti 93/2, 40122 Bologna, Italy\\
        \label{inst4}
         \and
            Hamburger Sternwarte, Gojenbergsweg 112, 21029 Hamburg, Germany\\
        \label{inst5}
         \and
            Istituto di Radioastronomia, INAF, Via Gobetti 101, 40122 Bologna, Italy\\
        \label{inst6}
        \and
            Max-Planck-Institut für extraterrestrische Physik, Giessenbachstrasse 1, 85748 Garching, Germany\\
        \label{inst7}
        \and
             Department of Computer Science, Aalto University, PO Box 15400, Espoo, FI-00100, Finland\\
        \label{inst8}
         \and
           Astrophysics, University of Oxford, Denys Wilkinson Building, Keble Road, Oxford OX1 3RH, UK\\
        \label{inst9}
        \and
        Cosmic Dawn Center (DAWN), Copenhagen N, Denmark \\
        \label{inst10}
        \and
        Niels Bohr Institute, University of Copenhagen, Jagtvej 128, 2200, Copenhagen N, Denmark\\
        \label{inst11}
        \and
        Department of Astronomy, University of Massachusetts, Amherst, MA, 01003, USA\\
        \label{inst12}
        \and
        Space Telescope Science Institute, 3700 San Martin Dr, Baltimore, MD 21218, USA\\
        \label{inst13}
        \and
            The Inter-University Institute for Data Intensive Astronomy, Department of Astronomy, University of Cape Town, Private Bag X3, Rondebosch 7701, South Africa\\
        \label{inst14}
        \and
            School of Science, Western Sydney University, Locked Bag 1797, Penrith NSW 2751, Australia\\
        \label{inst15}
        \and
            CSIRO Astronomy and Space Science, PO Box 1130, Bentley WA 6102, Australia\\
        \label{inst16}
        \and
            Department of Astronomy, University of Cape Town, Private Bag X3, Rondebosch 7701, South Africa\\
        \label{inst17}
        \and
            South African Radio Astronomy Observatory, 2 Fir Street, Observatory 7925, South Africa\\
        \label{inst18}
        \and
        Department of Physics and Electronics, Rhodes University, PO Box 94, Grahamstown 6140, South Africa\\
        \label{inst19}
        \and
            Department of Physics and Astronomy, University of the Western Cape, Robert Sobukwe Road, Bellville 7535, South Africa\\
        \label{inst20}
        \and
            National Radio Astronomy Observatory, 1003 Lopezville Road, Socorro NM 87801, USA\\
        \label{inst21}
        \and
        Centre for Astrophysics Research, University of Hertfordshire, College Lane, Hatfield, AL10 9AB, UK\\
        \label{inst22}
        \and
        ASTRON, the Netherlands Institute for Radio Astronomy, Oude Hoogeveensedijk 4, 7991 PD, Dwingeloo, The Netherlands\\ 
        \label{inst23}
        \and
        Leiden Observatory, Leiden University, PO Box 9513, 2300 RA, Leiden, The Netherlands\\
        \label{inst24}
        \and
        Purple Mountain Observatory, Chinese Academy of Sciences, 10 Yuanhua Road, Qixia District, Nanjing 210023, People’s Republic of China\\
        \label{inst25}
        \and
        Inter-University Institute for Data Intensive Astronomy, and Department of Physics and Astronomy, University of the Western Cape, Robert Sobukwe Road, 7535 Bellville, Cape Town, South Africa\\
        \label{inst26}
        \and
        School of Astronomy, Institute for Research in Fundamental Sciences (IPM), PO Box 19395-5531 Tehran, Iran \\
        \label{inst27}
        \and
        Max-Planck-Institut für Astronomie, Königstuhl 17, D-69117, Heidelberg, Germany\\
        \label{inst28}
        \and
        Max-Planck Institut für Radioastronomie, Auf dem Hügel 69, D-53121 Bonn, Germany\\
        \label{inst29}
        \and 
        South African Astronomical Observatory, PO Box 9, Observatory 7935, South Africa\\
        \label{inst30}
        \and 
        Department of Physics, University of Antananarivo, PO Box 906, Antananarivo 101, Madagascar\\
        \label{inst31}
             }
   \date{Received ; accepted }

  \abstract
    {We studied bent radio sources within X-ray galaxy groups in the COSMOS and XMM-LSS fields. The radio data were obtained from the MeerKAT International GHz Tiered Extragalactic Explorations data release 1 (MIGHTEE-DR1) at 1.2-1.3 GHz, with angular resolutions of 8.9" and 5", and median noise levels $rms_{\rm med} \sim$ = 3.5 and 5.5  $\mu$Jy/beam. Bent radio active galactic nuclei (AGN) were identified through visual inspection. 
    Our analysis included 19 bent radio AGN in the COSMOS field and 17 in the XMM-LSS field which lie within X-ray galaxy groups ($2\times10^{13} \lesssim M_{\rm 200c}/M_{\odot} = 3\times10^{14}$). We investigated the relationship between their bending angle (BA) — the angle formed by the jets or lobes of two-sided radio sources associated with AGN — and properties of their host galaxies and large-scale environment probed by the X-ray galaxy groups.     
    Our key findings are: a) In the XMM-LSS field, we observed a strong correlation between the linear projected size of the bent AGN, the group halo mass, and the projected distance from the group centre. This trend, consistent with previous studies, was not detected in the COSMOS sample. b) The BA is a function of environmental density, with the type of medium playing a significant role. Additionally, at $z \le 0.5$ we found a higher number of bent sources (BA $\leq 160^{\circ}$) compared to higher redshifts ($z \sim 1$), by a factor of $>1.5$. 
    This trend aligns with magnetohydrodynamic simulations, which suggest that denser environments and longer interaction times at lower redshifts contribute to this effect. Comparison with the literature suggests that jet bending in galaxy groups within the redshift range $0.1 < z < 1.2$ is primarily driven by ram pressure exerted on the jets, which occurs during quiescent phases of AGN activity. This study underscores the role of environmental interactions in shaping the morphology of radio AGN within galaxy groups, providing insights into the interplay between large-scale structure and AGN physics.}


   \maketitle
%

\section{Introduction}

Active galactic nuclei (AGN) in the radio come in many shapes and sizes. These puzzling astrophysical phenomena are related to large-scale structure and galaxy growth and evolution, while their shapes often reveal hints about their interaction with the large-scale environment.
\citep[e.g.][]{prestage1988cluster,smolvcic2017vla,croston2019environments}. Their jets, ejected in opposite directions from their supermassive black holes, interact with their surrounding environment, which can cause the jets to deviate from an expected straight morphology. New radio surveys add to the complexity of radio structures \citep[e.g.][]{Hurley-Walker2017,white2020gleamI,white2020gleamII,sejake2023meerkat}, as higher resolutions and sensitivities reveal detailed jet structures as well as faint emission that previously eluded observation \citep[e.g.][]{delhaize2021mightee,mahatma2023low}. Both radio AGN and star-forming galaxies (SFGs) emit non-thermal synchrotron radiation in the radio \citep[e.g.][]{Miley1980, Condon1992, Padovani2017, klein2018radio}, albeit as a result of different physical processes. Nevertheless, the radio signatures of AGN and SFGs can often be tangled and become indistinguishable without the use of ancillary multi-wavelength observations. Separating the radio AGN and SFG populations in radio continuum surveys is a difficult task as surveys probe deeper populations of the radio sky \citep[e.g.][]{white2015radio, white2017evidence, smolvcic2017vla, gurkan2018lofar, vardoulaki2019closer, vardoulaki_3GHz_FR, mingo2019revisiting, whittam2022mightee}. As this study investigates the jet distortion of extended radio AGN, and their deviation from a straight radio structure, we rely on the distinct jet features to select our samples from visual inspection.

Jet distortion is a complex phenomenon, as jets are observed from pc to Mpc scales and evolve over millions of years \citep{turner2015energetics}. Studies suggest jet distortion has a complex explanation and several causes. These include the jets' movement through the intergalactic medium \citep[IGM; e.g.][]{begelman1979twin,owen1976radio,garon2019radio}, buoyancy forces \citep[e.g.][]{sakelliou1996bent,smolvcic2007wide}, precession of jets \citep[e.g.][]{taylor1990vla,caproni2017jet}, gravitational interaction of companion galaxies \citep[e.g.][]{perley1979structure,begelman1984theory} or jets passing through an area with significant pressure gradients \citep[e.g.][]{Best1997AJI}.

Past studies which have investigated jet bending in relation to the large-scale environment have mainly employed surveys like FIRST \citep[beam size: 5", rms: \SI{150}{\micro Jy/beam},][]{becker1995first} or LoTSS \citep[beam size: 6", rms: \SI{83}{\micro Jy/beam},][]{shimwell2019lofar,Shimwell2022}, which cover large areas at the expense of sensitivity, resulting in samples consisting of millions of radio galaxies. Identifying bent radio AGN in large surveys and studying them in relation to their large-scale environment is not a trivial task. A plethora of good-quality multi-wavelength data is required for such studies. \cite{garon2019radio} studied the bending angle of 4304 radio galaxies, selected from FIRST, in optically selected galaxy clusters with masses\footnote{M$_{500}=$ is the mass of a cluster/group at a virial radius of 500 times the critical density of the Universe.} ranging from M$_{500}=$ \SI{5e14}{M_\odot} to \SI{3e15}{M_\odot}. They find that, statistically, the more the sources are bent, the closer they are to their cluster centre. Additionally, sources are more bent in more massive clusters, which is related to higher intracluster medium (ICM) pressures and galaxies moving through the ICM with higher velocities, which promotes jet bending due to ram pressure. Bent sources not located in known clusters are found in statistically overdense regions. \cite{mingo2019revisiting} find that the 459 bent radio galaxies, obtained from the LoTSS surveys, have a significantly higher rate of cluster association than their total sample of 5805 extended radio sources. While \cite{garon2019radio} and \cite{mingo2019revisiting} are limited to cluster redshifts up to 0.8 and 0.4, respectively, \cite{golden2021high} find 36 bent radio sources, selected from FIRST, in clusters up to redshift 2.2. They find that more bent sources tend to reside in richer clusters, which further supports that bent sources are found in the dense medium of massive clusters, even at higher redshifts. Additionally, in the LoTSS DR2 sample, \cite{golden2023} find that narrower sources lie inside clusters, which implies environmental differences in the populations of bent radio AGN. Simulations of galaxies in cluster environments also give insights into the relationship between jet morphology and cluster environments. In particular, \cite{mguda2015ram} investigated the likelihood of finding radio galaxies bent due to ram pressure in clusters of galaxies. They find that with increasing halo mass, the number of galaxies bent due to ram pressures increases, but since more massive clusters are rarer than less massive clusters, approximately the same number of galaxies bent due to ram pressure are found at halo masses above and below M$\rm_{halo}=3\times10^{14}$M$_{\odot}$. \cite{mguda2015ram} find that bent radio sources are found out to distances of \SI{1.5}{Mpc} for clusters with halo masses M$\rm_{halo}\ge10^{15}$M$_{\odot}$ from their cluster centre, whereas the bent sources in clusters with halo masses $10^{13}$M$_{\odot} \leq \text{M}\rm_{halo} \leq 10^{14}\text{M}_{\odot}$ are most likely found within \SI{400}{kpc} of their cluster centre.

This study investigates a different parameter space, extending the halo mass range of galaxy groups/clusters to lower halo masses ($4\times10^{12}$M$_{\odot} < \text{M}_{\rm 200c} <  3\times10^{14}$M$_{\odot}$). We choose two extragalactic fields, COSMOS and XMM-LSS, to study the radio population and produce samples of extended radio galaxies. This contrasts with studies like those of \cite{garon2019radio} and \cite{mingo2019revisiting}, who rely on citizen science projects like the Radio Galaxy Zoo or automated source detection pipelines to obtain large samples. Choosing to study well-known fields allows us to utilise deep radio surveys like the \SI{3}{GHz} VLA-COSMOS project \citep{smolvcic2017vlaSource} with a sensitivity of \SI{2.3}{\micro Jy/beam} and the $\sim$1.2-1.3 GHz MeerKAT International GHz Tiered Extragalactic Explorations - MIGHTEE - survey \citep{jarvis2017meerkat, heywood2022mightee, Hale2024} at roughly \SI{2}{\micro Jy/beam}. Furthermore, legacy fields like COSMOS are well-studied across the electromagnetic spectrum, which allows comprehensive source characterisation and direct comparisons to past and future studies. One such study is from \cite{vardoulaki_3GHz_FR,vardoulaki2021bent}, who previously investigated the population of bent radio sources in COSMOS with \SI{3}{GHz} VLA observations with sub-arcsecond resolution (0".75). Each source was classified based on the scheme by Fanaroff and Riley \citep{fanaroff1974morphology} to be an edge-darkened FRI-type source, an edge-brightened FRII-type source or a hybrid FRI/FRII, where one side is edge-darkened and the other edge-brightened. They investigate the relations of bent radio sources to their host properties, FR-type, the large-scale environment probed by the density fields and cosmic-web probes in COSMOS \citep{scoville2013evolution,darvish2015evolution,darvish2017cosmic}, and the group environments obtained from X-ray galaxy groups in COSMOS with halo masses M$_{500}=$ \SI{5e12}{M_\odot} to \SI{2e14}{M_\odot} \citep{gozaliasl2019chandra}. They also compared the bending angle to magneto-hydrodynamical simulations of radio sources in clusters from \cite{vazza2021}. While \cite{vardoulaki2021bent} found no strong correlations between jet bending and the large-scale environment, FR-type or host properties, they found indications that FRI type radio sources are found in filaments. Differences to other studies of bent radio sources \cite[e.g.][]{garon2019radio} are attributed to either low sample size or the different parameter space of the studies. Comparisons of \cite{vardoulaki2021bent} to the simulations of \cite{vazza2021} indicate that sources are more bent at lower redshifts, which may be attributed to a denser ambient medium at lower redshifts.

In this paper, we further investigate the jet bending of extended radio AGN in the COSMOS and XMM-LSS fields with the first data release of the MIGHTEE radio survey \citep{Hale2024}. This study is complementary to past studies and expands the investigation of bending angle of radio AGN to higher redshift (up to $z \sim$ 3.5) and lower halo mass ($\sim 10^{13-14.5}~M_{\odot}$). In Section \ref{DataSection}, we present the sample creation process and all relevant multi-wavelength observations utilised in this work. The methods are given in Section \ref{methods}. The analysis and discussion of our data are presented in Sections \ref{Results}, \ref{Discussion} \& \ref{T_expected}. Section \ref{Discussion} discusses the results in the context of past and current literature with a focus on sources in galaxy group environments. In Section \ref{T_expected}, we estimate the expected temperature that one would expect from the bending angle. We present our conclusions in Section \ref{Conclusions}. Throughout this work, we adopt a flat $\Lambda$CDM cosmology, using H$_{0}$ = 70 km s$^{-1}$ Mpc$^{-1}$, $\Omega_{m}$ = 0.3, and $\Omega_{\Lambda}$ = 0.7.

\section{Sample selection}
\label{DataSection}
\subsection{MIGHTEE}

The MeerKAT International Gigahertz Tiered Extragalactic Explorations \citep[MIGHTEE,][]{jarvis2017meerkat, heywood2022mightee,Hale2024} is a galaxy evolution survey currently underway, conducted by the MeerKAT radio telescope in South Africa \citep[][]{jonas2016meerkat}. With $\sim$1000 hours of observing time, the survey aims to image \SI{20}{deg^2} over four extragalactic fields: The European Large Area \emph{ISO} Survey South 1 (ELAIS-S1), the Extended Chandra Deep Field South (E-CDFS), and the fields that are the focus of this work: XMM-LSS and COSMOS. The survey aims for a depth of $\sim \SI{2}{\micro Jy/beam}$ at $\sim$1.2-1.3 GHz\footnote{We note that the frequency varies across the mosaics.}. This work uses the data release DR1 \citep[henceforth MIGHTEE-DR1][]{Hale2024}, providing a sky coverage of $\sim\SI{14.4}{deg^2}$ in XMM-LSS and $\sim\SI{4.2}{deg^2}$ in COSMOS.\\ 

Both radio mosaics from the MIGHTEE-DR1 have been primary beam-corrected and were imaged with two different visibility weighting schemes, resulting in two versions of radio maps for each field. The first version has higher resolution than the second, but is less sensitive. The second version downweights the short baselines resulting in a higher resolution but decreases the sensitivity of the data. The resulting radio maps have a resolution of 8.9" with a measured sensitivity (median rms) of $\sim$\SI{3.5}{\micro Jy/beam}, and a resolution of $\sim$5" with a sensitivity of $\sim$\SI{6}{\micro Jy/beam}. In detail, the median rms for the XMM-LSS field is 5.1 (3.2) $\mu$Jy/beam for the 5" (8.9") mosaic, while for the COSMOS mosaic is 5.6 (3.5) $\mu$Jy/beam. For the rest of this study, we will distinguish the different versions by their resolutions. For more information on the data reduction, we refer the reader to the related publications \citep{jarvis2017meerkat,heywood2022mightee,Hale2024}.

We identified 306 extended radio structures in XMM-LSS, and 254 extended radio structures in COSMOS after visual inspection of the MIGHTEE-DR1 mosaics. Although automated radio source identification methods have become sufficient in identifying simple radio structures \citep[e.g. ByBDSF][]{mohan2015,Polsterer2019}, even the more sophisticated automatic algorithms fail in identifying complex radio structures \citep[e.g.][]{vardoulaki_3GHz_FR, Boyce2023}. Additionally, although automatic algorithms such as PINK \citep{Galvin2020} are very useful in identifying radio structures, matching to the host galaxy and classifying sources, they need good training sample that depend on resolution and sensitivity \cite[e.g.][]{vardoulaki_3GHz_FR}. Since our project depends on the good identification of radio structures in the MIGHTEE mosaics and of their associated hosts, and since there was no extended source catalogue for MIGHTEE-DR1 at the beginning of the project and during the time the analysis took place, we chose the traditional way of visual inspection.  

Below we describe the process of cleaning up these samples to include only two-sided radio AGN, for which we could securely measure their bending angle (BA; see Section~\ref{efficacy BA}). For this reason, we used a large variety of multi-wavelength data. To the best of our ability, these sample of bent radio AGN include all sources for which we could securely measure the BA. The final samples, relevant to this analysis, contain extended radio AGN within the X-ray galaxy groups in the COSMOS and XMM-LSS fields (Table~\ref{tab:subsamples}). The radio properties of the final sample are presented in the appendix (Tables~\ref{tab:xmmlsssample}~\&~\ref{tab:cosmossample}; available via CDS).

\subsection{Multi-wavelength Data}
\label{ancData}

\subsubsection{VLA-COSMOS}

For the COSMOS field we utilise observations from the Very Large Array (VLA), which provide both excellent resolution and sensitivity, to improve the source characterisation in the COSMOS sample. The VLA-COSMOS \SI{1.4}{GHz} Large Project \citep{schinnerer2007vla}, was performed using the VLA and consists of 23 pointings covering the \SI{2}{deg^2} of the COSMOS field with a total observing time of 275 hours. The mean sensitivity reaches \SI{10.5}{\micro Jy/beam} (\SI{15}{\micro Jy/beam}) in the central deg$^2$ (\SI{2}{deg^2}), which is 2-3 times worse than the MIGHTEE-DR1, but the VLA data have much higher angular resolution than MIGHTEE-DR1. The beam size of the VLA mosaic is 1.4" $\times$ 1.5".\\

The VLA-COSMOS \SI{3}{GHz} Large Project \citep{smolvcic2017vlaSource} covers a sky area of \SI{2.6}{deg^2} with 64 pointings, fully covering the central \SI{2}{deg^2} of the COSMOS field, and expanding the area to \SI{2.6}{deg^2}. It reaches a median rms of \SI{2.3}{\micro Jy/beam} at the centre of the field and a sub-arcsecond resolution of 0".75, allowing us to study the sub-structures in high resolution and helps to disentangle sources. 

\subsubsection{GMRT 610 MHz}

In rare cases, we used the \SI{610}{MHz} GMRT observations for the XMM-LSS sample to get a better understanding of the sources morphologies. The GMRT \SI{610}{MHz} radio continuum survey \citep{smolvcic2018xxl} was conducted by the Giant Metrewave Radio Telescope at \SI{50}{cm} wavelength, covering \SI{25}{deg^2} over the XXL Northern field (XXL-North). The survey combined previous observations done with the GMRT at \SI{610}{MHz}, covering an area of \SI{12.66}{deg^2} within XXL-North, which also includes XMM-LSS \citep{tasse2007gmrt}. For the area that encloses XMM-LSS, \cite{smolvcic2018xxl} reports a median rms of \SI{200}{\micro Jy/beam}, improving from the reported rms of \SI{300}{\micro Jy/beam} from \cite{tasse2007gmrt}. The synthesised beam size of the final mosaic is 6.5" $\times$ 6.5".

\subsubsection{VLASS}

Because it was not always possible to determine the core region of the radio sources from the MIGHTEE data alone, we also used high frequency, high resolution data from the \SI{3}{GHz} Very Large Array Sky Survey \citep[VLASS]{lacy2020karl}, if necessary. VLASS is an all-sky radio survey that covers the entire sky observable by VLA north of a declination of \SI{-40}{deg}, covering completely both the XMM-LSS and COSMOS fields. The survey aims to cover an area of \SI{33885}{deg^2} with an angular resolution of 2".5 down to noise levels of \SI{70}{\micro Jy/beam} by 2024. For this study, we use the Epoch 1 Quick Look images \citep{Gordon2020} provided by the Canadian Initiative for Radio Astronomy Data Analysis (CIRADA\footnote{ \url{http://cutouts.cirada.ca/}}), offering radio cutouts with an rms $\sim \SI{0.12}{\milli Jy/beam}$.

\begin{table*}[h!]
\begin{center}
\caption{Radio data used in this work.}
\label{tab:RadioData}
\scalebox{0.8}{
\renewcommand{\arraystretch}{1.5}
\begin{tabular}[t]{c c c c c c}
\hline
Survey & Central Frequency & Sensitivity & Beam Size & Field & Reference\\
\hline
MIGHTEE-DR1 & 1.2-1.3 GHz & \SI{3.5}{\micro Jy/beam} & 8.9" $\times$ 8.9" & COSMOS & \cite{Hale2024}\\
MIGHTEE-DR1 & 1.2-1.3 GHz & \SI{3.2}{\micro Jy/beam} & 8.9" $\times$ 8.9" & XMM-LSS & \cite{Hale2024}\\
MIGHTEE-DR1 & 1.2-1.3 GHz & \SI{5.6}{\micro Jy/beam} & 5.2" $\times$ 5.2" & COSMOS & \cite{Hale2024}\\
MIGHTEE-DR1 & 1.2-1.3 GHz & \SI{5.1}{\micro Jy/beam} & 5" $\times$ 5" & XMM-LSS & \cite{Hale2024}\\
VLASS Epoch 1& \SI{3}{GHz} & \SI{0.12}{\milli Jy/beam} & 2.5" $\times$ 2.5" & XMM-LSS\&COSMOS & \cite{lacy2020karl}\\
\SI{3}{GHz} VLA & \SI{3}{GHz} & \SI{2.3}{\micro Jy/beam} & 0.75" $\times$ 0.75" & COSMOS & \cite{smolvcic2017vlaSource}\\
\SI{1.4}{GHz} VLA  & \SI{1.4}{GHz} & \SI{15}{\micro Jy/beam} & 1.4" $\times$ 1.5" & COSMOS & \cite{schinnerer2010vla}\\
GMRT \SI{610}{MHz} & \SI{610}{MHz} & \SI{200}{\micro Jy/beam} & 6.5" $\times$ 6.5" & XMM-LSS & \cite{smolvcic2018xxl}\\
\hline
\end{tabular}
}
\end{center}

\end{table*}

\subsubsection{HSC-SSP}

The Hyper Suprime-Cam Subaru Strategic Program \citep[HSC-SSP]{aihara2018hyper} provides deep optical data for both COSMOS and XMM-LSS with multi-band ($g,r,i,z,y$ plus four narrow-band filters) imaging. The survey was carried out by the wide-field camera HSC on the \SI{8.2}{m} Subaru telescope. The data is three-layered (wide, deep, ultradeep), covering an area and depth of about \SI{1200}{deg^2} ($r\sim$26), \SI{27}{deg^2} ($r\sim$27) and \SI{3.5}{deg^2} ($r\sim$28), respectively.

For this work, we utilise the optical wide $i$-band images from the third public data release \citep[PDR3]{aihara2022third}, as well as the photometric redshift catalogue computed from their data from the second public data release \citep[PDR2]{nishizawa2020photometric} to look for host positions and photometric redshifts for the XMM-LSS sample.

\subsubsection{COSMOS2020}

COSMOS2020 \citep{weaver2022cosmos2020} is the latest release of the photometric catalogue for the Cosmic Evolution Survey, building on the previous releases by \cite{capak2007first}, \cite{ilbert2008cosmos,ilbert2013mass}, \cite{muzzin2013public} and \cite{laigle2016cosmos2015}. The catalogue contains source detection with multi-wavelength photometry for over 1.7 million sources, providing two independent photometric redshift estimates (LePhare: \cite{arnouts2002measuring}, \cite{ilbert2006accurate}, EAZY: \cite{brammer2008eazy}) for all sources. For $i<21$ objects, the photometric redshift accuracy is better than 1\%, while the fainter objects $25<i<27$ reach a precision level of 5\%. Where available, we used the COSMOS2020 data to determine the host position and photometric redshifts \citep{weaver2022cosmos2020} for the COSMOS sample in this work.

\subsubsection{WISE}

WISE \SI{3.4}{\micro\meter} images \citep{wright2010wide} were first used to assign preliminary host positions for the XMM-LSS sample before using the higher resolved and deeper HSC-SSP images, and for the host positions for COSMOS sources positioned at the edge of the MIGHTEE-DR1 mosaic, where there is no coverage from COSMOS2020. Since WISE W1 is close to mid-IR, it samples a different galaxy population than HSC-SSP (mentioned below), which both use versions of $grizy$ passbands. Therefore, WISE images are still useful to find fainter galaxies that cannot be observed in the optical passbands of HSC-SSP.

\subsubsection{XMM-LSS}

For the 306 identified extended sources in XMM-LSS, overlays from both MIGHTEE resolutions were produced with background images of WISE W1 and HSC-SSP wide $i$-band to look for the host positions, using the VLASS and GMRT radio data when necessary (see Section~\ref{ancData}). 
For 282 of the 306 sources in XMM-LSS, we could assign a host position ($92\%$). 

\subsubsection{COSMOS}

For COSMOS, we also used the \SI{1.4}{GHz} and \SI{3}{GHz} VLA data, when available, to aid in the search for the correct host. The sample of extended radio sources from the \SI{3}{GHz} VLA data in COSMOS from \cite{vardoulaki_3GHz_FR} was used as a reference for the COSMOS sample in this work. We note that MeerKAT is sensitive to extended diffuse emission, due to the short baselines in its core, while the VLA resolves out some extended emission, causing extended sources to be missed (often only the compact parts are detected). Because of the difference in sensitivity and coverage of the MIGHTEE COSMOS data, visual inspection of the MIGHTEE data still yielded many extended radio sources that were previously not at the \SI{3}{GHz} data. We found 20 extended sources with jets from diffuse emission in MIGHTEE that is either not detected by the VLA or is at the noise level of the \SI{3}{GHz} survey. In some cases, the extended sources found in MIGHTEE could be seen in the \SI{3}{GHz} data by going below the $3\sigma$ noise level. This was a useful tool for the source characterisation in the COSMOS sample, as the \SI{3}{GHz} data at $1\sigma$ showed peaked emission along the jets and hotspots in the lobes, which is hidden in the noise. On the other hand, out of the 108 extended radio sources with bending angle presented in \cite{vardoulaki_3GHz_FR}, 48 ($44\%$) are not found in our sample from visual inspection of the MIGHTEE data. This is because the resolution of the MIGHTEE-DR1 data is insufficient to resolve the jet structures and substructures of radio sources $\lesssim$ 20". At redshift 1, the 5" beam size of MIGHTEE corresponds to $\approx \SI{40}{kpc}$, while the 0".75 beam of \SI{3}{GHz} VLA resolves $\approx \SI{6}{kpc}$. For a redshift of 2, these beam sizes correspond to $\approx \SI{42}{kpc}$ and $\approx \SI{6.3}{kpc}$, respectively.

By using the multi-wavelength data (optical, infrared and radio) and visual inspection we were able to assign a host position for 193 out of 254 radio sources in the COSMOS field ($76\%$). In COSMOS, many extended radio structures turned out to be blended point-like sources when analysed with multi-wavelength data, thus resulting in a lower host association percentage when compared to the XMM-LSS sample.

\subsection{Redshifts}

For both the XMM-LSS and COSMOS sample, we use spectroscopic redshifts provided by the HSC-SSP PDR3 data access website\footnote{\url{https://hsc-release.mtk.nao.ac.jp/doc/index.php/catalog-of-spectroscopic-redshifts__pdr3}}, which offers a collection of public spectroscopic redshift surveys. The spectroscopic redshifts surveys are from PRIMUS \citep{coil2011prism,cool2013prism}, VIPERS \citep{garilli2014vimos}, SDSS \citep{alam2015eleventh,ahumada202016th}, UDSz \citep{bradshaw2013high,mclure2013sizes}, GAMA \citep{liske2015galaxy}, 6dFGRS \citep{jones20096df}, VVDS \citep{fevre2013vimos}, VANDELS \citep{pentericci2018vandels}, DEIMOS-10k \citep{hasinger2018deimos}, 2dFGRS \citep{colless20032df}, zCOSMOS \citep{lilly2009zcosmos}, 3D-HST \citep{skelton20143d,momcheva20163d}, FMOS \citep{silverman2015fmos}, WiggleZ \citep{drinkwater2010wigglez}, DEEP2 \citep{newman2013deep2}, DEEP3 \citep{cooper2011deep3}, C3R3 \citep{masters2017complete,masters2019complete} and LEGA-C \citep{straatman2018large}. In addition, we use the spectroscopic redshifts from the IMACS survey \citep{kelson2014carnegie} in COSMOS. We also utilise the near position search from the NASA/IPAC Extragalactic Database\footnote{\url{https://ned.ipac.caltech.edu/}} (NED) for a handful of sources to obtain a spectroscopic redshift value. Spectroscopic redshifts are available for 47\% of the XMM-LSS sample (89 out of 189 sources) and for 34\% of the COSMOS sample (39 out of 116 sources).

For photometric redshifts in XMM-LSS, we use the Mizuki photometric redshift wide catalogue from the HSC-SSP second public data release \cite[][]{nishizawa2020photometric}, which uses template fitting with Bayesian priors on physical properties of galaxies to compute the most probable redshift, and which completely covers the area of XMM-LSS in the MIGHTEE-DR1. Only objects that have been observed with at least three bands are included in the catalogue and we only consider redshifts that have a reduced $\chi^2_{\nu}<5$ from the best-fit model \citep{nishizawa2020photometric}. We have investigated the photometric redshift catalogue of \cite{Hatfield2022} and compared to the Mizuki photometric redshifts. We find that the redshifts agree within the errors in most cases ($>$95\%). Inside galaxy groups, which is important for this study, there is no difference. 

For COSMOS, we use the photometric redshifts from COSMOS2020 computed with LePhare \citep{weaver2022cosmos2020}, if secure spectroscopic redshift were not available. We note that the quality of the photometric redshifts outside the region that is covered by UltraVISTA is worse compared to the inner region of the field. Because the sources from outside the UltraVISTA region are missing the $YJHK_s$ bands and are only selected from $i$ and $z$ bands, we expect to lose redshift accuracy as the optical rest-frame emission from galaxies gets redshifted into the near-IR range not visible in $i$ and $z$ bands at higher redshifts.

By comparing the values of photometric and spectroscopic redshifts available from the photometric catalogues, we calculate the median accuracy of the photometric redshifts for both samples: 
the photometric redshift accuracy for COSMOS\footnote{This value agrees with the photometric redshift accuracy reported in \cite{laigle2016cosmos2015}.}  is $(z_s-z_p)/(1+z_s) = 0.007$ , and for XMM-LSS  $(z_s-z_p)/(1+z_s) = 0.018$. The photometric precision of the COSMOS2020 catalogue is 1\% at $i \approx 20$AB and 4\% at $i \approx 26$AB \citep{weaver2022cosmos2020}. For the XMM-LSS HSC-SSP photometric redshift catalogue the photometric accuracy is $\approx 3\%$ \citep{nishizawa2020photometric}. 

\subsection{Samples}

For the upcoming analysis of bent radio sources in Section \ref{Results}, we reduce our samples to only include sources where it is possible to measure a bending angle, given the MIGHTEE-DR1 data at hand. We exclude sources without a host association, which we need for assigning the bending angle, the initial number of sources is reduced from 306 to 282 for XMM-LSS, and from 254 to 193 for COSMOS. We only include radio sources with two-sided jets/lobes where measuring the bending angle is possible. For the rest of the analysis, we will only take into account these sub-samples of objects for which we can securely measure their bending angles, i.e. including only two-sided radio AGN for which the BA could be measured reliably. The final sample for XMM-LSS includes 217 sources, 189 ($87\%$) of which have redshift available. The final sample for COSMOS includes 142 sources, 116 ($82\%$) of which have redshift available (see Table \ref{tab:subsamples}).

\begin{table}[h!]
\begin{center}
\caption[]{Sample size overview. The succeeding line is always a sub-sample of the preceding line. The last line presents the final sample used in this analysis}
\label{tab:subsamples}
\scalebox{1}{
\renewcommand{\arraystretch}{1.5}
\begin{tabular}[t]{l c c}
\hline
Sample & XMM-LSS & COSMOS\\
\hline
From visual inspection & 306 & 254\\
With host & 282 & 193\\
With bending angle & 217 & 142\\
With redshift & 189 & 116\\
Within X-ray coverage & 183 & 76\\
Inside X-ray Galaxy Groups & 17& 19\\
\hline
\end{tabular}
}
\end{center}

\end{table}

\subsubsection{X-ray}
\label{X-ray Data}
We cross-match our radio samples with the X-ray galaxy groups in the XMM-LSS and COSMOS fields. As a proxy of the environment, we take the X-ray galaxy groups' mass and temperature (see Section~\ref{BA environ}).

For the COSMOS field, we use X-ray galaxy groups identified by \emph{XMM-Newton} and Chandra in the 0.5-\SI{2}{keV} band \citep[and in preparation]{gozaliasl2019chandra}, which provides coverage for the central $\sim\SI{2.3}{deg^2}$ of the MIGHTEE-DR1 COSMOS mosaic with robust group identification up to a redshift of $\sim$2. The X-ray galaxy group catalogue features 322 groups with group masses M$_{200}$ ranging from $\SI{4e12}{M_\odot}$ to $\SI{3e14}{M_\odot}$, obtained with the X-ray luminosity $L_X$ halo mass, $L_X -M_{200}$ scaling relation \citep[][]{leauthaud2009weak}. Similarly, the mean group temperature $T$ was calculated with the $L_X -T$ scaling relation \citep[][]{finoguenov2007xmm}. The X-ray flux limit for the 0.5-\SI{2}{keV} band is \SI{3e-16}{erg s^{-1} cm^{-2} s^{-1}}.

\begin{figure*}[ht!]
\centering
\includegraphics[width=\textwidth]{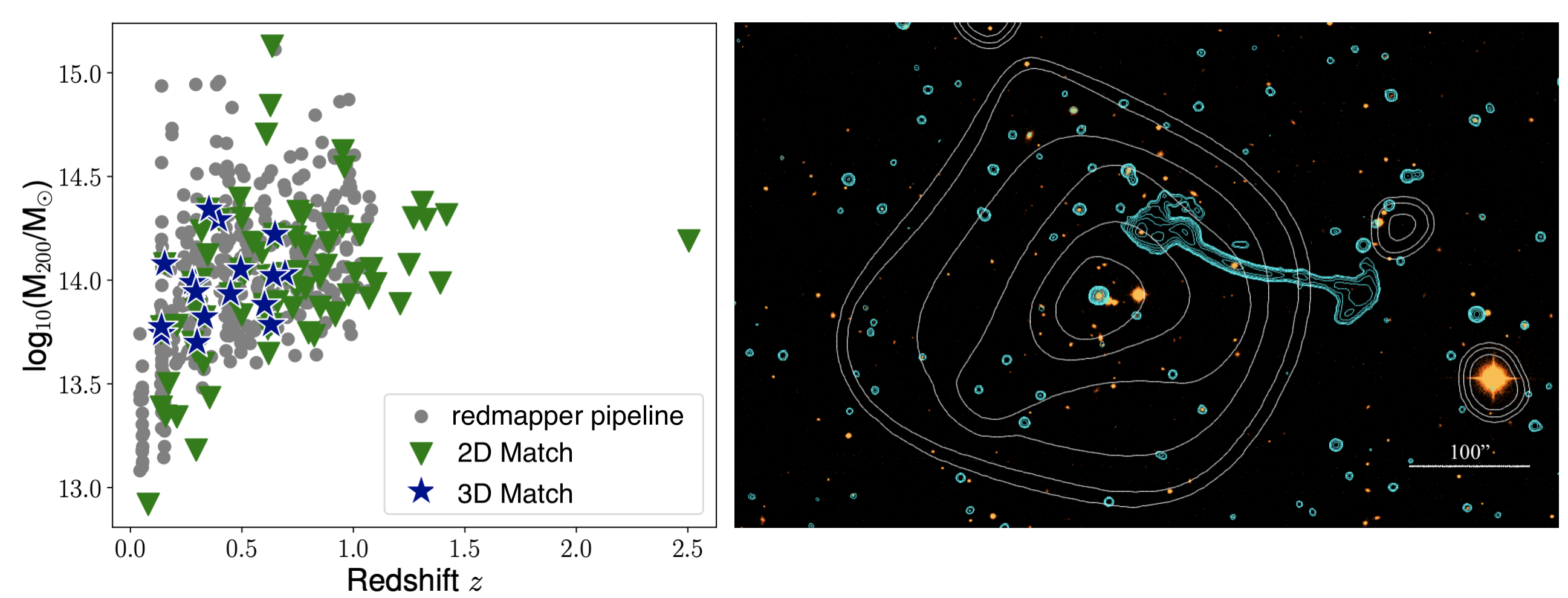}
\caption{(Left): X-ray galaxy group masses M$_{200}$ as a function of redshift in XMM-LSS, calculated with redMaPPer (grey circles), calculated with the redshift of sources in the XMM-LSS sample that visually coincide with the 0.5-\SI{2}{keV} extended X-ray emission (green triangles, 2D Match) and from cross-matching the X-ray galaxy groups from the "2D Match" to the ones from redMaPPer, given $\Delta z\leq0.018$ (blue stars, 3D Match). (Right): Example of a radio galaxy spatially coinciding with extended X-ray emission. The background is the SDSS $i$-band data, overlaid with MIGHTEE 8.9" radio data at $3\sigma$ (cyan contours) and 0.5-\SI{2}{keV} extended X-ray emission (white contours). A scale of 100 arcsec in length is shown on the bottom right. This corresponds to 267 kpc at the redshift of the source 178, $z$ = 1.54.}
\label{fig:xrayXMMLSS}
\end{figure*}

To determine which galaxies in the COSMOS sample are X-ray galaxy group members, we search for all sources that are located within the virial radius $r_{200}$ of the galaxy groups in COSMOS. This radius defines a sphere with an interior mean gas density 200 times the critical density of the Universe at the redshift of the group. We then check if the redshift of the radio galaxy $z_{galaxy}$ from our sample is at the same redshift range as the redshift of the galaxy group $z_{group}$, given by the redshift accuracy of the COSMOS sample: $\Delta z = (z_{galaxy}-z_{group})/(1+z_{group}) \leq 0.007$. Finally, we cross-match the R.A. and Dec. of all host galaxies that fulfill $r\leq r_{200}$ and $\Delta z\leq 0.007$ with the R.A. and Dec. of the known group members within 1". From this procedure, we find that, out of the 76 radio sources of the COSMOS sample which are inside the \emph{XMM-Newton} and Chandra coverage, 19 are X-ray galaxy group members (25\%).

For the XMM-LSS sample, we utilise \emph{XMM-Newton} data from the 0.5-\SI{2}{keV} band that roughly covers the northern two-thirds of the MIGHTEE-DR1 XMM-LSS mosaic. We analysed all \emph{XMM-Newton} observations in overlap with the radio data, that became public prior to 2023. We used XMMSAS\footnote{https://www.cosmos.esa.int/web/xmm-newton/sas} version 21.0.0 for the initial data reduction. For the XMM data screening we followed the prescription outlined in \cite{finoguenov2007xmm} on data screening and background evaluation, with updates described in \cite{bielby2010wircam}. To detect and study faint extended sources, we first remove the flux produced by the point sources, following \cite{finoguenov2009roadmap}. We detect the extended emission in the 0.5-2 keV mosaic image using the wavelet scales from 0.5 to 2 arcminutes. To identify X-ray galaxy groups, redMaPPer \citep{rykoff2014redmapper} is run in scanmode \citep[e.g. as in][]{ider2020cosmological,Kluge2024} was employed (grey circles in Figure~\ref{fig:xrayXMMLSS}), which utilised the photometric data from the 10th Data Release of the DECam Legacy Survey \citep[DECaLS,][]{dey2019overview} and version 8 of the red-sequence Matched-filter Probabilistic Percolation cluster-finding algorithm code \citep[redMaPPer,][]{rykoff2014redmapper}. In contrast to the COSMOS field, which offers excellent spectrophotometric coverage, the quality of the photometric redshifts from DECaLS is insufficient to ensure robust group identification. We therefore visually confirmed which radio sources in the XMM-LSS field are located within the extended X-ray emission from the 0.5-\SI{2}{keV} \emph{XMM-Newton} data (e.g. right panel of Figure~\ref{fig:xrayXMMLSS}) and calculated rough group properties using the redshifts of the host galaxies in our sample.

We perform a 2D spatial match between the locations of the radio sources in XMM-LSS and the X-ray extended data from \emph{XMM-Newton} (green triangles in Figure~\ref{fig:xrayXMMLSS}) and find that 79 out of the 183 ($43\%$) radio sources of our bending angle sample lie within the X-ray coverage. By matching also in redshift space, using the extended X-ray sources from the redMaPPer pipeline, we find that 17 out of 183 radio sources (9\%) lie inside X-ray galaxy groups (blue stars in Figure~\ref{fig:xrayXMMLSS}). We note, that for a radio source to be considered an X-ray galaxy group member, the redshift of the host galaxy must lie within the redshift range of the extended X-ray source calculated from the redMaPPer pipeline, given by the redshift accuracy of the XMM-LSS sample: $\Delta z = (z_{galaxy}-z_{redMaPPer})/(1+z_{redMaPPer}) \leq 0.018$. Additionally, we excluded all group members associated with a group with richness $\lambda<10$. 

\begin{figure}[h!]
\centering
\begin{subfigure}{.9\columnwidth}
  \centering
  \includegraphics[width=\linewidth]{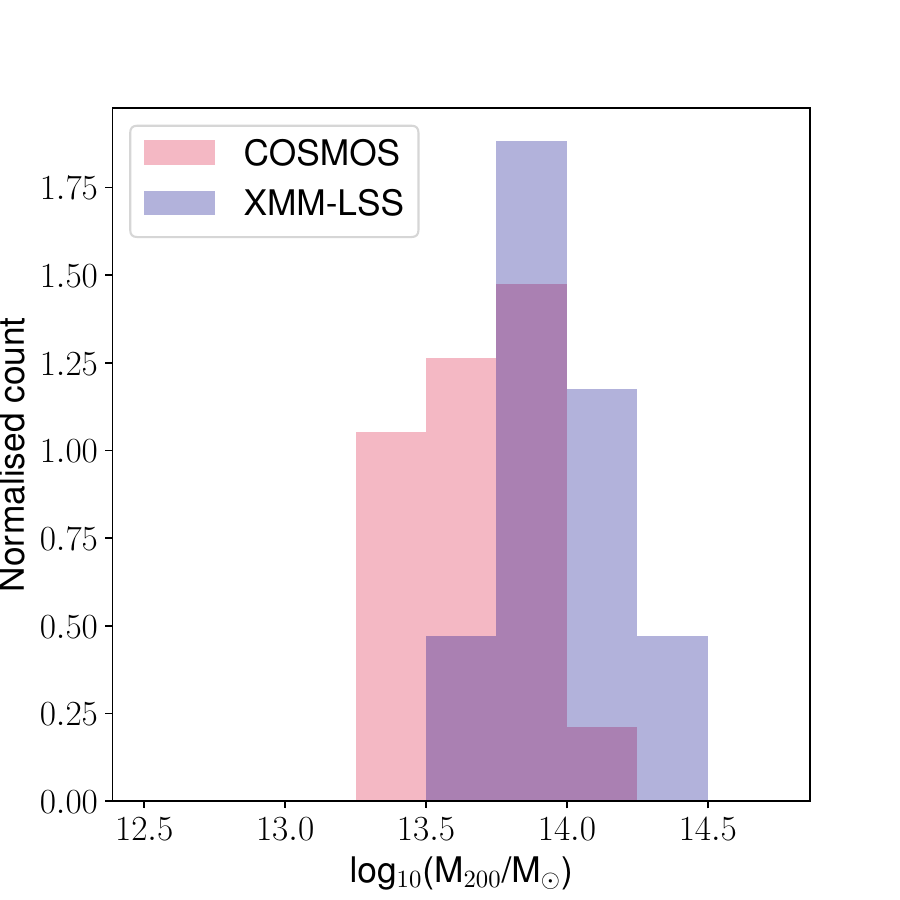}
\end{subfigure}\\
\begin{subfigure}{.9\columnwidth}
  \centering
  \includegraphics[width=\linewidth]{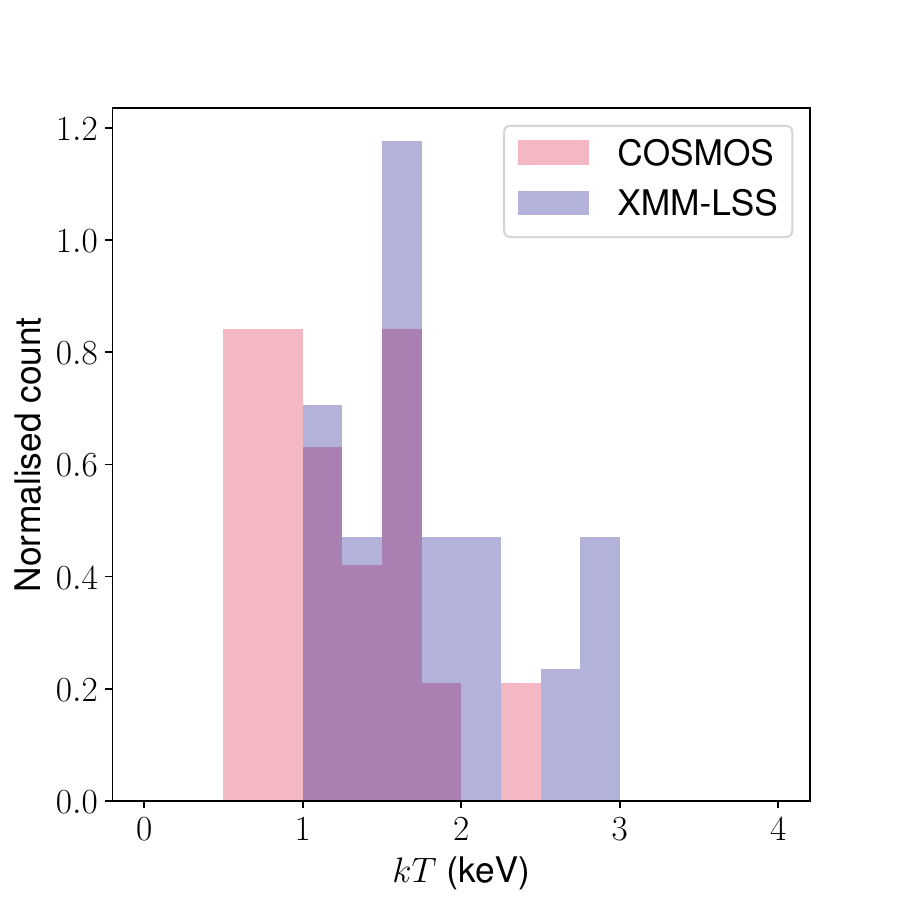}
\end{subfigure}
\caption{Normalised count of the 19 extended radio AGN inside X-ray galaxy groups found in COSMOS \citep[and in prep.]{gozaliasl2019chandra} and the 17 galaxy groups in XMM-LSS associated with extended radio AGN and presented in Section \ref{BA environ}. (Top): Group masses M$_{200}$ in M$_{\odot}$. The bin size is $0.25 \times$ log$_{10}$(M$_{200}$/M$_{\odot}$). (Bottom): Group temperatures $kT$ in keV. The bin size is \SI{0.25}{keV}.}
\label{fig:kTM200Histo}
\end{figure}

The halo masses M$_{200c}$ for the 17 members in the XMM-LSS sample range from $\SI{5e13}{M_\odot}$ to $\SI{2e14}{M_\odot}$. The group with the lowest flux is found at \SI{4e-15}{erg s^{-1} cm^{-2} s^{-1}}. Redshifts range from 0.34 to 0.7. The properties of the X-ray galaxy group of XMM-LSS and COSMOS are shown in Figure~\ref{fig:kTM200Histo} and discussed further in Section~\ref{Discussion}. 

We note that objects inside the X-ray coverage which are not members of galaxy groups might lie in mass halos below $<1.5(1 + z) \times 10^{13} M_{\odot}$ (for $z > 1$), not probed by our current X-ray data \citep[see][]{gozaliasl2019chandra}.

\section{Methods}
\label{methods}

\subsection{Largest Angular Size}

To obtain the values for largest angular size for each source in our samples, we add the angular distances between the edges of the $3\sigma$ contours of the 5" MIGHTEE data and the host position. For sources that have no clear lobe structure or no host information, we use the distance from edge to edge of the $3\sigma$ contours. We chose the 5"  MeerKAT map over the 8.9" one, for more accurate measurements and to reduce blending effects. 

Using the 5" MIGHTEE data for angular size determination comes at the cost of potentially missing diffuse emission picked up by the more sensitive 8.9" data. We find that on average the difference in angular size between the same objects of the two radio maps is $\approx$ 8", close to the 8.9" MIGHTEE beam size, suggesting that the difference is related to the beam. Thus we are confident in using the angular size measurements from the 5" map for our analysis.

\subsection{The bending angle}
\label{Method: bending angle}

Building on previous studies \citep[e.g.][]{silverstein2017increased,vardoulaki_3GHz_FR,vardoulaki2021bent,garon2019radio,golden2021high}, we use the bending angle (BA), defined as the angle between the jets or lobes of two-sided radio sources, to study the distortion of the jet structure in relation to their large-scale environment. The bending angle thus provides a quantitative way to measure the deviation from a straight line.

A completely straight source corresponds to a bending angle of $\SI{180}{\degree}$, while bent sources have an angle $<\SI{180}{\degree}$. We do not distinguish between upwards and downwards bending in the projected plane to the observer. Thus, the bending angle is a positive definite quantity between \SI{0}{\degree} and $\SI{180}{\degree}$, where radio sources are more bent the closer the bending angle is to \SI{0}{\degree}.

We measure the bending angle of each source in two ways: The first method, which we call the \emph{peak flux method}, measures the angle between the vectors that originate in the host position and go to the peak flux in each jet, choosing the brightest pixel as the endpoint. We define the peak flux as the hotspot in lobe structures, as typically seen in FRII-type sources, or as the peak surface brightnesses, which are typically close to the centre in FRI-type jets. The second method, called the \emph{edge method}, measures the angle between the two vectors going from the host position to the end of each $3\sigma$ contour, where the pixel that maximises the length of each vector is chosen at the $3\sigma$ contour. The quantitative differences between the two methods are presented in Section \ref{Results}. In the following section we discuss the usefulness and limitations of the bending angle methods as well as qualitative differences between the two methods of obtaining the bending angle.

\subsection{The Efficacy of the bending angle}
\label{efficacy BA}

We first want to address the limitations of using the bending angle, as described in Section \ref{Method: bending angle}, as a method to investigate the distortion of jetted radio AGN. Similarly to the angular size, the measurement of the bending angle is affected by projection effects. Because we can only see radio sources as 2D projections on the sky, we cannot accurately account for the true shape the radio source has in 3D space. For example, a galaxy with a large bending angle seen from earth could be seen as having a small bending angle for an observer from another direction as the projected distance and angle between the jets change. This has nothing to do with the intrinsic or extrinsic properties of the radio source, but is of purely geometrical nature, limited by the line-of-sight of the observer. An argument can be made that in an isotropic Universe, the error of the bending angle due to projection effects will average out over a large enough sample. An important consequence of this is that the bending angle is better suited for a statistical analysis over large samples rather than a source-by-source approach. For this reason, we refrain from making strong statements about the most bent sources of our samples in our analysis and mostly distinguish between straight or slightly bent sources (BA $>160^{\circ}$) and moderately or very bent sources (BA $\leq160^{\circ}$) in our statistics. The very bent sources (BA $\leq100^{\circ}$) are discussed in more detail in the appendix. The value of $160^{\circ}$ was chosen to allow comparisons to the literature (see Section~\ref{Discussion}).

Another source of uncertainty from geometrical arguments is that positional errors of the peak surface brightness (or edge position) that is used to determine the bending angle will result in bigger errors for the bending angle the closer the peak surface brightness (or edge position) is to the core. This is a concern for radio galaxies with small angular size, where small positional changes will result in larger changes of the bending angle. We therefore investigate the relation of size and bending angle. While the sources larger than \SI{1}{Mpc} in our samples are typically not bent below $130^{\circ}$, we find no correlations between bending angle and angular or linear size. We should note that large sources (>\SI{1}{Mpc}) are rare, and consequently, there are very few in our samples that cover small sky areas. Nevertheless, we do not observe a correlation between linear projected size and bending angle.

As we showed in Section~\ref{DataSection}, the characterisation of extended radio sources can be highly dependent on the radio survey's resolution, sensitivity and frequency. For this reason, we measure the bending angle both from edges to radio core (\emph{edge method}) and peak fluxes to radio core (\emph{peak flux method}) for each source where possible. Using the \emph{edge method} allows to include more diffuse emission from FRI-type sources, which are subject to interaction with the environment. While sources with a typical FRII morphology will not show much difference between the two methods of obtaining the bending angle, FRI sources can show a stark difference between the two methods. This is because FRI sources have their peak surface brightness closer to the radio core, while their extended, diffuse emission can be subjected to deformation due to environmental effects. An advantage of measuring the bending angle with the \emph{peak flux method} is that we expect it to be less affected by the selection effects related to observed frequency, sensitivity, and angular resolution of the survey, since the positions of the peak surface brightness should not change greatly over different radio data sets, in the case of well-defined radio jets. Nevertheless, for surveys with sub-arcsecond resolution, like the \SI{3}{GHz}-COSMOS \citep{smolvcic2017vlaSource} or LOFAR-VLBI observations \citep{Sweijen2022}, changes in the peak surface brightness are observed between $\sim$ 6" and sub-arcsecond resolutions. We note that with the \emph{edge method} we should expect differences between low frequency and high frequency observations, with the former probing more extended and diffuse emission. Different telescope baselines will also have an effect on this. The bending angles measured from the two methods follow the same distribution for both fields, suggesting that the bending angle is statistically consistent between the two methods. The median of the absolute deviations between the angles of the \emph{peak flux method} and the \emph{edge method} are $5^{\circ}$ and $6^{\circ}$ for XMM-LSS and COSMOS, respectively.

\section{bending angles: Analysis and Results of Observational XMM-LSS and COSMOS MIGHTEE Data}
\label{Results}

For sources where we can measure the bending angle with both methods, we calculate $\Delta BA=(BA_{PeakFlux}-BA_{Edge})/(1+BA_{Edge})$ and find $\Delta BA<0.01$ for 20\% for all sources, $\Delta BA<0.1$ for $\sim$ 84\% of sources in XMM-LSS, as well as $\Delta BA<0.1$ for $\sim$ 78\% of sources in COSMOS. The median values for the objects with bending angles from the two methods are listed in Tables \ref{tab:BA_Table_flux} and \ref{tab:BA_Table_edge}. The scatter of the median is given by the 16th and 84th percentile. We note that the number of objects in Tables \ref{tab:BA_Table_flux} and \ref{tab:BA_Table_edge} differ because it was not possible to use both methods of BA measurement on all objects (e.g. lack of prominent peak flux for the peak flux method measurement).

\begin{table}[h!]
\begin{center}
\caption{Median bending angles from the \emph{peak flux method}. Here we present all sources that we could measure a bending angle, including those outside X-ray galaxy groups.}
\label{tab:BA_Table_flux}
\scalebox{1}{
\renewcommand{\arraystretch}{1.5}
\begin{tabular}[t]{l c c c c}
\hline
\textbf{Sample} & \textbf{$N$} &  \multicolumn{3}{c}{\textbf{bending angle (deg.)}}\\
 &   & \textbf{Median$^{84\%}_{16\%}$} & \textbf{Min} & \textbf{Max}\\
\hline
XMM-LSS & 214 & $168.0^{175.0}_{146.1}$ & 55 & 180\\
COSMOS & 112 & $168.0^{175.2}_{141.3}$ & 46 & 180\\
Combined & 326 & $168.0^{175.0}_{145.0}$ & 46 & 180\\
\hline
\end{tabular}
}
\end{center}

\end{table}

\begin{table}[h!]
\begin{center}
\caption{Median bending angles from the \emph{edge method}. Here we present all sources that we could measure a bending angle, including those outside X-ray galaxy groups.}
\label{tab:BA_Table_edge}
\scalebox{1}{
\renewcommand{\arraystretch}{1.5}
\begin{tabular}[t]{l c c c c}
\hline
\textbf{Sample} & \textbf{$N$} &  \multicolumn{3}{c}{\textbf{bending angle (deg.)}} \\
 &  & \textbf{Median$^{84\%}_{16\%}$} & \textbf{Min} & \textbf{Max}\\
\hline
XMM-LSS & 217 & $168.0^{176.0}_{138.6}$ & 16 & 180\\
COSMOS & 142 & $166.0^{177.0}_{135.9}$ & 45 & 180\\
Combined & 359 & $167.5^{176.0}_{137.0}$ & 16 & 180\\
\hline
\end{tabular}
}
\end{center}

\end{table}

The median values from Tables \ref{tab:BA_Table_flux} and \ref{tab:BA_Table_edge} show that the bending angles from the two methods yield similar median and scatter values. A Kolmogorov-Smirnov test (K-S test) with a significance level of 0.05 confirms that the bending angles from the two methods come from the same distribution for both samples (XMM-LSS: K-S statistic $=0.1$, $p$-value $=0.25$; COSMOS: K-S statistic $=0.09$, $p$-value $=0.77$). In Table \ref{tab:BA_Table_bentness}, we show the number of sources in each sample that are straight or slightly bent (BA $>160^{\circ}$), moderately bent ($100^{\circ}<$ BA $<160^{\circ}$) and very bent (BA $\leq100^{\circ}$), as well as their median BA values. We find that for both samples, well over $50\%$ of sources are straight or slightly bent, with only a few very bent sources in each sample. For the rest of this work, we will use the bending angles obtained from the \emph{edge method} unless stated otherwise.

\begin{table}[h!]
\begin{center}
\caption{Degree of bending for sources in the XMM-LSS and COSMOS samples. Here we present all sources that we could measure a bending angle, including those outside X-ray galaxy groups.}
\label{tab:BA_Table_bentness}
\scalebox{0.8}{
\renewcommand{\arraystretch}{1.3}
\begin{tabular}[t]{l | c c c c}
\hline
 & \multicolumn{2}{c}{\textbf{XMM-LSS}} & \multicolumn{2}{c}{\textbf{COSMOS}}\\
 & \textbf{N} & \textbf{Median$^{84\%}_{16\%}$} & \textbf{N} & \textbf{Median$^{84\%}_{16\%}$}\\
\hline
\textbf{Straight/slightly bent}  & 139 & $174.0^{177.0}_{167.0}$ &  88  & $174.0^{178.0}_{167.0}$\\
(BA $>160^{\circ}$) & ($\sim$ 64\%) & & ($\sim$ 57\%)\\
\textbf{Moderately bent}  & 71 & $144.0^{155.0}_{120.4}$ & 54 & $147.0^{158.0}_{120.5}$\\
($100^{\circ}<$ BA $<160^{\circ}$) & ($\sim$ 33\%) & & ($\sim$ 38\%)\\
\textbf{Very bent} & 7 & $83.0^{97.0}_{22.7}$ &6 & $63.0^{73.4}_{50.6}$\\
(BA $\leq100^{\circ}$) & ($\sim$ 3\%)& & ($\sim$ 4\%)\\
\hline
\end{tabular}
}
\end{center}

\end{table}

\subsection{bending angle vs Large-Scale Environment}
\label{BA environ}

To investigate the relation between bending angle and large-scale environment, we cross-correlate the sources in our sample to the X-ray galaxy groups (see Section~\ref{X-ray Data}) with the aim to find relations between bending angle and group properties, such as group mass and temperature, and to understand the role the large-scale environment probed by galaxy groups plays to shaping the radio structure of extended radio AGN. We constrain the radio-source sample to the same area coverage as the X-ray observations, which cover $\sim$\SI{2.3}{deg^2} in COSMOS and $\sim$\SI{7}{deg^2} in XMM-LSS.

We find a trend with redshift for objects that are members of X-ray galaxy groups, that is, a larger number of bent sources (BA $\leq 160^{\circ}$) at lower redshifts. We apply a halo mass cut of $log_{10}(M_{200}/M_{\odot}) > 13.5$, to probe the same group population at all redshifts \citep[see][]{vardoulaki2023evolution}, and find that in the COSMOS sample at $z \leq$ 0.5 we have $<BA> = 141^{\circ} \pm 39^{\circ}$ (5 objects), while at $z > 0.5$ the $<BA> = 154^{\circ} \pm 16^{\circ}$ (2 sources). For XMM-LSS, at $z \leq$ 0.5, we have $<BA> = 95^{\circ} \pm 49^{\circ}$ (7 objects), while at $z > 0.5$ the $<BA> = 130^{\circ} \pm 8^{\circ}$ (2 sources). Below we discuss separately the bent sources in COSMOS and XMM-LSS. Accounting for very bent sources ($<100^{\circ}$), these are located at $z \leq$ 0.5 in both samples. Only source 252 in COSMOS is below the halo mass cut, while the other very bent source in COSMOS (source 247) and the two in XMM-LSS (sources 1 and 200) are above the cut.

\subsubsection{Group Members in COSMOS}

We find 19 ($25\%$) bent sources inside X-ray galaxy groups and 57 outside (see Table~\ref{tab:subsamples}). For the 19 sources in the COSMOS sample that are inside X-ray groups, we find that the median bending angle (with the 16th and 84th percentile) is $156.0^{171.1}_{117.7}$ degrees, while the median bending angle for the 57 sources that are not considered group members and are in the same area coverage as \emph{XMM-Newton} and Chandra is $168.0^{177.0}_{140.8}$ degrees.

\begin{figure}[h!]
\centering
\includegraphics[width=0.83\linewidth]{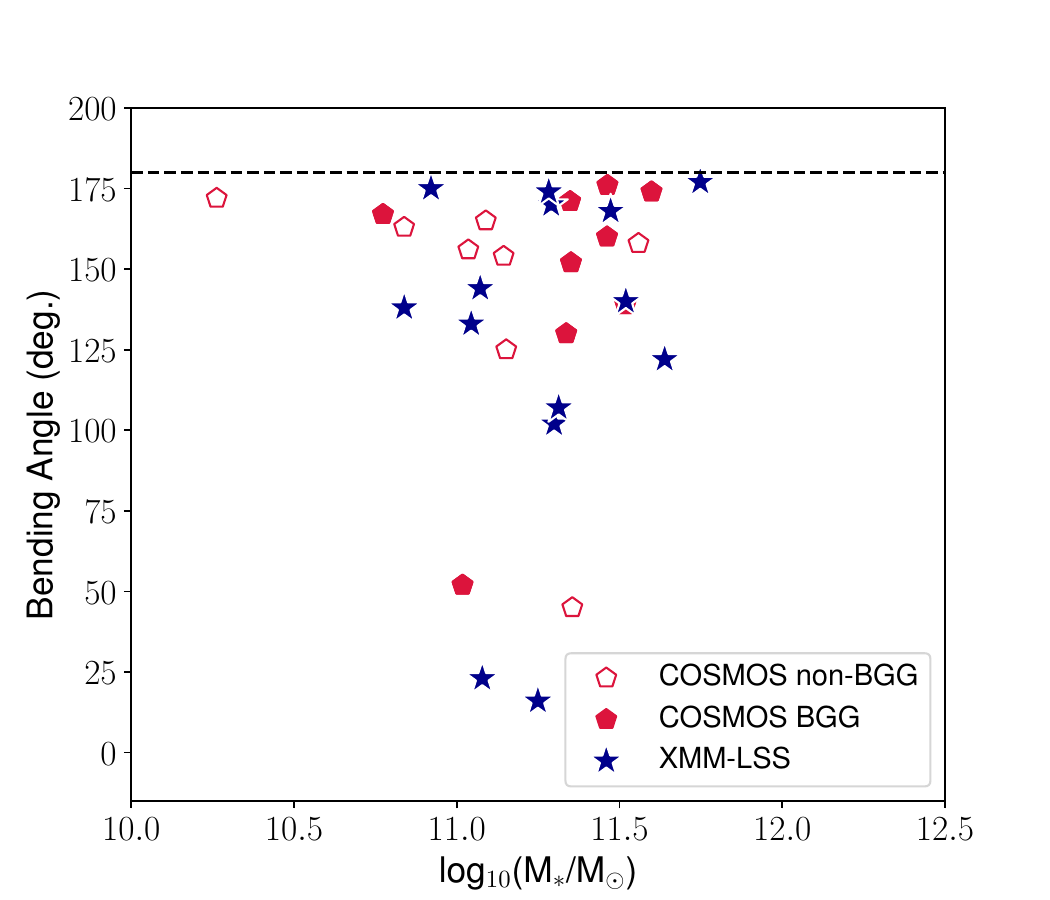}
  \includegraphics[width=0.83\linewidth]{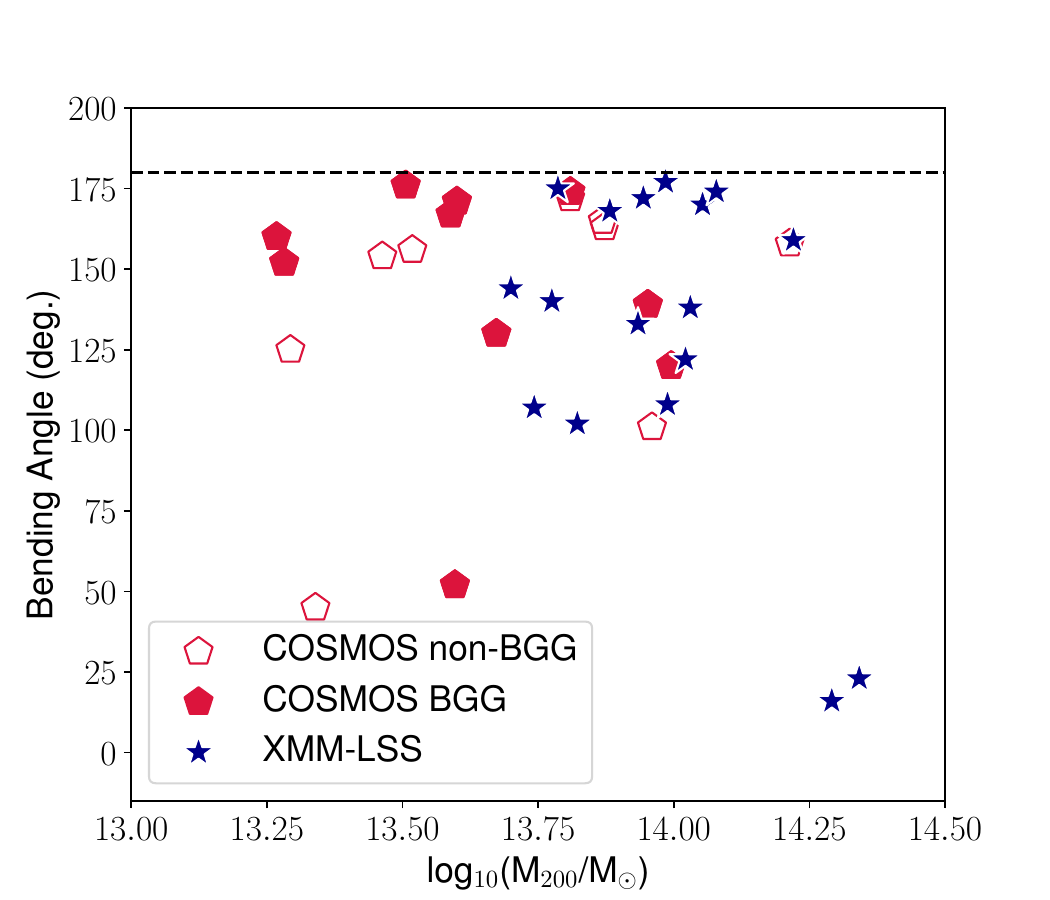}
\includegraphics[width=0.83\linewidth]{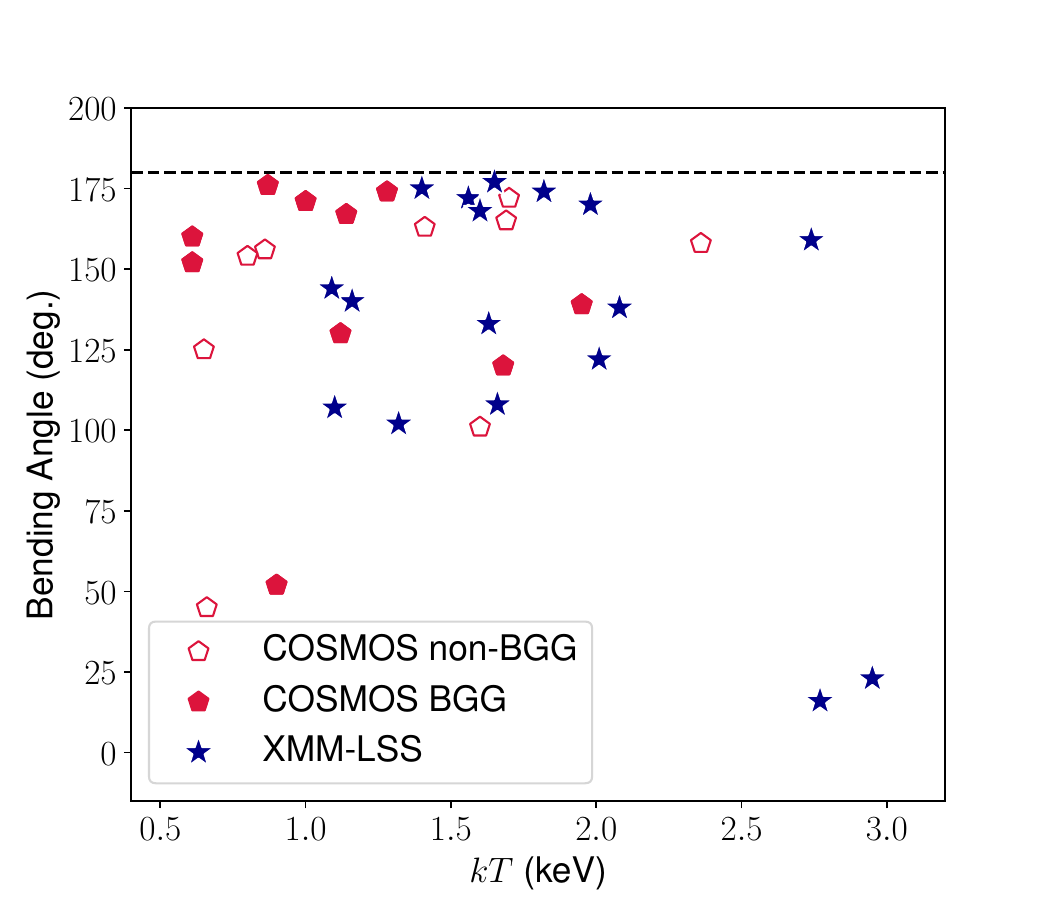}
\caption{{bending angle in degrees as a function of the stellar mass M$_{\star}$ (top), of the X-ray galaxy group mass M$_{200}$ (middle), and of X-ray galaxy group temperature $kT$ in keV (bottom). In all panels, red pentagons denote COSMOS objects and filled blue stars denote XMM-LSS objects. In the COSMOS sample, BGGs are shown as filled symbols. For the XMM-LSS sample, the BGG information is not available at the time of writing due to the different methods the X-ray groups were defined (see Section~\ref{X-ray Data}). The dashed line at 180$^{\circ}$ indicates a straight source.}}
\label{fig:BA_Mstar_M200_kT_COSMOS_XMM-LSS}
\end{figure}

At the top panel of Figure \ref{fig:BA_Mstar_M200_kT_COSMOS_XMM-LSS}, we plot the bending angle for radio sources in the COSMOS X-ray galaxy groups in relation to host stellar mass as red pentagons. The stellar mass was obtained from \cite{gozaliasl2019chandra}. We do not see a correlation between bending angle and stellar mass for the COSMOS X-ray galaxy group members, although bent and very bent sources have M$_{\star}>10^{11}$ M$_{\odot}$. Brightest group galaxies (BGGs) are highlighted by filled-out symbols. The 19 members from our sample tend to occupy the high stellar mass end at their respective redshift. This is expected since radio AGN are more likely to be hosted by more massive galaxies \citep[e.g.][]{magliocchetti2022hosts}.

At the middle panel of Figure \ref{fig:BA_Mstar_M200_kT_COSMOS_XMM-LSS}, we plot the bending angles of X-ray group members in COSMOS against the corresponding halo mass, expressed in terms of M$_{200}$, which is the mass of the group inside the virial radius $r_{200}$. We see no clear trend between group mass and the bending angle, possibly due to the low sample size of 19 objects. Also, we do not observe a significant difference between the bending angles of BGGs and non-BGGs.

We also see that there is no source in COSMOS, for which we could robustly measure a bending angle, located in a galaxy group beyond a redshift of 1.2. This is likely because of the low numbers of high redshift sources in our sample and the low number of high redshift X-ray galaxy groups. We acknowledge the small sample sizes for some subsets (e.g., high-redshift or very bent sources in galaxy groups). Expanding the dataset to additional fields (e.g. ELAIS-S1) or wider sky areas (assuming a wealth of multi-wavelength observations) could address this limitation, but this is out of the scope of the current work. Nevertheless, literature studies, show that bent sources in clusters exist at high redshifts. For example, the study of \cite{golden2021high} in an area of \SI{300}{deg^2}, finds 36 bent radio sources in clusters up to $z\sim2.2$. Furthermore, \cite{hale2018clustering} suggest that AGN could occupy less massive groups at $z>1$, which require high sensitivity X-ray observations \citep[also see][for further discussion]{vardoulaki2023evolution}.

The bottom panel in Figure \ref{fig:BA_Mstar_M200_kT_COSMOS_XMM-LSS} shows the bending angle as a function of the mean group temperature $kT$ in keV. We find no strong correlation between the bending angle and $kT$. Very bent sources show low temperatures, while there is a lack of very bent sources at higher temperatures. The Spearman test between bending angle and temperature gives only a correlation coefficient of r$_s=0.19$ with a $p$-value of 0.44. This corresponds to a weak to no correlation with no evidence to reject the null hypothesis, suggesting that the correlation is not physical.

\begin{figure}[h!]
\centering

\includegraphics[width=\linewidth]{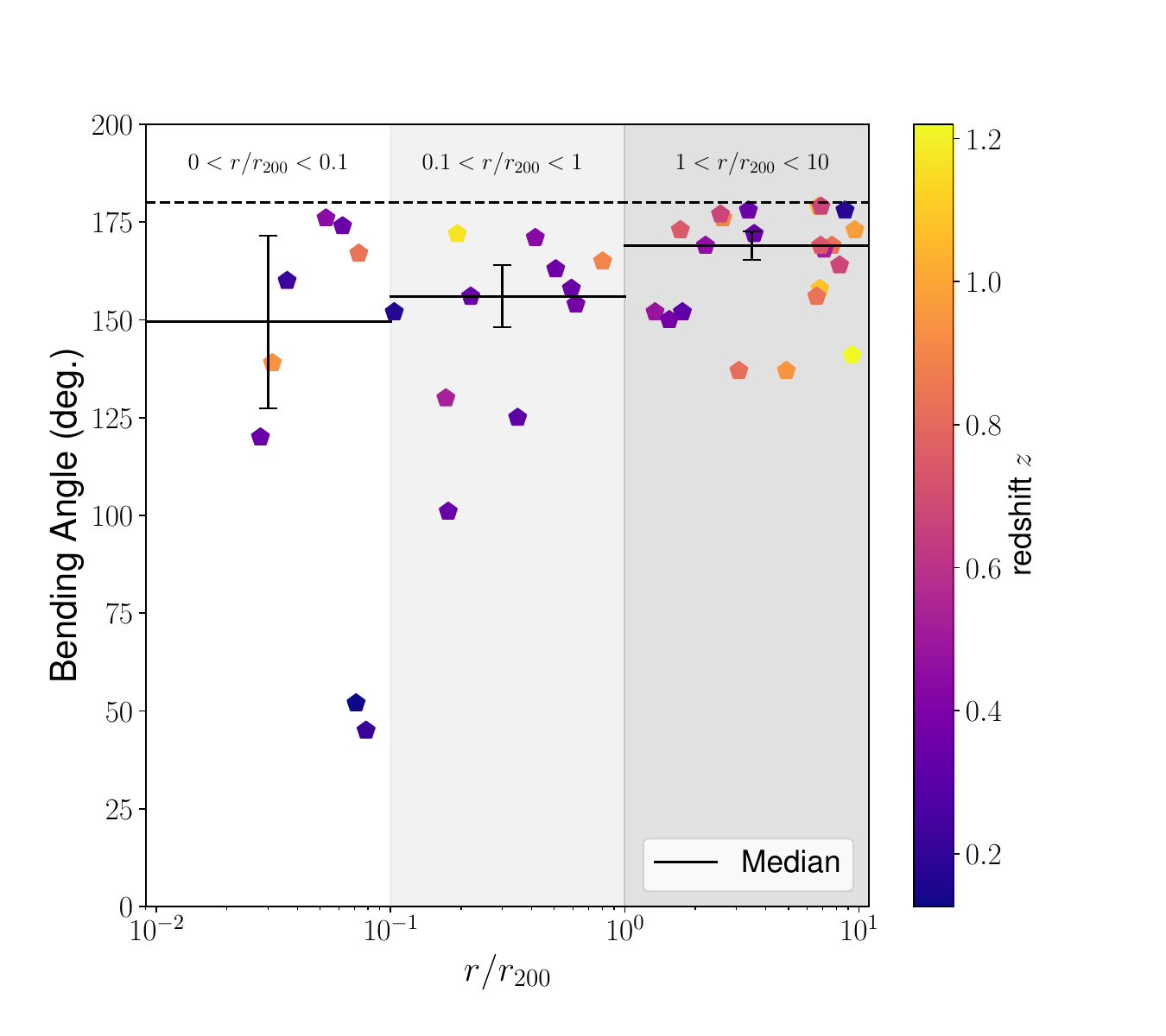}
\caption{bending angle in degrees of radio sources in X-ray galaxy groups as a function of distance to the group centre in units of $r_{200}$ for the COSMOS field. We distinguish between the \emph{core region} ($0<r/r_{200}<0.1$), the \emph{inner region} ($0.1<r/r_{200}<1$) and the \emph{outer region} ($1<r/r_{200}<10$) of the groups. The black lines and error bars show the median and standard median error of the bending angles in the regions. The redshift of the radio sources is shown with a colour bar. The dashed line at 180$^{\circ}$ indicates a straight source.}
\label{fig:BA_r200_COSMOS}
\end{figure}

To investigate the relation between bending angle and distance from the X-ray group centre in the COSMOS field, we plot in Figure \ref{fig:BA_r200_COSMOS} the bending angles against the projected distance $r$ of the radio sources to the group centre, normalised by the virial radius $r_{200}$. The \emph{core region} typically covers the range of $0<r/r_{200}<0.1$ \citep[e.g.][]{navarro1995simulations,navarro1996structure,navarro1997universal}. The range $0.1<r/r_{200}<1$ is deemed the \emph{inner region} of the X-ray galaxy group. We also include all sources out to $r/r_{200}<10$ (and $\Delta z\pm0.007$) from the centre, named the \emph{outer region}. We do not consider these galaxies as group members, because they lie beyond the virial radius of $r_{200}$ and were also not assigned a membership based on the studies of \cite{gozaliasl2019chandra}. The reason we include them in this plot is to investigate trends in the periphery of the X-ray galaxy groups. We find that the two most bent sources in the COSMOS sample are located in the \emph{core region} of their corresponding X-ray galaxy group. We also find that the bending angle moderately correlates with distance to the group centre, with strong evidence to reject the null hypothesis (Spearman test: r$_s=0.4$, $p$-value $=0.01$). Similarly, we find that the redshift for sources in galaxy groups both moderately correlate with the bending angle (r$_s=0.55$, $p$-value $=0.02$) and distance to the group centre (r$_s=0.46$, $p$-value $=0.0003$). 

To look for emerging trends in the relatively low sample size of X-ray galaxy group members, we split the 19 sources into two sub-samples of sources that are straight or slightly bent (BA $>160^{\circ}$) and moderately or very bent sources (BA $\leq160^{\circ}$). In Table \ref{tab:X-ray properties COSMOS}, we compile the median values of the group properties for straight versus bent radio sources in groups.

\begin{table*}[h!]
\begin{center}
\caption{Median X-ray group properties and percentiles for straight (BA $> 160$ deg.) and bent (BA $\leq$ 160 deg.) group members in the COSMOS sample. BA is the bending angle in degrees; $z$ the redshift; $M_{200}$ is the mass of the galaxy group in $M_{\odot}$; $r$ is the distance from the X-ray galaxy groups centre normalised to $r_{200}$; $kT$ is the temperature of the group in keV; and $N_{\rm gal}$ is the number of galaxies that are members of the X-ray galaxy group. }
\label{tab:X-ray properties COSMOS}
\renewcommand{\arraystretch}{1.5}
\begin{tabular}[t]{c c c c c c c c}
\hline
\textbf{BA} &\textbf{$N$} & \multicolumn{6}{c}{\textbf{Median$^{84\%}_{16\%}$}}\\
(deg.) & & BA/deg. & $z$ & log$_{10}$(M$_{200}$/M$_{\odot}$) & $r/r_{200}$ & $kT$/keV & $N_{gal}$\\
\hline
$>160$ & 7 & $171.0^{174.1}_{164.9}$ & $0.44^{0.90}_{0.36}$ & $13.81^{13.87}_{13.59}$ & $0.19^{0.52}_{0.06}$ & $1.28^{1.69}_{0.99}$ & $20.0^{83}_{11}$\\
$\leq160$ & 12 & $134.5^{156.5}_{89.2}$ & $0.34^{0.41}_{0.21}$ & $13.56^{13.97}_{13.29}$ & $0.14^{0.41}_{0.03}$ & $0.88^{1.74}_{0.64}$ & $22.5^{117}_{7}$\\
\hline
\end{tabular}
\end{center}

\end{table*}

We report that 12 out of 19 ($63\%$) sources in groups in the COSMOS field have a BA $\leq160^{\circ}$. For these sources, we observe lower median values for X-ray group redshift, halo mass and mean group temperature, as well as a smaller distance to the group centre, but only within the scatter values of the medians. There is no clear divide between the properties of bent and straight sources in groups; 
however there is large overlap in the distributions of the two samples. The median number of group members with M$_{\star}>10^9$ M$_{\odot}$ for straight and bent sources are comparable, with the number of galaxies that are members of the groups being N$_{gal}=20.0^{83}_{11}$ for straight sources and N$_{gal}=22.5^{117}_{7}$ for bent sources.

\subsubsection{Group Members in XMM-LSS}

For XMM-LSS field, we obtain 17 ($10\%$) sources inside X-ray galaxy groups and 149 outside X-ray galaxy groups. 
For the 17 sources in the XMM-LSS sample that are inside X-ray groups, we find that the median bending angle is $140.0^{172.9}_{104.8}$ degrees, while the median bending angle for sources that are not considered group members and that are also covered by \emph{XMM-Newton} is $169.0^{176.0}_{146.0}$ degrees. 

In Figure \ref{fig:BA_Mstar_M200_kT_COSMOS_XMM-LSS}, we show the bending angle as a function of stellar (top panel) and halo mass (middle panel), where XMM-LSS sources are plotted as filled blue stars. We see that the two most bent sources in XMM-LSS are associated with more massive groups (M$_{200}>10^{14}$M$_{\odot}$) and with massive hosts (M$_{\star}>10^{11}$M$_{\odot}$) while the other sources do not show any trend between bending angle and group or stellar mass. We note that most sources, barring two (sources 49 and 57), are associated with massive hosts (M$_{\star}>10^{11}$M$_{\odot}$), similar to COSMOS, which has three group members below $10^{11}$M$_{\odot}$. We note that the halo mass parameter space is different from that of the COSMOS field, with the latter probing halo masses below $10^{13.6}$M$_{\odot}$.

The bottom panel in Figure \ref{fig:BA_Mstar_M200_kT_COSMOS_XMM-LSS} shows the bending angle in XMM-LSS as a function of the mean group temperature $kT$ in keV. We find no strong correlation between the bending angle and $kT$. Very bent sources show high temperatures, contrary to what is seen from the COSMOS sources. We discuss this further in Section~\ref{T_expected} and in the appendix.

\begin{figure}[h!]
\centering

  \includegraphics[width=\linewidth]{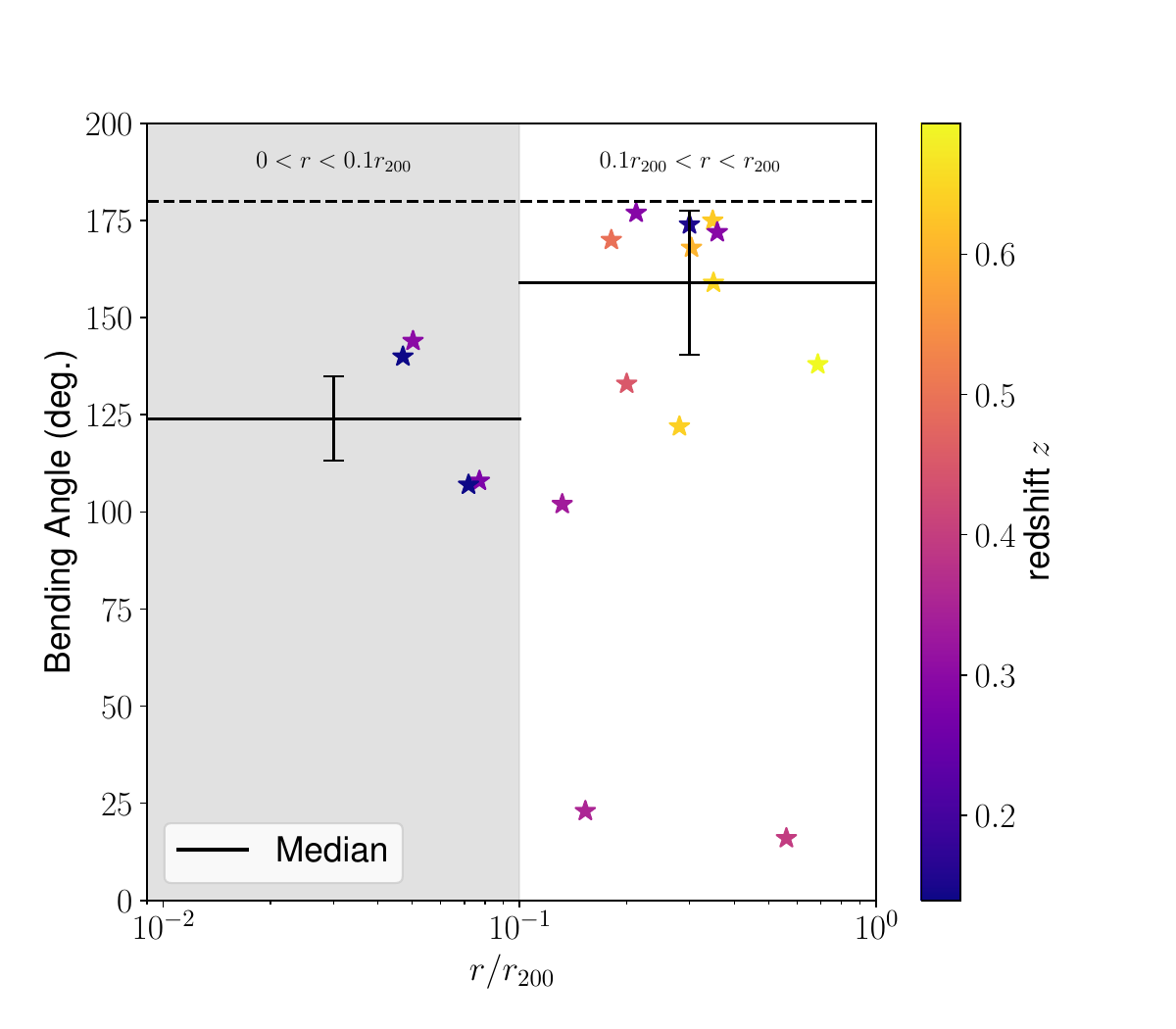}
\caption{{bending angle in degrees of radio sources in X-ray galaxy groups as a function of distance to the group centre in units of $r_{200}$ for the XMM-LSS field. We distinguish between the \emph{core region} ($0<r/r_{200}<0.1$) and the \emph{inner region} ($0.1<r/r_{200}<1$) of the groups. The black lines and error bars show the median and standard median error of the bending angles in the regions. The redshift of the radio sources is shown with a colour bar. The dashed line at 180$^{\circ}$ indicates a straight source.}}
\label{fig:BA_r200_XMMLSS}
\end{figure}

In Figure \ref{fig:BA_r200_XMMLSS}, we show the bending angle as a function of distance to the group centre in units of $r_{200}$. Similarly to Figure \ref{fig:BA_r200_COSMOS}, we distinguish between the \emph{core} and \emph{inner region} of the X-ray group, but do not include sources beyond $r_{200}$, due to the poorer photometric data available for XMM-LSS, compared to COSMOS (see Section~\ref{X-ray Data}). This means we could not define in a similar manner, and robustly, the sources at the \emph{outer region} of galaxy groups in XMM-LSS, as we did in COSMOS. We find that the median bending angle of group members of the \emph{core region} is BA$_{med}=124.0^{142.1}_{107.5}$ and BA$_{med}=159.0^{174.1}_{95.7}$ for members of the \emph{inner region}. In contrast to the COSMOS group members, the two most bent sources in XMM-LSS are located in the \emph{inner region} rather than the \emph{core region} of groups. We note that the two very bent sources are narrow-angle tail (NAT) sources, and probably in-falling to the group centre. We discuss these further in the appendix.

\begin{table*}[h!]
\begin{center}
\caption{Median X-ray group properties and percentiles for straight (BA $> 160$ deg.) and bent (BA $\leq$ 160 deg.) group members in the XMM-LSS sample. BA is the bending angle in degrees; $z$ the redshift; $M_{200}$ is the mass of the galaxy group in $M_{\odot}$; $r$ is the distance from the X-ray galaxy groups centre normalised to $r_{200}$; $kT$ is the temperature of the group in keV; and $N_{\rm gal}$ is the number of radio sources. }
\label{tab:X-ray properties XMMLSS}
\renewcommand{\arraystretch}{1.5}
\begin{tabular}[t]{c c c c c c c}
\hline
\textbf{BA} &\textbf{$N$} & \multicolumn{5}{c}{\textbf{Median$^{84\%}_{16\%}$}}\\
(deg.) & & BA/deg. & $z$ & log$_{10}$(M$_{200}$/M$_{\odot}$) & $r/r_{200}$ & $kT$/keV\\
\hline
$>160$ & 6 & $173.0^{175.4}_{169.6}$ & $0.39^{0.61}_{0.26}$ & $13.96^{14.06}_{13.86}$ & $0.30^{0.35}_{0.21}$ & $1.63^{1.85}_{1.53}$\\
$\leq160$ & 11 & $122.0^{141.6}_{70.4}$ & $0.35^{0.65}_{0.22}$ & $13.99^{14.25}_{13.76}$ & $0.15^{0.43}_{0.06}$ & $1.66^{2.75}_{1.14}$\\
\hline
\end{tabular}
\end{center}

\end{table*}

In Table \ref{tab:X-ray properties XMMLSS}, we show the median group properties for sources in XMM-LSS that are straight or slightly bent (BA $>160^{\circ}$) and for bent sources (BA $\leq160^{\circ}$). We see that 11 out of 17 group members have a BA $\leq160^{\circ}$. We find only slight differences between the median redshift, halo mass and mean group temperature for straight and bent radio sources, with overlapping distributions. The median distance to the group centre for bent sources is $\SI{0.15}{r/r_{200}}$, which is smaller by a factor of 2 when compared to the value for straight and slightly bent sources.

In Table \ref{tab:BA group vs field} we compile the median bending angles for the group members in the core and inner regions of X-ray galaxy groups and also compare the bending angles of all group members to the radio sources that are not considered X-ray group members (field sources). For both fields, we find that sources located in the \emph{core region} are more bent ($22\%$ more bent in XMM-LSS and $4\%$ more bent in COSMOS) than the sources in the \emph{inner region} of galaxy groups. Similarly, group members are more bent ($17\%$ more bent in XMM-LSS and $7\%$ more bent in COSMOS) than the sources located in the field. The two most bent sources in each sample are located in galaxy groups. These four sources in particular will be further discussed in the appendix.

\begin{table}[h!]
\begin{center}
\caption{bending angles for group and field sources in the same area coverage.}
\label{tab:BA group vs field}
\renewcommand{\arraystretch}{1.5}
\begin{tabular}[t]{ c c c c c}
\hline\hline
\textbf{Field} & \textbf{$N$} &  \multicolumn{3}{c}{\textbf{bending angle (deg.)}}\\
 &  & \textbf{Median$^{84\%}_{16\%}$}  & \textbf{Min} & \textbf{Max}\\
\hline\hline
\multicolumn{5}{c}{\textbf{\emph{core region} ($0<r/r_{200}<0.1$)}}\\
\hline
XMM-LSS & 4 & $124.0^{142.1}_{107.5}$ &  107 & 144\\
COSMOS & 8 & $149.5^{167.4}_{128.0}$ & 45 & 176\\
\hline
\multicolumn{5}{c}{\textbf{\emph{inner region} ($0.1<r/r_{200}<1$)}}\\
\hline
XMM-LSS & 13 & $159.0^{174.1}_{95.7}$ &  16 & 177\\
COSMOS & 11 & $156.0^{167.4}_{128.0}$ & 101 & 172\\
\hline
\multicolumn{5}{c}{\textbf{Group Members}}\\
\hline
XMM-LSS & 17 & $140.0^{172.9}_{104.8}$ &  16 & 177\\
COSMOS & 19 & $156.0^{171.1}_{117.7}$ & 45 & 176\\
\hline
\multicolumn{5}{c}{\textbf{Field Sources}}\\
\hline
XMM-LSS & 149 & $169.0^{176.0}_{146.0}$ &  83 & 180\\
COSMOS & 57 & $168.0^{177.0}_{140.8}$ &  102 & 180\\
\hline
\end{tabular}
\end{center}

\end{table}

\section{Discussion}
\label{Discussion}

\subsection{Radio Size and Luminosity of X-ray Galaxy Group Members}
\label{Size,Luminosity in Groups}

We investigated relations between the physical properties of X-ray galaxy group members in XMM-LSS and COSMOS. The sky coverage of MIGHTEE-DR1 observations is \SI{14.4}{deg^2} for XMM-LSS and \SI{4.2}{deg^2} for COSMOS, from our total samples, i.e. all sources within the DR1 mosaics for which we measured their BA, we obtain 15.07 sources/deg$^{2}$ for XMM-LSS and 33.81 sources/deg$^{2}$ in COSMOS. Our results suggest that COSMOS is a more densely populated field than XMM-LSS. Literature studies of bent radio AGN, and in particular the study of \cite{golden2019high}, who targeted clusters of galaxies selected from the VLA FIRST radio survey \citep[][beam size: 5", rms: $\sim$\SI{150}{\micro Jy/beam}]{Helfand2015,becker1995first} and the study of \cite{wing2011galaxy}, indicates an expected number of bent double sources (BA $\leq160^{\circ}$) of the order of 646 in \SI{300}{deg^2}, or 2.12 sources/deg$^{2}$. The increased source count per square degree in our samples is attributed to the high sensitivity of the MIGHTEE survey, the inclusion of sources with BA $>160^{\circ}$ and the larger redshift range ($0.01<z<3.2$). If we constrain our samples to bent group members (BA $\leq160^{\circ}$) in the redshift range of \cite{golden2021high} of $0.35<z<2.2$, we get 0.71 sources/deg$^{2}$ in XMM-LSS and 1.74 sources/deg$^{2}$ in COSMOS, compared to the 0.12 sources/deg$^{2}$ \cite{golden2021high} find from the sample of 36 high-$z$ bent sources in clusters. 

\cite{golden2021high} investigate the parameter space $36 > \rm BA (deg) \leq 160$, linear size of $\sim$120-600 arcsec, $10^{24.7}< L_{\rm 1.4 GHz} / [\rm W Hz^{-1}] <10^{27.7}$.  
Other studies, such as that of \cite{garon2019radio} find 988 bent radio sources (BA $\leq160^{\circ}$) in 10575 deg$^{2}$ below redshift $z <0.8$, which gives $\sim$0.1 sources/deg$^{2}$. \cite{garon2019radio} investigate the parameter space $0.02 > z > 0.8$, $0.2 > \rm BA (deg) > 180$ (values changed to match our conversion), angular size 0.2-1.3 arcmin, and $L_{\rm min} = 2\times10^{23}\rm W~Hz^{-1}$, while they probe clusters with masses $M_{500} > 5 \times 10^{14} M_{\odot}$. We attribute the discrepancy in the findings in the different parameter space probed. Restricting our sample to $z \leq 0.8$ yields $\sim$ 1 sources/deg$^{2}$ for XMM-LSS, and $\sim$ 4 sources/deg$^{2}$ for COSMOS. Finally, \cite{mingo2019revisiting} identify 459 bent-tailed in LoTSS below redshift $z < 0.4$, covering 424 deg$^{2}$, which gives $\sim$ 1 source/deg$^{2}$. Considering only sources within clusters and their match fraction of $\sim$50\%, \cite{mingo2019revisiting} find 0.54 bent-tailed sources/deg$^{2}$, reportedly WATs and NATs, including core-jet sources. A direct comparison to our sample is not possible. The interesting result is in COSMOS we find a larger number of bent sources per square degree than in other fields/studies. COSMOS is known to have several overdensities in the redshift range covered by our study \citep[see][]{scoville2013evolution}. We discuss this point further down.

We further investigate the reason we do not find bent radio AGN above $z=1.2$ in our samples of group members. \cite{golden2021high} find a total of 9 bent, double-lobed radio galaxies above $z=1.2$ in clusters in \SI{300}{deg^2}, 
which corresponds to 0.03 bent double sources/deg$^{2}$ for $z\ge1.2$. From this, we should expect to find 0.43 bent double sources/deg$^{2}$ in XMM-LSS and 0.13 bent double sources/deg$^{2}$ in COSMOS for $z\ge1.2$. From this comparison, we conclude that bent radio AGN at $z\ge1.2$ are rare and large sky coverage is required to find them.

By comparing the linear projected size and radio luminosity of the COSMOS and XMM-LSS X-ray galaxy group members, we find a moderate correlation between those quantities, where r$_s=0.46$, $p$-value $=0.05$ for COSMOS and r$_s=0.44$, $p$-value $=0.08$ for XMM-LSS. This also corresponds to a moderate correlation between radio luminosity and linear size, but with weaker evidence to reject the null hypothesis, likely owed to the smaller sample size. If we compare the whole COSMOS and XMM-LSS samples, the correlation between linear size and luminosity is stronger for galaxy group members than for the whole sample. Our results agree with the literature and the moderate correlation found by \cite{golden2021high}.

We note that except for a giant radio galaxy (GRG) we find in COSMOS \citep[Source 178; see also][]{delhaize2021mightee}, the bent sources in groups and clusters have radio sizes between 100 - 800 kpc\footnote{We define GRGs as sources with linear projected size $\geq$ 1 Mpc}. The GRG in COSMOS recently reported in \cite{Charlton2024} is located at the north edge of the coverage of the DR1 mosaic and missing half the jet structure towards the north, and is thus not fulfilling our criteria for the measurement of the BA, i.e. two-sided radio structures. \cite{malarecki2015giant} report that hosts of GRGs are usually found in environments of higher galaxy density, similar to group environments. However, we note that out of the 11 GRGs (2 sources in COSMOS and 9 sources in XMM-LSS) we find in our samples inside the X-ray coverage, only Source 178 in COSMOS is inside a group environment (M$_{\rm 200}\sim 2\times10^{13}\text{M}_{\odot}$). Recently, \cite{neronov2024} explored the reason why GRGs such as Porphyrion with a size of 7 Mpc, grow so large. They argue that such systems can expand inside filaments and their jets trace a very high-energy $gamma$-ray beam emitted by AGN.

\begin{figure}[h!]
\centering
\begin{subfigure}{.9\columnwidth}
  \centering
  \includegraphics[width=\linewidth]{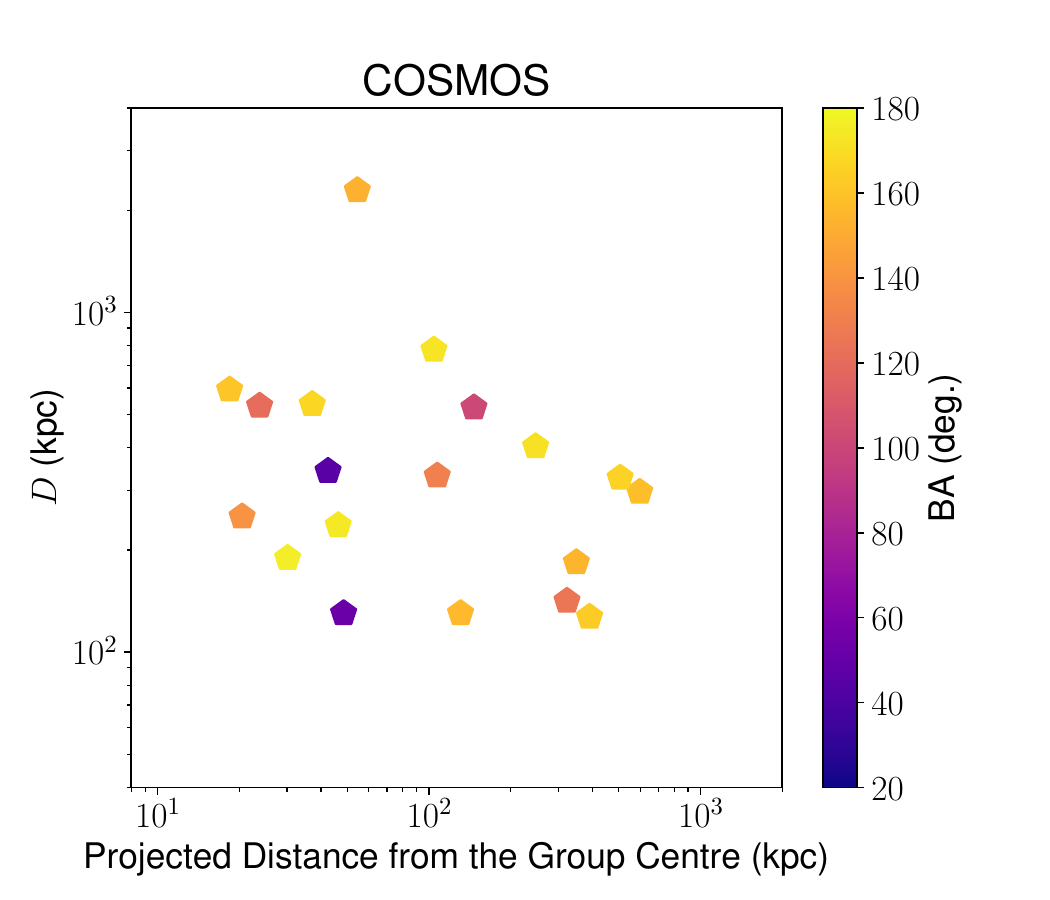}
\end{subfigure}\\
\begin{subfigure}{.9\columnwidth}
  \centering
  \includegraphics[width=\linewidth]{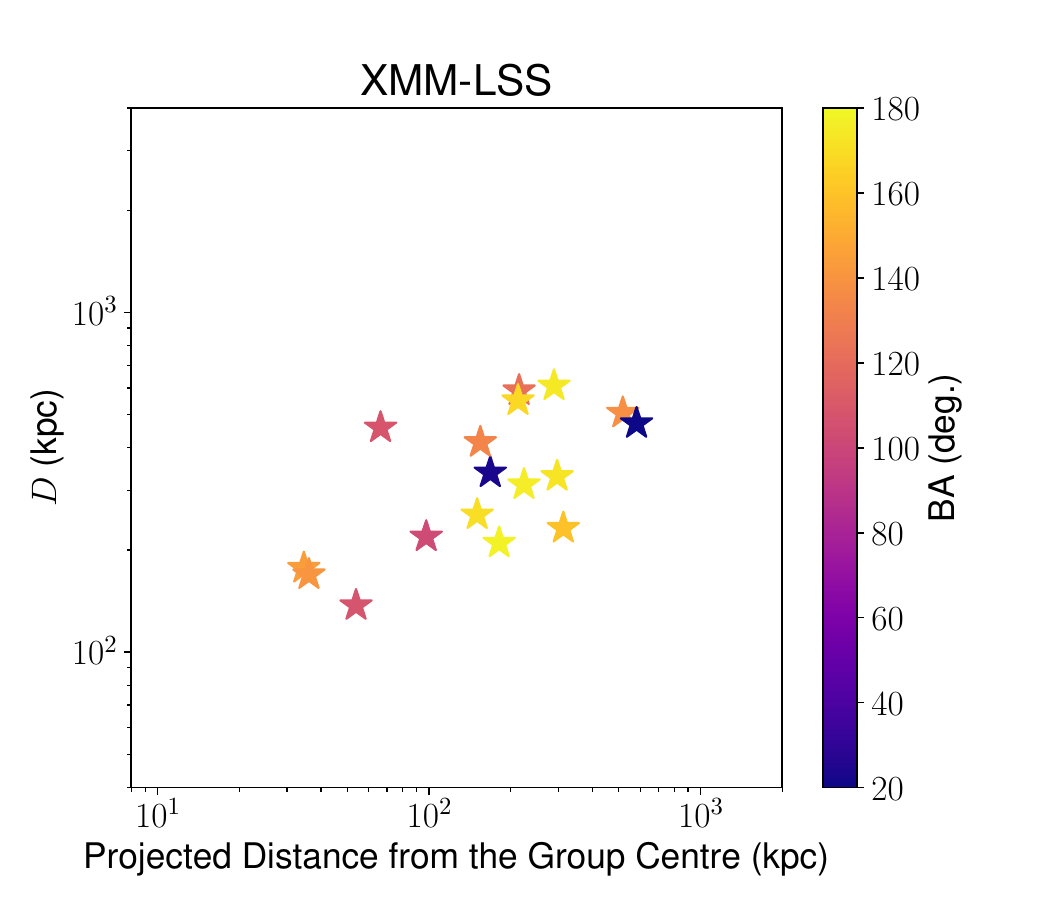}
\end{subfigure}
\caption{Projected linear size $D$ of group members in kpc as a function of projected distance from the group centre in kpc. The BA of the radio sources are shown by a colour scale. (top): COSMOS group members. (bottom): XMM-LSS group members.}
\label{fig:dist_kpc_D_groups}
\end{figure}

Interestingly, sources larger than \SI{500}{kpc}, which are members of X-ray galaxy groups, are less bent by $\sim25\%$ on average compared to sources smaller than that. While there is no statistically significant correlation between bending angle and size, we note that we find a moderate correlation with strong evidence to reject the null hypothesis between the distance from the group centre and linear size for the XMM-LSS group members (r$_s=0.60$, $p$-value $=0.01$, see bottom panel of Figure \ref{fig:dist_kpc_D_groups}). This could indicate that, as the ICM density increases towards the centre of the group, the expansion of radio jets is hindered. \cite{moravec2019massive,moravec2020massive} find such a relationship between radio size and distance from cluster centres from observations and the self-similar jet model from \cite{falle1991self}, arguing that the radio size depends on jet power, lifetime of the source and the density of the surrounding medium:

\begin{equation}
    D = c\left(\frac{t^3 Q\rm_{jet}}{\rho}\right)^{1/(5-\alpha)}
    \label{eq:size_power_density}
\end{equation}

\noindent
with $D$ the size of the radio source, $c$ a dimensionless constant encompassing the adiabatic index of the surrounding gas and the opening angle of the jets, $Q\rm_{jet}$ the jet power, $\rho$ the density of the environment and $t$ the lifetime of the source. $\rho$ is typically modelled by a radial profile of the form $\rho = \rho_0 r^{-\alpha}$, where $r$ is the distance from the source.

Under the assumption of no strong radial density gradients on jet scales, $\alpha$ becomes 0 and $\rho$ only depends on the distance to the cluster centre. However, as seen in the top panel of Figure \ref{fig:dist_kpc_D_groups}, no correlation between distance from the group centre and radio size is found for our COSMOS group members. \cite{golden2021high} also find no agreement with the relationship of \cite{moravec2019massive}, suggesting that the relationship between size and distance from group/cluster centre is not straightforward. We will further discuss Equation \ref{eq:size_power_density} by estimating $Q\rm_{jet}$ and $\rho\rm_{ICM}$ in Section \ref{Jet Size, power, rhoICM}.

\subsection{X-ray Galaxy Group Properties}

While we can compare the intrinsic host properties of the members of X-ray galaxy groups, it is not trivial to compare the group properties from X-ray galaxy groups found for COSMOS and XMM-LSS. In addition to the differences in spectrophotometric data quality available for the fields (see Section~\ref{X-ray Data}), the surveys we utilise for the COSMOS groups \citep[see][and in prep.]{gozaliasl2019chandra} have a lower flux limit than the survey conducted for XMM-LSS \citep[see][]{gozaliasl2014mining}, resulting in lower halo mass and temperature ranges in COSMOS.

Since we have a robust catalogue of 322 X-ray galaxy groups available for COSMOS \citep[and in prep.]{gozaliasl2019chandra}, we can compare the ranges of all galaxy groups in COSMOS to the 17 groups in XMM-LSS that host bent radio sources of our samples, which we show in Figure \ref{fig:kTM200Histo}. While this comparison cannot show the differences in distribution for all galaxy groups of the two fields, we can confirm that the COSMOS groups are distributed at lower halo masses and temperatures compared to those in XMM-LSS which host bent radio sources, owing to the difference in flux limit.

In a relaxed group, the gas density will increase as the distance to the group centre decreases \citep[e.g.][]{ascasibar2003radial}. Thus, jet bending due to the movement of a radio galaxy through the group medium will be more pronounced for galaxies that are in closer proximity to the group centre. This is shown in the study of \cite{garon2019radio}, who investigated 4304 radio galaxies in optically selected galaxy clusters, and find that the jet bending becomes less severe the further the galaxy is from the cluster core. In our study, we see, on average, a similar behaviour (with the exception of the two NATs in XMM-LSS). As we summarise in Table \ref{tab:BA group vs field}, we find that the median bending angle for sources located in X-ray galaxy groups is lower than for the sources we do not consider group members, with the lowest median bending angles found in the \emph{core region} ($r/r_{200}<0.1$) of the groups. The scatter of these values is large, ranging from $30^{\circ}$ to $80^{\circ}$ between the 16th and 84th percentiles, and straight or slightly bent objects (BA $>160^{\circ}$) are also found in X-ray galaxy groups. A K-S test between the BAs for group members and field sources shows that the BA distribution is different between members of the field and of groups, (K-S statistic $=0.53$, $p$-value $=0.02$ in XMM-LSS and K-S statistic $=0.42$, $p$-value $=0.07$ in COSMOS), though this is limited by the small number of group members. Even so, for both the XMM-LSS and COSMOS sample we find that $\sim64\%$ of all group members have a BA$\leq160^{\circ}$, while only $\sim34\%$ of all field sources have a BA$\leq160^{\circ}$.

As Table \ref{tab:BA group vs field} shows, we still find sources with bending angles down to 83$^{\circ}$ in the field. A reasonable question would be why we find bent sources at all if they are not located in a dense group or cluster environment. The reason can be attributed to the sensitivity of the current X-ray observations. In particular, for COSMOS, we can only detect X-ray galaxy groups with halo masses $\approx1.5 (1+z)\times10^{13}$M$_{\odot}$ \citep{vardoulaki2019closer}. Additionally, a good photometric catalogue plays an important role for the robust membership assignment of the galaxies in groups. As we have discussed in Section \ref{X-ray Data}, the method of identifying galaxies as members of a group in XMM-LSS only allows us to assign a secure membership to 17 bent radio AGN. Thus, bent sources that are not members of groups can be used as tracers for groups and clusters \citep[e.g.][]{hintzen1984wide,blanton2000first,smolvcic2007wide,mingo2019revisiting,vardoulaki2019closer}. Another reason could be that a radio galaxy interacted with a group in the past and now is located outside the virial radius of the group \citep[e.g.][]{wetzel2014galaxy}. Bent sources like WATs can also be located in filaments of the cosmic web \cite[e.g.][]{edwards2010first,garon2019radio,vardoulaki2021bent,morris2022does} instead.

Ignoring the very bent sources located in the X-ray galaxy groups (BA $\leq100^{\circ}$), which are discussed in more detail in the appendix, we find no trend between bending angle and group halo mass or temperature for either the COSMOS or the XMM-LSS samples (see Figs \ref{fig:BA_Mstar_M200_kT_COSMOS_XMM-LSS}, \ref{fig:BA_r200_COSMOS} and \ref{fig:BA_r200_XMMLSS}). We again compare this to the large sample of 4304 bent radio galaxies located in optically selected galaxy clusters \citep{garon2019radio}, who find that more bent sources are located in more massive clusters with higher ICM pressures. One of the reasons we might not obtain this trend is because our sample size of group members is too small to find any significant correlations. Another explanation is the different parameter space of \cite{garon2019radio}, who examine sources in galaxy clusters ranging from M$_{500}=$ \SI{5e14}{M_\odot} to \SI{30e14}{M_\odot}, while the galaxy groups that host the bent sources of our samples have halo masses M$_{500}=$ \SI{2e13}{M_\odot} to \SI{1e14}{M_\odot} in COSMOS and M$_{500}=$ \SI{4e13}{M_\odot} to \SI{2e14}{M_\odot} in XMM-LSS. The galaxy group masses M$_{200}$ from the X-ray galaxy groups are converted to M$_{500}$ using the COLOSSUS code \citep[COsmology, haLO and large-Scale StrUcture toolS]{diemer2018colossus}. The halo mass of groups and clusters is related to the dispersion velocities of their members \citep[e.g.][]{saro2013toward}, so we expect galaxies to move faster through the ICM in more massive groups or clusters. Similarly, scaling relations between dispersion velocities of galaxies and group temperature show that statistically, hotter group environments are indicative of members that move through the ICM with high velocities \citep[e.g.][]{lubin1993relation}.

As we discuss in the upcoming section, the halo mass is correlated to the ICM pressure of the group or cluster. Both higher galaxy velocities through the ICM and higher ICM pressures should promote jet bending through ram pressure that is exerted on the jets \citep{begelman1979twin}. At first glance, it is therefore unexpected to find no correlation between bending angle and halo mass or temperature. This begs the question if the halo masses and temperatures we observe in XMM-LSS and COSMOS are too small to cause ram pressure-induced jet bending. \cite{mguda2015ram} find from simulations that radio sources bent due to ram pressure are equally found in halo masses above and below 10$^{14.5}$M$_\odot$, but that this comes from the fact that the lower mass clusters far outnumber higher mass clusters. In other words, more massive clusters are more likely to host bent sources due to ram pressure, but are rare, while less massive clusters are less likely to host bent sources due to ram pressure, but are not rare. Since all of the X-ray galaxy groups in our study have masses below 10$^{14.5}$M$_\odot$, we are disproportionately affected by the small sample size of groups. However, studies have shown that the difference in properties of clusters and groups is not a simple matter of up- or downscaling \citep[e.g.][]{sanderson2003birmingham,borgani2004x,gaspari2011agn}. The heating due to feedback from AGN \citep[see][for a review]{fabian2012observational} has a bigger impact on the smaller halos of groups, resulting in a steeper $L_X-T$ scaling relation than for clusters \citep[e.g.][]{helsdon2000intragroup, magliocchetti2022hosts}. This is connected to a flattening in the gas density profile in groups with temperatures below 3-\SI{4}{keV} \citep{ponman1999thermal}, which applies to the groups that host the bent sources of our samples. This can be related to the result of \cite{smolvcic2011occupation} and \cite{vardoulaki2023evolution}, which show that sources remain active inside galaxy groups compared to the field. In more massive clusters, the effects of heating from feedback will be less severe than for groups. This makes a direct comparison to massive cluster environments difficult. We therefore investigate the relation of jet bending and ram pressure more directly.

\subsection{Ram Pressure as a Reason for Jet Bending in Galaxy Groups}
\label{ram pressure}

As discussed above, we expect jets to bend in group environments because of the ram pressure the ICM exerts on the jets as they move through the dense group medium \citep{begelman1979twin, jones1979hot}. This pressure is expressed as P$_{\text{ram}} = \rho_{\text{ICM}}\,v^2_{\text{gal}}$, where  $\rho\rm_{ICM}$ is the density of the ICM and $v\rm_{gal}$ is the relative velocity between the galaxy and the ICM gas particles \citep{jones1979hot}. The curvature of the jets in relation to the ram pressure can be expressed by:

\begin{equation}
\label{eq:jetbending}
    \frac{\rho_{\text{ICM}}v^2_{\text{gal}}}{h}=\frac{\rho_jv^2_j}{R}
\end{equation}

\noindent
where $\rho_{j}$ and $v_j$ are the gas density and velocity of the jet particles, $h$ is the scale height, i.e. the radius of the jet, and $R$ is the radius of the jet curvature \citep{begelman1979twin}.

Assuming that the groups are approximately virialised, meaning the groups are in dynamical equilibrium, we can use $P_{\rm ICM}$ as a proxy for $P_{\rm ram}$ \citep[e.g.][]{garon2019radio}. To estimate $P_{\rm ICM}$, we adopt the formula from \cite{arnaud2010universal}, which uses simulations from \cite{nagai2007effects} and observations of 33 local clusters observed by \emph{XMM-Newton} to calculate a universal galaxy cluster pressure profile:

\begin{equation}
  \label{eq:P_icm}
  \begin{split}
P_{\rm ICM}(d) = 1.65\times10^{-3} E(z)^{8/3} \times \left( \frac{M_{500}}{3\times10^{14}M_{\odot}} \right)^{2/3+\alpha_P+\alpha^{'}_{P}(d)} \\ \times \mathbb{P}(d)\:\text{keV cm}^{-3}  
\end{split}
\end{equation}

\begin{figure}[t!]
\includegraphics[width=\linewidth]{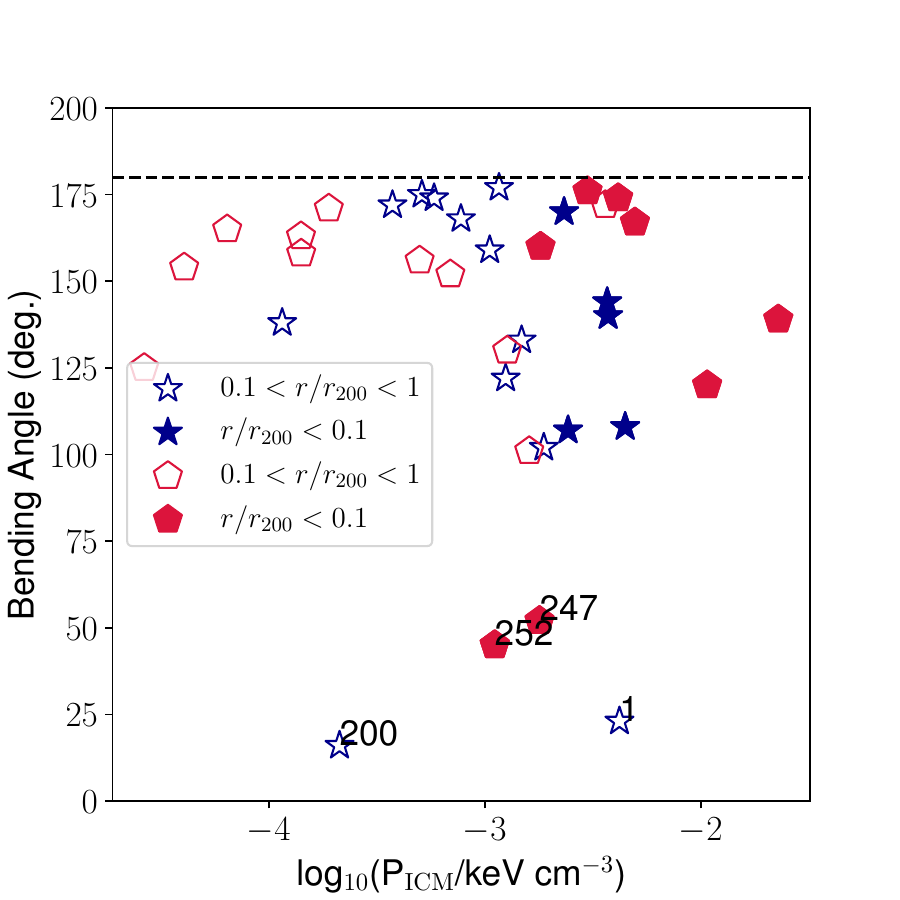}
    \caption{$P_{\rm ICM}$ calculated from Equation \ref{eq:P_icm} for the X-ray galaxy group members of the COSMOS (red pentagons) and XMM-LSS (blue stars) sample as a function of bending angle. The source IDs for very bent sources are annotated. The dashed line at 180$^{\circ}$ in both plots indicates a straight source.}
    \label{fig:BA_Picm}
\end{figure}

\noindent
where $d=\frac{r}{r_{500}}$, $E(z) = \Omega_M(1+z)^3+\Omega_\Lambda$, M$_{500}$ is the group mass within $r_{500}$, $z$ is the redshift and $\mathbb{P}(d)$ the generalised NFW model adopted from \cite{nagai2007effects}. $\alpha_P$ and $\alpha^{'}_{P}(d)$ are fit parameters adopted from \cite{arnaud2010universal}. While the universal pressure profile accounts for halo mass and redshift evolution, the model from \cite{arnaud2010universal} is based on local clusters assuming self-similar evolution for higher redshifts, which might not be strictly applicable to the galaxy groups we probe, though \cite{hernandez2022madpsz} show that universal pressure profiles work well up to redshifts of $z\sim1$.

Figure \ref{fig:BA_Picm} shows the $P_{\rm ICM}$ calculated from Equation \ref{eq:P_icm} for the X-ray galaxy group members of the COSMOS and XMM-LSS sample, presented in Section \ref{BA environ}, as a function of bending angle. As expected, we find that sources in the \emph{core region} ($r/r_{200}<0.1$) are in higher pressure environments than sources of the \emph{inner region} ($0.1<r/r_{200}<1$). For COSMOS, we do not observe a correlation between the bending angle and the ICM pressure. For the XMM-LSS sample, there is no correlation between bending angle and ICM pressure (r$_s=-0.35$, $p$-value $=0.16$). A negative correlation between bending angle and pressure would be expected as it corresponds to smaller bending angles at higher pressures.

We find that for group environments where $P_{\rm ICM}\ge10^{-3}$ keV cm$^{-3}$, we observe a lower median bending angle compared to lower pressures. For $P_{\rm ICM}\ge10^{-3}$ keV cm$^{-3}$, the median bending angles are BA$_{\rm med}=139.0^{172.8}_{81.4}$ and BA$_{\rm med}=133.0^{163.4}_{105.0}$ for the COSMOS and XMM-LSS X-ray galaxy group members, respectively. For $P_{\rm ICM}\leq10^{-3}$ keV cm$^{-3}$, the median bending angles are BA$_{\rm med}=157.0^{164.8}_{152.2}$ and BA$_{\rm med}=170.0^{174.2}_{113.6}$ for COSMOS and XMM-LSS, respectively. This is consistent with the findings of \cite{garon2019radio}, who estimate that $P_{\rm ram}$, for which we use $P_{\rm ICM}$ as a proxy, has to be at least of order magnitude $10^{-3}$ keV cm$^{-3}$ to induce jet bending. This threshold is based on an AGN triggering model from \cite{marshall2018triggering}, who compare simulations of galaxies moving through cluster environments to observational data. For the less massive galaxy groups, such as those used in this work, the semi-analytic model of \cite{marshall2018triggering} predicts that, given the same velocities, galaxies must be closer to the galaxy group centre compared to galaxy clusters to be in the regime of $P_{\rm ram}$ induced AGN triggering. The AGN that are triggered in groups from \cite{marshall2018triggering} are typically found in the range $0<r<1r_{\rm vir}$, with $r_{\rm vir}$ the virial radius of the galaxy group, which is in line with our definition of group membership up to the virial radius $1r_{200}$. Even though our limited sample size of X-ray galaxy group members does not show any significant correlation between bending angle and $P_{\rm ICM}$, we can confirm that $P_{\rm ICM}\sim10^{-3}$ keV cm$^{-3}$ works well as a threshold for ram pressure induced jet bending. We find 11 bent group members in both samples (58\% in COSMOS and 65\% in XMM-LSS) to be at $P_{\rm ICM}\ge10^{-3}$ keV cm$^{-3}$. This includes all WATs we find in groups in COSMOS and XMM-LSS, which are expected to be predominantly bent by ram pressure \citep[e.g.][]{smolvcic2007wide,o2023wide}. We make note of one outlier, Source 200 in the XMM-LSS sample, which is a NAT but estimated to be at an ICM pressure of $2\times10^{-4}$ keV cm$^{-3}$. A possible explanation for this is discussed in Appendix~\ref{peculiar Sources}, but we mention here that $P_{\rm ICM}$ as a proxy for $P_{\rm ram}$ is based on the assumption that the groups are relaxed, which might not be the case for all groups \citep{gozaliasl2020kinematic}.

Furthermore, \cite{marshall2018triggering} find that higher redshift galaxy clusters ($z\sim1$) are less constrained on the distance to the cluster centre ($0.5\leq r/r_{\rm vir}\leq2$) when it comes to AGN activity triggered by ram pressure. This means that for galaxy groups at higher redshift ($z\sim1$), sources bent due to ram pressure are found further away from the galaxy group centre compared to low redshift galaxy groups \citep{marshall2018triggering}. This is consistent with our results shown in the bottom panel of Figure \ref{fig:BA_Mstar_M200_kT_COSMOS_XMM-LSS}, where the bent X-ray galaxy group members show a moderate correlation between redshift and distance from the galaxy group centre (Spearman test for COSMOS sample: r$_s=0.46$, $p$-value $=0.003$; XMM-LSS sample: r$_s=0.63$, $p$-value $=0.007$). Denser cluster environments at lower redshifts fit into the accepted picture of hierarchical structure formation, shown here also for group environments.

\begin{figure}[h!]
\centering
\begin{subfigure}{.99\columnwidth}
  \centering
  \includegraphics[width=\linewidth]{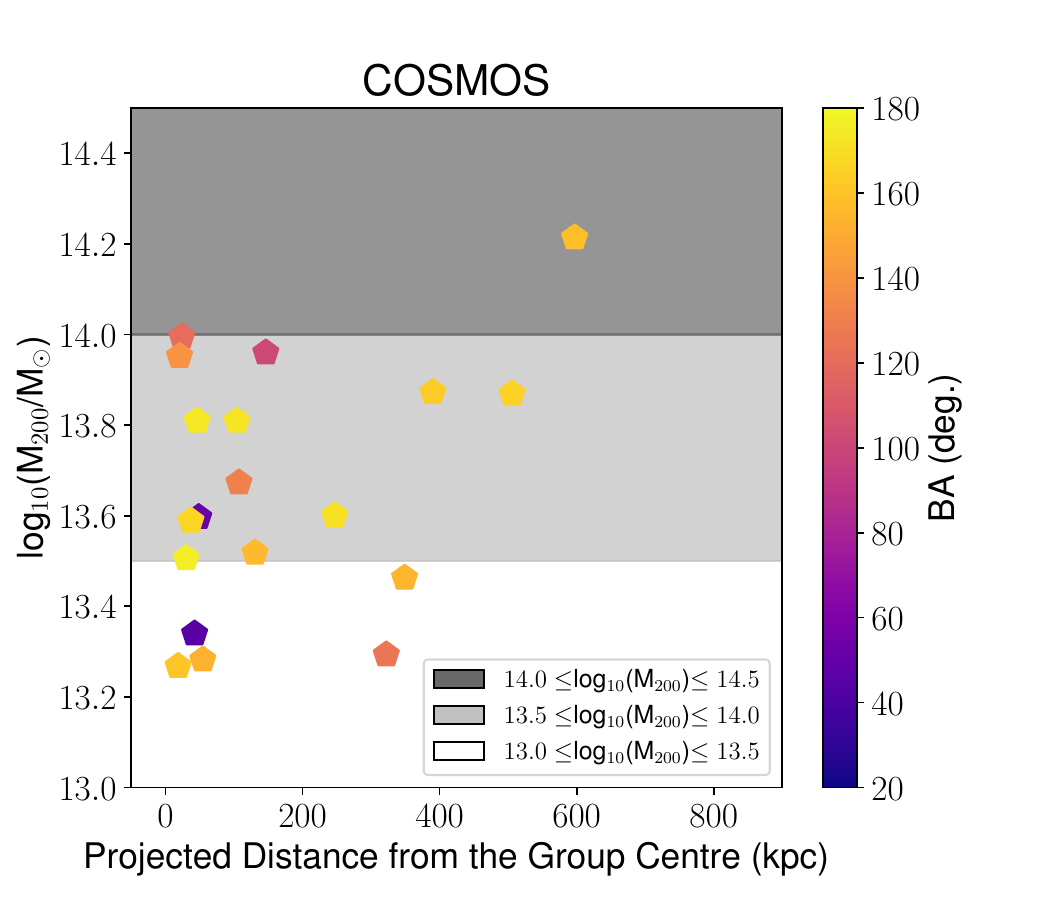}
\end{subfigure}\\
\begin{subfigure}{.99\columnwidth}
  \centering
  \includegraphics[width=\linewidth]{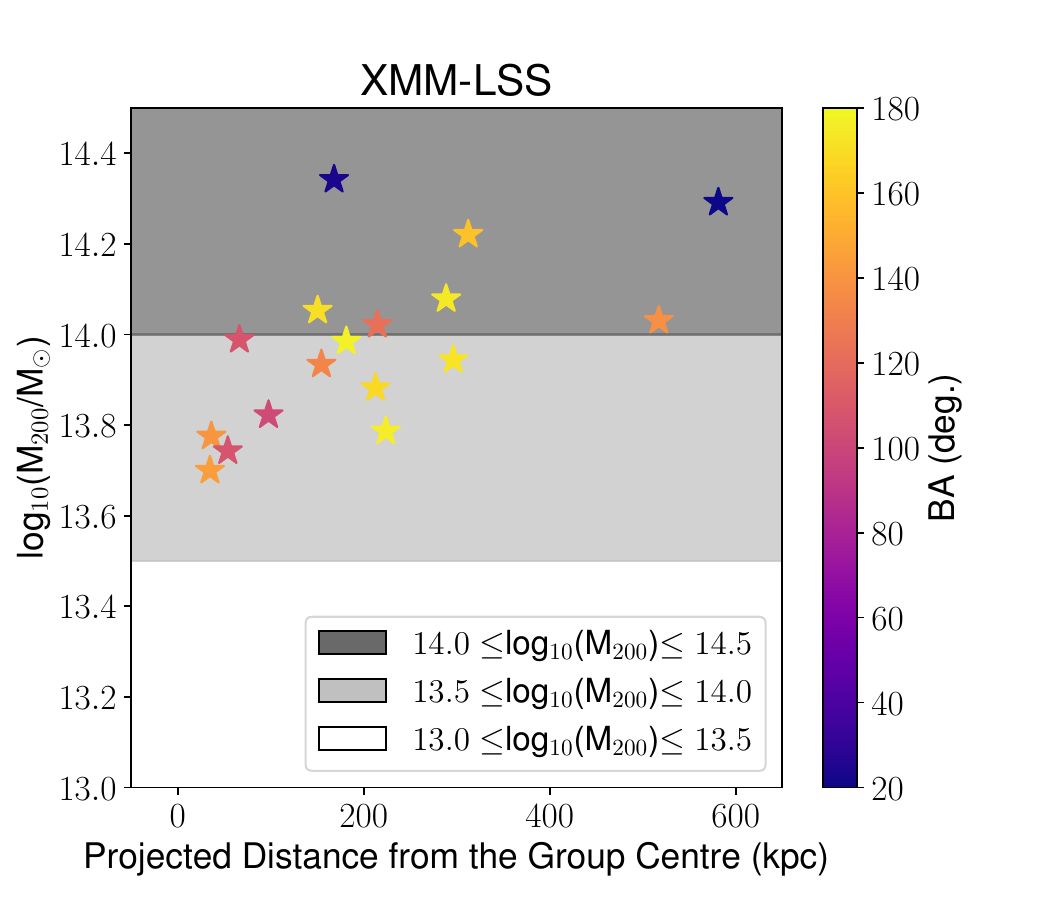}
\end{subfigure}
\caption{Group mass M$_{200}$ in M$_{\odot}$ as a function of projected distance from the galaxy group centre in kpc. The bending angles of the sources that are found in the galaxy groups are shown by a colour scale. (top): COSMOS. (bottom): XMM-LSS.}
\label{fig:M200projdistBA}
\end{figure}

A study by \cite{mguda2015ram} looks into simulations of radio sources in cluster environments that are bent due to ram pressure. They investigate the fraction of radio galaxies that surpass the ram pressure threshold for jet bending, derived from Equation \ref{eq:jetbending} with values from \cite{freeland2011intergalactic}. They find that for low-mass clusters, which \cite{mguda2015ram} define as log$_{10}$(M$_{halo} / $M$_{\odot})\leq14.5$, radio sources bent due to ram pressure are most likely found within \SI{400}{kpc} from the cluster centre. Bent sources in clusters with halo masses $14.0\leq$ log$_{10}$(M$_{halo} / $M$_{\odot})\leq14.5$ are likely found within \SI{800}{kpc} from the cluster centre.

Figure \ref{fig:M200projdistBA} shows that this relationship between group mass and projected distance from the group centre is in good agreement with the sources of our samples located in X-ray galaxy groups, where bent and very bent sources are found up to 400 kpc from the group centre in both COSMOS and XMM-LSS. The sources that are located beyond \SI{400}{kpc} from the group centre are only found in halo masses log$_{10}$(M$_{200} / $M$_{\odot})\ge13.9$.\\
We demonstrate that the jet bending of the sources in the X-ray galaxy groups in COSMOS and XMM-LSS is well explained by ram pressure that is exerted on jets for the halo masses in the range M$_{200}=$ \SI{2e13}{M_\odot} to \SI{2.2e14}{M_\odot}. The lack of correlation between halo mass and bending angle for group members can be due to the small sample size, as finding bent sources in galaxy groups could be rare, as indicated by the number of WATs we find: for COSMOS X-ray galaxy group members, we have robustly identified 2 WATs with BA$\leq100^{\circ}$ ($\sim0.9$ WATs / deg$^{2}$) and none in the X-ray galaxy group members of XMM-LSS (where for BA$\leq160^{\circ}$ we have $\sim1.3$ WATs / deg$^{2}$ in COSMOS and $\sim0.6$ WATs / deg$^{2}$ in XMM-LSS in galaxy groups). \cite{mingo2019revisiting} find $\sim0.22$ WATs / deg$^{2}$ in galaxy clusters from coverage of \SI{424}{deg^2}, selected from LoTSS \citep{shimwell2017lofar,shimwell2019lofar}, highlighting that large area coverage is required to create large samples of bent radio sources.

\subsection{Relating Jet Size, ICM density and Jet Power}
\label{Jet Size, power, rhoICM}
We expect that a radio source in a large cluster will experience a more or less homogeneous ICM density between the host and the jets, as the radio source itself is small compared to the size of the cluster \citep[e.g.][]{garon2019radio}. Conversely, in small galaxy groups, where the radio size can be comparable to the group size, the ICM density could be significantly higher or lower at the jets compared to the host. To investigate this phenomenon in our sample of group members, we divide the values of $P\rm_{ICM}$ from Equation \ref{eq:P_icm} by the mean group temperature to obtain the expected ICM density $\rho\rm_{ICM}$ at the host position and the end-points of both jets for each source. We then sub-divide all group members into sources where the ICM density at the end-points of the jets is within or outside a factor of 5 $\rho\rm_{ICM}$ at the host. This arbitrary threshold is chosen to generate two roughly equally sized sub-samples of sources where $\rho\rm_{ICM}$ is comparable at the jets and the host and where $\rho\rm_{ICM}$ is different between the jets and the host.

\begin{figure}[h!]
    \centering
    \includegraphics[width=1\columnwidth]{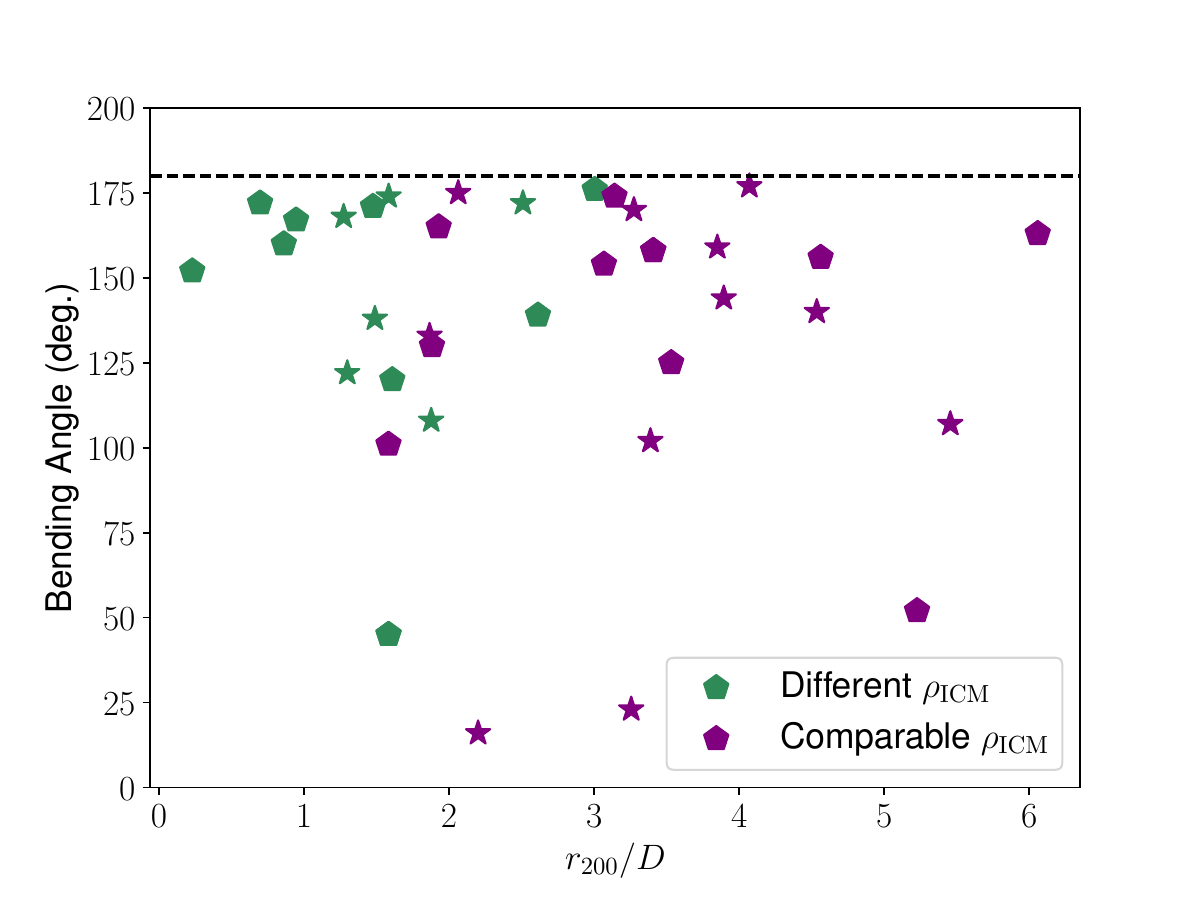}
    \caption{bending angle in degrees as a function of $r_{200}$ of the galaxy groups over the radio size $D$ of the bent group member in the group. The COSMOS group members are pentagons and the XMM-LSS group members are stars. If the ICM density at the end-points of the jets are within a factor of 5 $\rho\rm_{ICM}$ at the host, the symbols are purple; if they are outside a factor of 5 $\rho\rm_{ICM}$ at the host, they are green. The dashed line indicates a straight source.}
    \label{fig:r200overD vs BA}
\end{figure}

Figure \ref{fig:r200overD vs BA} shows the bending angles of the group members in XMM-LSS and COSMOS as a function of $r_{200}$ of the galaxy groups over the radio size $D$ of the bent group member. The sub-populations described above are highlighted with different colours. The sub-sample of group members where the ICM density of the jets is within a factor of 5 $\rho\rm_{ICM}$ at the host is overwhelmingly found at $r_{200}/D > 3$, where the virial radius $r_{200}$ is larger than the radio source itself and a more homogeneous ICM around the source is expected. Conversely, at $r_{200}/D < 3$ we find the population of sources where the ICM density of the jets is outside a factor of 5 from $\rho\rm_{ICM}$ at the host. Here, the group size and radio size are comparable and the ICM density gradient is apparent along the radio source.

While we can confirm that the ICM density varies between the jets and the host depending on how comparable the radio size is to the group size, we find no correlation to the bending angle. A K-S test shows that the bending angles from the two populations in Figure \ref{fig:r200overD vs BA} come from the same parent distribution (K-S statistic $=0.35$, $p$-value $=0.57$). This means that the difference between the two populations is most likely driven by the jet size, as $\rho\rm_{ICM}$ from Equation \ref{eq:P_icm} is a function of distance from the group centre, and longer jets will therefore experience larger density gradients than shorter jets.

We combine the results regarding the intrinsic properties of our samples, namely size and luminosity, and the extrinsic medium of galaxy groups, given by the ICM density. As discussed in Section \ref{Size,Luminosity in Groups}, Equation \ref{eq:size_power_density} suggests that the size of our sources should be dependent on the jet power, the density of the surrounding medium and the lifetime of the source. We have not performed an in-depth analysis to obtain the sources' lifetimes, but we estimate the other quantities: for each group member, we estimate the jet power $Q\rm_{jet}$ from the radio luminosity\footnote{We assume the convention $S_{\nu}\propto \nu^{\alpha}$ for the radio spectral index.} at \SI{1.4}{GHz} by employing Equation 4 from \cite{smolvcic2017vla},  $\text{log}_{10}Q_{\rm jet}(L_{\rm 1.4GHz})=0.86 \text{log}_{10}L_{\rm 1.4GHz}+14.08+1.5 \text{log}_{10}f_W$, with $f_W\approx4$, an uncertainty parameter. To account for the ICM density gradients our sources experience, we take the mean ICM density value $\overline{\rho}\rm_{ICM}$ at the host and the end points of each jet, as described above.

\begin{figure}[h!]
    \centering
    \includegraphics[width=1\columnwidth]{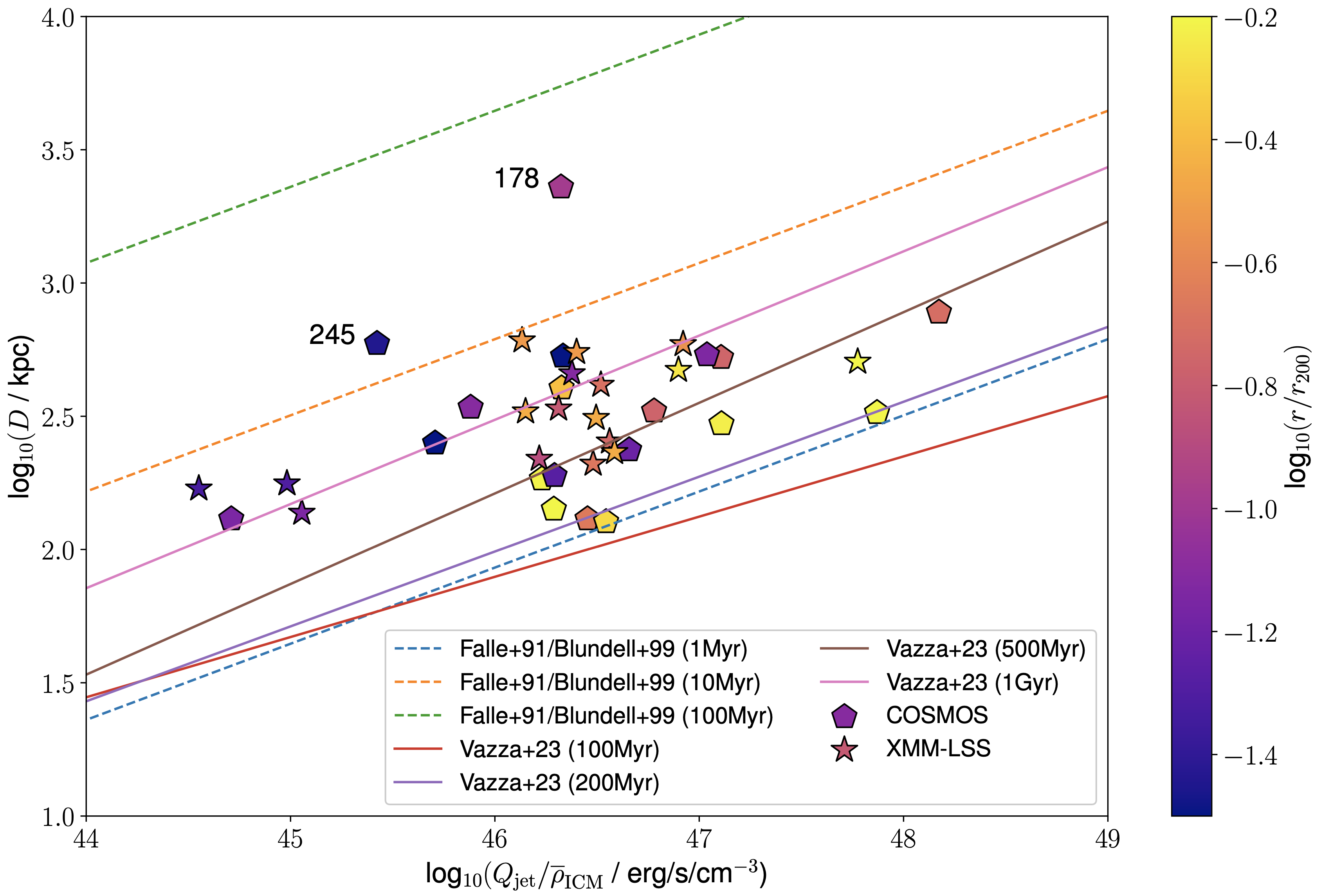}
    \caption{Radio size $D$ in kpc as a function of the jet power $Q\rm_{jet}$ over the mean ICM density $\overline{\rho}\rm_{ICM}$ for the group members of COSMOS (pentagons) and XMM-LSS (stars). The projected distance to the group centre in kpc, normalised to $r_{200}$, is given by a colour scale. We plot dashed lines at constant $t$ from Equation \ref{eq:size_power_density} of \cite{falle1991self}, with values from \cite{blundell1999inevitable}; $Q\rm_{jet}$ is normalised with the average $\overline{\rho}\rm_{ICM}$ of our samples. Solid lines denote the relation between size of the radio source and jet power, normalised with the median environmental density ($10^{-29} \rm g/cm^{-3}$) from the MHD simulations of \cite{vazza2023life}. We note the minimum $Q\rm_{jet}$ value used in the simulation is $3\times10^{43} \rm erg/s$, corresponding to X-axis values of $5\times10^{45} \rm erg/s/cm^{-3}$, thus a simple extrapolation to lower values should be done with caution.}
    \label{fig:D_QoverRHO}
\end{figure}

From Figure \ref{fig:D_QoverRHO} we can confirm that the radio size $D$ increases with larger $Q\rm_{jet}/\overline{\rho}\rm_{ICM}$, as Equation \ref{eq:size_power_density} suggests, with two notable outliers in the COSMOS sample (Source 178 \& 245). $Q\rm_{jet}/\overline{\rho}\rm_{ICM}$ is larger for sources further away from the group centre, indicating that more powerful jets and/or a less dense medium results in larger jet sizes. A Spearman test between $D$ and $Q\rm_{jet}/\overline{\rho}\rm_{ICM}$ gives a moderate correlation for the XMM-LSS group members (r$_s=0.50$, $p$-value $=0.04$). This correlation is weaker and not robust for the COSMOS group members (r$_s=0.26$, $p$-value $=0.29$). We note that the two outliers are the BGG of their group and are found in groups with halo masses M$_{200}\lesssim 10^{13.3}$ M$_{\odot}$, the two least massive galaxy groups that host the bent sources of our samples. Both of these sources are found at $r_{200}/D < 1$ (see Figure~\ref{fig:r200overD vs BA}), i.e. where the radio size is larger than the virial radius of the galaxy group. For these sources, the assumption of a homogeneous ICM density around the source is not fulfilled. The sources' lifetimes are also not accounted for by our study, which could have a large impact on the jet power and size \citep{hardcastle2019radio}, contributing to the scatter observed in Figure \ref{fig:D_QoverRHO}. We therefore compare to models and simulations: we use Equation \ref{eq:size_power_density} from the model of \cite{falle1991self} with characteristic values taken from \cite{blundell1999inevitable} to calculate the radio size at constant $t$ over varying jet power (where $\rho=2\times10^3$ m$^{-3}$, $\alpha=1.5$, $c=3.5$). The theoretical lines from Equation \ref{eq:size_power_density} have been normalised by the median ICM density of our samples for the appropriate scaling of the X-axis in Figure \ref{fig:D_QoverRHO}. We also show the tracks of the average $D-Q_{jet}$ relation obtained by fitting the evolution, at 4 different epochs,  of five resimulations of radio sources at the centre of a small galaxy cluster from \cite{vazza2023life}, who investigated the role of a varying jet power ($Q\rm_{jet}$ ranging from $3\times10^{43} \rm erg/s$ to $1.5\times10^{45} \rm erg/s$) on the overall circulation of electrons injected by the central radio galaxy. The fits from \cite{vazza2023life} have been normalized by the expected ICM density calculated from the median environmental density along the jet propagation ($\sim10^{-29} \rm g/cm^{-3}$) from the MHD simulations of \cite{vazza2023life}. To estimate the density of the medium the jets are expanding into, we use Equation \ref{eq:jetbending}, where we take $h/R=0.05$ as an upper limit \citep{begelman1979twin} and $(v_j/v_{\rm gal})^2
\approx(10/3)^2$, estimated from the average galaxy velocity of our group members (see Section~\ref{T_expected}) and average jet velocities from \cite{vazza2023life}. This gives an average $\rho_{\rm ICM}$ of $\approx6\times10^{-3}$cm$^{-3}$ for the simulated sources in the centres of small clusters, which is consistent with what we find for group members in the \emph{core region} of galaxy groups. We stress that \cite{vazza2023life} are simulating evolving sources and environments, where the moving ICM becomes a dominant factor after a few tens of Myr, causing the $D-Q\rm_{jet}/\overline{\rho}\rm_{ICM}$ lines to deviate from analytical models. After $t>500$ Myr, the fitting formula give the average distance from the cluster centre of cosmic rays injected by jets, even if by that time they have become entirely undetectable in the radio band.   While the $D-Q\rm_{jet}/\overline{\rho}\rm_{ICM}$ relations from Equation \ref{eq:size_power_density} with values from \cite{blundell1999inevitable} are derived from broad assumptions, we are in good agreement with \cite{SiddhantRLAGN}, who found that the spectral ages of 28 extended radio sources in XMM-LSS (all of which are part of our XMM-LSS sample) are mostly found between 1 and \SI{10}{Myr}. 
While we found no good agreement with the relation of \cite{moravec2019massive} between radio size and distance from the group centre for the COSMOS group members (see Figure~\ref{fig:M200projdistBA}), we show that, taking into account the ICM density at a given distance from the group centre, the group members of our samples grow with larger $Q\rm_{jet}/\overline{\rho}\rm_{ICM}$. Scatter is introduced by projection effects, the sources' lifetimes and the fact the ICM density gradients are not negligible on the scales of the jets in small galaxy groups. In Section~\ref{T_expected} \& Appendix~\ref{peculiar Sources} we discuss the different types of environments our sources might be interacting with. The theoretical model discussed here is most likely too simplistic to probe the diversity of these environments.

\section{Estimating ICM temperature from Jet Bending}
\label{T_expected}

In the previous sections, we investigated the jet bending due to the movement of the radio AGN through the ambient medium (see Eq. \ref{eq:jetbending}). We further explore this picture to include a strong magnetic field from a radio AGN that is injected into the ambient plasma by the jet, being subsequently shaped by the bulk motion of the ambient medium. 
New jet particles, accelerated from the central black hole, will experience a change in the direction of the magnetic field and will follow the path dictated by the magnetic field as this interacts with the surrounding medium. The bending angle then indicates the motion of the particles frozen to the magnetic field. \cite{mendygral2012mhd} used MHD simulations to show that even in a relaxed galaxy cluster, the bulk motion of the ICM significantly distorts jets and lobes injected by multiple AGN bursts. Cosmological ENZO-MHD simulations from \cite{vazza2021} of two radio jetted AGN inside clusters at $z$ = 0.5 and $z$ = 1, and the analysis of them by \cite{vardoulaki2021bent}, suggest that sources at lower redshifts, that also lie in hotter and denser environments, are more bent. Additionally, these have had more opportunities for jet interaction with the IGM and a bigger volume to expand into. 

\begin{figure}[h!]
    \centering
    \includegraphics[width=0.65\columnwidth]{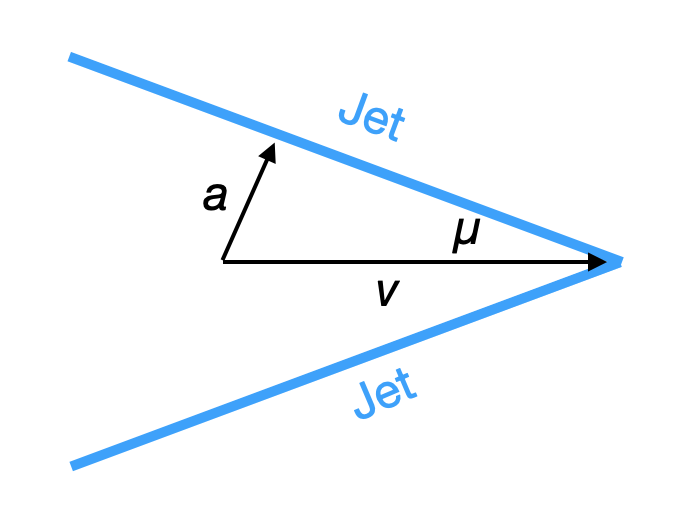}
    \caption{Schematic of a supersonic radio galaxy moving with velocity $v$ through a medium with sound speed $a$. The Mach angle $\mu$ is equal to $BA/2$.}
    \label{fig:MachAngle}
\end{figure}

We estimate the ambient temperature that is expected in order to explain the jet bending relative to a given angle, and from there we test whether the ratio of expected temperature and the mean group temperature deviates from a typical radial profile expected for galaxy groups. We utilise the known relation between the Mach number $\mathcal{M}$ and the angle $\mu$ that disturbances of supersonic flows produce with respect to the flow velocity $v$:

 \begin{equation}
     \label{eq:Mach angle}
     sin(\mu) = \frac{1}{\mathcal{M}} = \frac{a}{v}
  \end{equation}

\noindent
where $\mu$ is the Mach angle and $a$ is the speed of sound of a given medium \citep[see for e.g.][ for discussion on supersonic flows]{Springel2007, Massey2011}. In our framework, the radio source moves with velocity $v$ through a medium with sound speed $a$, where the Mach angle of the jets is then given by $\mu = \text{BA}/2$, with BA being the bending angle we assign to the source (see Figure~\ref{fig:MachAngle}). A straight source with BA$=180^{\circ}$ will result in a Mach number $\mathcal{M}=1$, whereas severely bent sources with BA$=40^{\circ}$ will result in a Mach number $\mathcal{M}\approx3$. Shocks are created from the bulk motion of the medium. 

From the ideal gas law, we can also estimate the temperature $T$ of a given medium from the sound speed $a$:

\begin{equation}
    \label{eq:a_T}
    a = \sqrt{\frac{\gamma k T}{m}}
\end{equation}

\noindent
where we use $\gamma = 5/3$ as the adiabatic index of the ICM, $k$ the Boltzmann constant and $m = 1.66\times10^{-27}$ kg the mass of ionised hydrogen. By combining Equations \ref{eq:Mach angle} \& \ref{eq:a_T} we can estimate the expected temperature of the medium with which the jets of a source at a given velocity are interacting:

\begin{equation}
    \label{eq:Texptected}
    T\rm_{expected} = \frac{\textit{m} \textit{a}^2}{\gamma \textit{k}} = \frac{\textit{m} \textit{v}^2}{\gamma \textit{k} \mathcal{M}^2} = \frac{\textit{m} \textit{v}^2}{\gamma \textit{k}} \times sin^2(BA/2)
\end{equation}

To obtain the velocity values of the sources, we calculate the velocity difference for sources in the COSMOS sample, where robust spectroscopic redshifts are available within $10r_{200}$ from the group centre, given by the relative difference between the spectroscopic redshifts of the galaxy and the group, $ \Delta v = \frac{|z_{\rm gal}-z_{\rm group}|}{1+z_{\rm group}}\times c$, 
with $c$ the speed of light. Due to the lack of robust spectroscopic group redshifts, we omit the XMM-LSS group members here. To reduce the inclusion of interlopers, we restrict to sources where $\Delta v / \sigma_{\rm disp}<2.7$ \citep[following][]{mamon2013mamposst}, with $\sigma_{\rm disp}$ the velocity dispersion of the galaxy group which has been scaled by the radial velocity dispersion profile, adopted from \cite{more2009satellite}.

\begin{figure}[h!]
    \centering
    \includegraphics[width=\columnwidth]{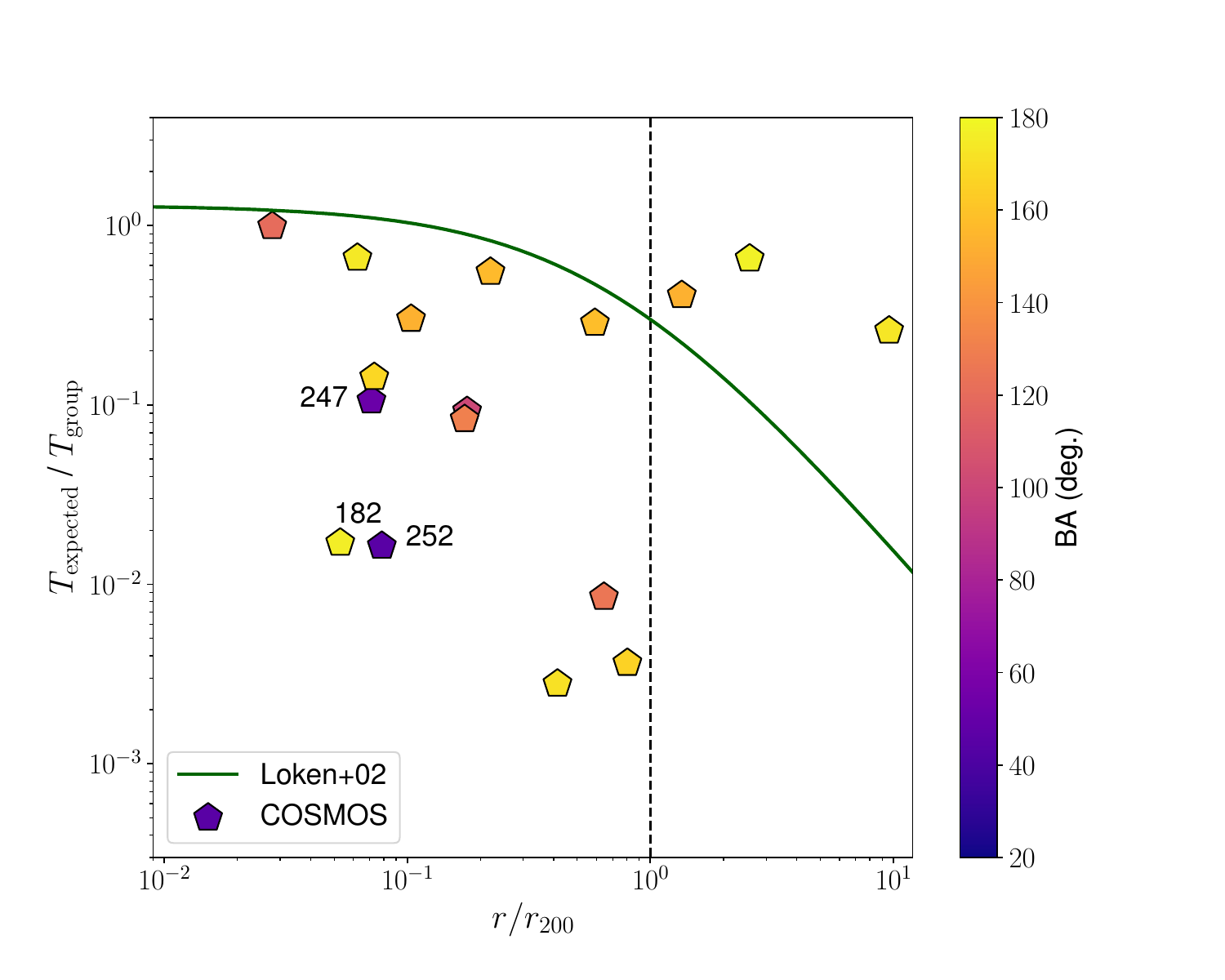}
    \caption{Ratio of the expected temperature from Equation \ref{eq:Texptected} over the mean group temperature as a function of distance from the group centre for COSMOS sources within $10r_{200}$ and $\Delta v / \sigma_{\rm disp}<2.7$. We show the universal galaxy cluster temperature profile of \cite{loken2002universal}, given by $T/T_0=1.3(1+1.5r/r_{200})^{-1.6}$. The dashed line shows the distance corresponding to the virial radius $r_{200}$ from the group centre. The IDs of some sources are annotated. The BA of the sources is given by a colour scale.}
    \label{fig:T_exp_groups_r_r200}
\end{figure}

Figure \ref{fig:T_exp_groups_r_r200} shows the ratio of the expected temperature from Equation \ref{eq:Texptected} over the mean group temperature as a function of distance from the group centre for COSMOS sources within $10r_{200}$ and $\Delta v / \sigma_{\rm disp}<2.7$. Under the assumption that the jet bending is caused by the galaxy moving across the group medium, one would expect the sources to follow a typical temperature profile of groups and clusters, here shown by the universal galaxy cluster temperature profile of \cite{loken2002universal}, given by $T/T_0=1.3(1+1.5r/r_{200})^{-1.6}$. Compared to \cite{loken2002universal}, the profile we calculate for the COSMOS sources within $r_{200}$ shows a steeper temperature gradient, while this is not seen for the three sources outside $r_{200}$. We cannot assign these sources securely to groups, thus we cannot rule out the possibility that their $\Delta v$, and therefore $T_{\rm expected}$ values are overestimated. On the other hand, the steep negative temperature gradient observed for sources within $r_{200}$ could be evidence that the jet bending of these sources is a result of interactions with large-scale structures such as the medium of superclusters or the warm-hot intergalactic medium (WHIM), which would be at lower temperatures than the group medium. Typical temperatures for the WHIM are in the range $10^{5}–10^{7}$ K \citep[see][ and references therein]{zhao2024}. The WHIM's influence on the thermal properties of galaxy clusters is supported by both observations and simulations \citep[see][ and references therein]{popping2007}, though it is one of several factors contributing to the complex temperature profiles observed in these massive structures. 

As we previously mentioned, Source 252, which strongly deviates from the expected temperature profile, is part of a formation process of a large galaxy cluster, where material is accreted from the filaments of superclusters \citep{smolvcic2007wide}, further supporting that the media of large-scale structures can play a crucial role in shaping the jets of radio sources, which could also explain the jet bending we observe for Source 247 (see Appendix~\ref{peculiar Sources}). Additionally, in their work on simulating radio jet distortion in cluster mergers, \cite{paola2024} discuss the creation of WAT sources as part of the early evolution
of the jet and cluster merger, only forming if the burst precedes the first core passage. They note that the distortion of the radio jets is primarily influenced by bulk motions rather than the presence of substructure, and WAT type radio AGN arise regardless of the merger’s mass ratio, but depend on the jet’s initial burst timing and the position of the minor cluster. In contrast, the straight Source 182 seen in Figure \ref{fig:T_exp_groups_r_r200}, which is the BGG of its group, is moving slowly relative to its group ($\Delta v\approx50$ km/s), thus resulting in a low $T_{\rm expected}$ estimation from Equation \ref{eq:Texptected}. Since jet bending is a complex phenomenon that is not explained by a single mechanism, this scenario will not explain all sources in our sample. However, the action of the temperature profile is evident as defining the upper limit of the bending angles. Stronger bending is left unaccounted, since it can be the result of the projection effects which can only reduce the observed bending angles but cannot increase it.

\section{Conclusions}
\label{Conclusions}
In this work, we investigated the bent radio sources located in the COSMOS and XMM-LSS fields, selected from visual inspection of the MIGHTEE-DR1 radio survey at $\sim$1.2-1.3 GHz with beam sizes of 8.9" and $\sim$5", and median central $rms$ of $\sim$3.2-3.5 $\mu$Jy/beam and $\sim$5.1-5.6 $\mu$Jy/beam, respectively. We found 217 objects in XMM-LSS and 142 objects in COSMOS where we could robustly measure the bending angle, i.e. the angle formed between the jets/lobes of a two-sided source. From these, we studied the bent radio AGN that lie within X-ray galaxy groups, 17 sources in XMM-LSS and 19 sources in COSMOS. The latter lie within groups with halo masses $2\times10^{13} \lesssim M_{\rm 200c}/M_{\odot} = 3\times10^{14}$. We thus investigated the relations between the bending angle and the large-scale environment probed by the X-ray galaxy groups. We compared two methods of obtaining the bending angle and further compared to studies with other methodologies and radio data of different frequencies, sensitivities and resolution to confirm that the bending angle is a good method to characterise jet bending. We summarise our findings in the following:\\

\begin{enumerate}

    \item There is an indication that a larger number of bent sources (BA$\leq160^{\circ}$) are found at lower redshifts, in particular for objects inside X-ray galaxy groups and for halo masses $\geq 10^{13.5}M_{\odot}$, bar the small number statistics. This trend of BA with redshift persist for COSMOS and for XMM-LSS sources when we apply a halo mass cut.
  
    Furthermore, we find all very bent sources (BA$\leq100^{\circ}$) at $z\leq1$ and only straight or slightly bent sources (BA $>160^{\circ}$) at $z\ge1.5$. 
    We speculate that lower redshift radio galaxies are statistically more bent, as they had more time for interactions with the environment. The latter is supported by the study of \cite{vardoulaki2021bent} of MHD simulated radio sources at $z = 0.5$ and $z = 1$ using the COSMOS bent radio AGN.
    
    \item From comparisons to simulations \citep{mguda2015ram,marshall2018triggering}, we find that the jet bending of the sources in the X-ray galaxy groups in COSMOS and XMM-LSS can be explained by ram pressure that is exerted on jets for the halo masses in the range M$_{200}=$ \SI{2e13}{M_{\odot}} to \SI{2.2e14}{M_\odot}. No correlation between bending angle and halo mass or group temperature is found for our samples, either because finding bent sources in these halo mass ranges is rare or because interactions with superclusters play a dominant role for the jet bending. 

    \item We found a strong correlation between bending angle and projected distance from the X-ray galaxy groups centre in the XMM-LSS field. This relation is seen in some studies in literature \citep{moravec2019massive}, but not in others \citep{golden2021high}, and is also not seen in the COSMOS field, where we observe a large scatter. The relationship between BA and distance from the group centre is not straightforward. One of the parameters in play is the halo mass of the galaxy groups/clusters, but the type of IGM/ICM medium also plays an important role in jet bending.
       
    \item We estimate the expected temperature of the medium that the jets are interacting with from the bending angle. For COSMOS group members, we find a steeper temperature profile than one would expect from galaxy groups, suggesting the sources are interacting with the colder medium of superclusters or the warm-hot intergalactic medium (WHIM). While the role of the WHIM in influencing jet bending is an intriguing possibility supported by environmental conditions and indirect evidence, direct observational confirmation remains elusive, warranting further investigation.
    
    \item We found jets to be $\sim25\%$ less bent for sources larger than \SI{500}{kpc} in X-ray galaxy groups compared to smaller sources, which is attributed to denser group environments that hinder jet expansion and promote ram pressure induced jet bending. Less powerful sources in dense group environments are smaller compared to more powerful sources in less dense group environments. Out of 11 giant radio galaxies (2 sources in COSMOS and 9 sources in XMM-LSS) in the X-ray coverage of our samples, only 1 is found inside a group environment with halo mass of M$_{\rm 200}\sim 2\times10^{13}\text{M}_{\odot}$. Incidentally, this is the largest source in the COSMOS sample.
       
    \item The median bending angle for members of X-ray galaxy groups is smaller (COSMOS: $156.0^{171.1}_{117.7}$ deg.; XMM-LSS: $140.0^{172.9}_{104.8}$ deg.) than for objects in the field (COSMOS: $168.0^{177.0}_{140.8}$ deg.; XMM-LSS: $169.0^{176.0}_{146.0}$ deg.). Likewise, we find that sources located in the \emph{core region} ($r/r_{200}<0.1$) are more bent (COSMOS: $149.5^{167.4}_{128.0}$ deg.; XMM-LSS: $124.0^{142.1}_{107.5}$ deg.) than the sources in the \emph{inner region} ($0.1<r/r_{200}<1$) of galaxy groups (COSMOS: $156.0^{167.4}_{128.0}$ deg.; XMM-LSS: $159.0^{174.1}_{95.7}$ deg.). The differences are larger for the XMM-LSS group members, either due to small number statistics or because the COSMOS groups are selected from lower group halo masses and temperatures. Larger samples of bent source in galaxy groups are needed for statistically robust results. We propose that the bent sources of our samples can be used as tracers for galaxy groups below the X-ray detection limit (with fluxes below $\approx$ \SI{3e-16}{ergs^{-1}cm^{-2}s^{-1}} and masses below $\approx1.5 (1+z)\times10^{13}$M$_{\odot}$ \citep{vardoulaki2019closer} or outside the X-ray coverage.
    
    \item The very bent sources of our samples (BA$\leq100^{\circ}$) show no intrinsic differences to the rest of our samples. In galaxy groups, we find 2 very bent WATs (BA<$55^{\circ}$) in COSMOS and 2 very bent NATs (BA<$25^{\circ}$) in XMM-LSS. The 2 WATs are dominant group members (brightest and second brightest group galaxy) in groups with lower halo masses and temperatures compared to the 2 NATs, who are farther away from the group centre (infalling). This indicates that the groups of the 2 WATs are not relaxed and that the 2 NATs are moving at high velocities through the ICM.

\end{enumerate}

Although the bending angle is a good approximation for studying the distortion of the radio structure of two-sided radio AGN, one of the biggest limitations of the bending angle is the projection of radio jets on the sky, as bending from projection effects is disconnected from physical properties of the radio galaxy and its environment. One can model the projection of jets assuming the viewing angle with respect to the line of sight \citep[e.g.][]{sawant2022jetcurry}. This can be useful for investigating individual sources, but is not a feasible approach for large samples. Neural networks show promising results for deprojection tasks \citep[e.g.][]{balakrishnan2019visual} and could be used in the future to correct for the projection effects of radio jets. Nevertheless, deprojecting all sources of our samples is not trivial, and is out of the scope of this paper.

We have demonstrated the necessity of small sample studies in deep fields, with state-of-the-art multi-wavelength data sets to investigate in depth populations of bent radio AGN, their properties, host stellar mass and large-scale environment. As radio astronomy is evolving and going all-sky, studies like these are crucial for training machine learning algorithms to identify bent radio AGN and investigate relations with their large scale environment.

\begin{acknowledgements}
      VB would like to thank Heinz Andernach, Aritra Basu, Shubham Bhagat, Vijay Mahatma and Gülay Gürkan for their support. Sections of the this paper are part of the Master's thesis of VB. The writing of this manuscript was finalised by the supervisor with the valuable support of the co-authors. EV acknowledges support from Carl Zeiss Stiftung with the project code KODAR. We gratefully acknowledge the contributions of the entire COSMOS collaboration consisting of more than 200 scientists. AF acknowledges the hospitality of the National Observatory of Athens and Prof. Manolis Plionis during the final stages of preparation of this work. FV acknowledges the financial support from the Cariplo "BREAKTHRU" funds Rif: 2022-2088 CUP J33C22004310003, and the usage of computing power the Gauss Centre for Supercomputing e.V. (www.gauss-centre.eu) for supporting this project by providing computing time through the John von Neumann Institute for Computing (NIC) on the GCS Supercomputer JUWELS at J\"ulich Supercomputing Centre (JSC), under projects "radgalicm2". CLH, IHW and MJJ acknowledge support from the Oxford Hintze Centre for Astrophysical Surveys which is funded through generous support from the Hintze Family Charitable Foundation. The MeerKAT telescope is operated by the South African Radio Astronomy Observatory (SARAO; \url{www.ska.ac.za}), which is a facility of the National Research Foundation (NRF), an agency of the Department of Science and Innovation. We acknowledge the use of the ilifu cloud computing facility – \url{www.ilifu.ac.za}, a partnership between the University of Cape Town, the University of the Western Cape, Stellenbosch University, Sol Plaatje University and the Cape Peninsula University of Technology. The Ilifu facility is supported by contributions from the Inter-University Institute for Data Intensive Astronomy (IDIA – a partnership between the University of Cape Town, the University of Pretoria and the University of the Western Cape, the Computational Biology division at UCT and the Data Intensive Research Initiative of South Africa (DIRISA). The authors acknowledge the Centre for High Performance Computing (CHPC), South Africa, for providing computational resources to this research project. The Hyper Suprime-Cam (HSC) collaboration includes the astronomical communities of Japan and Taiwan, and Princeton University. The HSC instrumentation and software were developed by the National Astronomical Observatory of Japan (NAOJ), the Kavli Institute for the Physics and Mathematics of the Universe (Kavli IPMU), the University of Tokyo, the High Energy Accelerator Research Organization (KEK), the Academia Sinica Institute for Astronomy and Astrophysics in Taiwan (ASIAA), and Princeton University. Funding was contributed by the FIRST program from Japanese Cabinet Office, the Ministry of Education, Culture, Sports, Science and Technology (MEXT), the Japan Society for the Promotion of Science (JSPS), Japan Science and Technology Agency (JST), the Toray Science Foundation, NAOJ, Kavli IPMU, KEK, ASIAA, and Princeton University.
\end{acknowledgements}

\noindent
\texttt{Software: scipy \citep{scipy}, ipython \citep{ipython}, matplotlib \citep{matplotlib},  astropy \citep{astropy}, COLOSSUS \citep{diemer2018colossus}, halotools \citep{hearin2017forward}, CASA \citep{mcmullin2007casa}, TOPCAT \citep{taylor2011topcat}}.

%
%

\bibliographystyle{aa}
\bibliography{bibliography}

\begin{thebibliography}{174}
\expandafter\ifx\csname natexlab\endcsname\relax\def\natexlab#1{#1}\fi

\bibitem[{Ahumada {et~al.}(2020)Ahumada, Prieto, Almeida, Anders, Anderson,
  Andrews, Anguiano, Arcodia, Armengaud, Aubert, {et~al.}}]{ahumada202016th}
Ahumada, R., Prieto, C.~A., Almeida, A., {et~al.} 2020, The Astrophysical
  Journal Supplement Series, 249, 3

\bibitem[{Aihara {et~al.}(2022)Aihara, AlSayyad, Ando, Armstrong, Bosch, Egami,
  Furusawa, Furusawa, Harasawa, Harikane, {et~al.}}]{aihara2022third}
Aihara, H., AlSayyad, Y., Ando, M., {et~al.} 2022, Publications of the
  Astronomical Society of Japan, 74, 247

\bibitem[{Aihara {et~al.}(2018)Aihara, Arimoto, Armstrong, Arnouts, Bahcall,
  Bickerton, Bosch, Bundy, Capak, Chan, {et~al.}}]{aihara2018hyper}
Aihara, H., Arimoto, N., Armstrong, R., {et~al.} 2018, Publications of the
  Astronomical Society of Japan, 70, S4

\bibitem[{Alam {et~al.}(2015)Alam, Albareti, Prieto, Anders, Anderson,
  Anderton, Andrews, Armengaud, Aubourg, Bailey, {et~al.}}]{alam2015eleventh}
Alam, S., Albareti, F.~D., Prieto, C.~A., {et~al.} 2015, The Astrophysical
  Journal Supplement Series, 219, 12

\bibitem[{Arnaud {et~al.}(2010)Arnaud, Pratt, Piffaretti, B{\"o}hringer,
  Croston, \& Pointecouteau}]{arnaud2010universal}
Arnaud, M., Pratt, G., Piffaretti, R., {et~al.} 2010, Astronomy \&
  Astrophysics, 517, A92

\bibitem[{Arnouts {et~al.}(2002)Arnouts, Moscardini, Vanzella, Colombi,
  Cristiani, Fontana, Giallongo, Matarrese, \& Saracco}]{arnouts2002measuring}
Arnouts, S., Moscardini, L., Vanzella, E., {et~al.} 2002, Monthly Notices of
  the Royal Astronomical Society, 329, 355

\bibitem[{Ascasibar {et~al.}(2003)Ascasibar, Yepes, M{\"u}ller, \&
  Gottl{\"o}ber}]{ascasibar2003radial}
Ascasibar, Y., Yepes, G., M{\"u}ller, V., \& Gottl{\"o}ber, S. 2003, Monthly
  Notices of the Royal Astronomical Society, 346, 731

\bibitem[{{Astropy Collaboration} {et~al.}(2013){Astropy Collaboration},
  {Robitaille}, {Tollerud}, {Greenfield}, {Droettboom}, {Bray}, {Aldcroft},
  {Davis}, {Ginsburg}, {Price-Whelan}, {Kerzendorf}, {Conley}, {Crighton},
  {Barbary}, {Muna}, {Ferguson}, {Grollier}, {Parikh}, {Nair}, {Unther},
  {Deil}, {Woillez}, {Conseil}, {Kramer}, {Turner}, {Singer}, {Fox}, {Weaver},
  {Zabalza}, {Edwards}, {Azalee Bostroem}, {Burke}, {Casey}, {Crawford},
  {Dencheva}, {Ely}, {Jenness}, {Labrie}, {Lim}, {Pierfederici}, {Pontzen},
  {Ptak}, {Refsdal}, {Servillat}, \& {Streicher}}]{astropy}
{Astropy Collaboration}, {Robitaille}, T.~P., {Tollerud}, E.~J., {et~al.} 2013,
  \aap, 558, A33

\bibitem[{Balakrishnan {et~al.}(2019)Balakrishnan, Dalca, Zhao, Guttag, Durand,
  \& Freeman}]{balakrishnan2019visual}
Balakrishnan, G., Dalca, A.~V., Zhao, A., {et~al.} 2019, in Proceedings of the
  IEEE/CVF International Conference on Computer Vision, 171--180

\bibitem[{Becker {et~al.}(1995)Becker, White, \& Helfand}]{becker1995first}
Becker, R.~H., White, R.~L., \& Helfand, D.~J. 1995, Astrophysical Journal v.
  450, p. 559, 450, 559

\bibitem[{Begelman {et~al.}(1984)Begelman, Blandford, \&
  Rees}]{begelman1984theory}
Begelman, M.~C., Blandford, R.~D., \& Rees, M.~J. 1984, Reviews of Modern
  Physics, 56, 255

\bibitem[{Begelman {et~al.}(1979)Begelman, Rees, \&
  Blandford}]{begelman1979twin}
Begelman, M.~C., Rees, M.~J., \& Blandford, R.~D. 1979, Nature, 279, 770

\bibitem[{Best {et~al.}(1997)Best, Longair, \& Rottgering}]{Best1997AJI}
Best, P.~N., Longair, M.~S., \& Rottgering, H. J.~A. 1997, Monthly Notices of
  the Royal Astronomical Society, 286, 785

\bibitem[{Bielby {et~al.}(2010)Bielby, Finoguenov, Tanaka, McCracken, Daddi,
  Hudelot, Ilbert, Kneib, Le~Fevre, Mellier, {et~al.}}]{bielby2010wircam}
Bielby, R., Finoguenov, A., Tanaka, M., {et~al.} 2010, Astronomy \&
  Astrophysics, 523, A66

\bibitem[{Blanton {et~al.}(2000)Blanton, Gregg, Helfand, Becker, \&
  White}]{blanton2000first}
Blanton, E., Gregg, M., Helfand, D., Becker, R., \& White, R. 2000, The
  Astrophysical Journal, 531, 118

\bibitem[{Blundell \& Rawlings(1999)}]{blundell1999inevitable}
Blundell, K.~M. \& Rawlings, S. 1999, Nature, 399, 330

\bibitem[{Borgani {et~al.}(2004)Borgani, Murante, Springel, Diaferio, Dolag,
  Moscardini, Tormen, Tornatore, \& Tozzi}]{borgani2004x}
Borgani, S., Murante, G., Springel, V., {et~al.} 2004, Monthly Notices of the
  Royal Astronomical Society, 348, 1078

\bibitem[{{Boyce} {et~al.}(2023){Boyce}, {Hopkins}, {Riggi}, {Rudnick},
  {Ramsay}, {Hale}, {Marvil}, {Whiting}, {Venkataraman}, {O'Dea}, {Baum},
  {Gordon}, {Vantyghem}, {Dionyssiou}, {Andernach}, {Collier}, {English},
  {Koribalski}, {Leahy}, {Micha{\l}owski}, {Safi-Harb}, {Vaccari}, {Alexander},
  {Cowley}, {Kapinska}, {Robotham}, \& {Tang}}]{Boyce2023}
{Boyce}, M.~M., {Hopkins}, A.~M., {Riggi}, S., {et~al.} 2023, \pasa, 40, e028

\bibitem[{Bradshaw {et~al.}(2013)Bradshaw, Almaini, Hartley, Smith, Conselice,
  Dunlop, Simpson, Chuter, Cirasuolo, Foucaud, {et~al.}}]{bradshaw2013high}
Bradshaw, E., Almaini, O., Hartley, W., {et~al.} 2013, Monthly Notices of the
  Royal Astronomical Society, 433, 194

\bibitem[{Brammer {et~al.}(2008)Brammer, van Dokkum, \&
  Coppi}]{brammer2008eazy}
Brammer, G.~B., van Dokkum, P.~G., \& Coppi, P. 2008, The Astrophysical
  Journal, 686, 1503

\bibitem[{Capak {et~al.}(2007)Capak, Aussel, Ajiki, McCracken, Mobasher,
  Scoville, Shopbell, Taniguchi, Thompson, Tribiano, {et~al.}}]{capak2007first}
Capak, P., Aussel, H., Ajiki, M., {et~al.} 2007, The Astrophysical Journal
  Supplement Series, 172, 99

\bibitem[{Caproni {et~al.}(2017)Caproni, Abraham, Motter, \&
  Monteiro}]{caproni2017jet}
Caproni, A., Abraham, Z., Motter, J.~C., \& Monteiro, H. 2017, The
  Astrophysical Journal Letters, 851, L39

\bibitem[{{Charlton} {et~al.}(2024){Charlton}, {Delhaize}, {Thorat}, {Heywood},
  {Jarvis}, {Hardcastle}, {An}, {Delvecchio}, {Hale}, {Whittam}, {Br{\"u}ggen},
  {Marchetti}, {Morabito}, {Randriamanakoto}, {White}, \&
  {Taylor}}]{Charlton2024}
{Charlton}, K.~K.~L., {Delhaize}, J., {Thorat}, K., {et~al.} 2024, arXiv
  e-prints, arXiv:2411.06813

\bibitem[{Coil {et~al.}(2011)Coil, Blanton, Burles, Cool, Eisenstein,
  Moustakas, Wong, Zhu, Aird, Bernstein, {et~al.}}]{coil2011prism}
Coil, A.~L., Blanton, M.~R., Burles, S.~M., {et~al.} 2011, The Astrophysical
  Journal, 741, 8

\bibitem[{Colless {et~al.}(2003)Colless, Peterson, Jackson, Peacock, Cole,
  Norberg, Baldry, Baugh, Bland-Hawthorn, Bridges, {et~al.}}]{colless20032df}
Colless, M., Peterson, B.~A., Jackson, C., {et~al.} 2003, arXiv preprint
  astro-ph/0306581

\bibitem[{{Condon}(1992)}]{Condon1992}
{Condon}, J.~J. 1992, \araa, 30, 575

\bibitem[{Cool {et~al.}(2013)Cool, Moustakas, Blanton, Burles, Coil,
  Eisenstein, Wong, Zhu, Aird, Bernstein, {et~al.}}]{cool2013prism}
Cool, R.~J., Moustakas, J., Blanton, M.~R., {et~al.} 2013, The Astrophysical
  Journal, 767, 118

\bibitem[{Cooper {et~al.}(2011)Cooper, Aird, Coil, Davis, Faber, Juneau, Lotz,
  Nandra, Newman, Willmer, {et~al.}}]{cooper2011deep3}
Cooper, M.~C., Aird, J.~A., Coil, A.~L., {et~al.} 2011, The Astrophysical
  Journal Supplement Series, 193, 14

\bibitem[{Croston {et~al.}(2019)Croston, Hardcastle, Mingo, Best, Sabater,
  Shimwell, Williams, Duncan, R{\"o}ttgering, Brienza,
  {et~al.}}]{croston2019environments}
Croston, J., Hardcastle, M., Mingo, B., {et~al.} 2019, Astronomy \&
  Astrophysics, 622, A10

\bibitem[{Darvish {et~al.}(2017)Darvish, Mobasher, Martin, Sobral, Scoville,
  Stroe, Hemmati, \& Kartaltepe}]{darvish2017cosmic}
Darvish, B., Mobasher, B., Martin, D.~C., {et~al.} 2017, The Astrophysical
  Journal, 837, 16

\bibitem[{Darvish~Sarvestani(2015)}]{darvish2015evolution}
Darvish~Sarvestani, B. 2015, PhD thesis, UC Riverside

\bibitem[{Delhaize {et~al.}(2021)Delhaize, Heywood, Prescott, Jarvis,
  Delvecchio, Whittam, White, Hardcastle, Hale, Afonso,
  {et~al.}}]{delhaize2021mightee}
Delhaize, J., Heywood, I., Prescott, M., {et~al.} 2021, Monthly Notices of the
  Royal Astronomical Society, 501, 3833

\bibitem[{Dey {et~al.}(2019)Dey, Schlegel, Lang, Blum, Burleigh, Fan, Findlay,
  Finkbeiner, Herrera, Juneau, {et~al.}}]{dey2019overview}
Dey, A., Schlegel, D.~J., Lang, D., {et~al.} 2019, The Astronomical Journal,
  157, 168

\bibitem[{Diemer(2018)}]{diemer2018colossus}
Diemer, B. 2018, The Astrophysical Journal Supplement Series, 239, 35

\bibitem[{{Dom{\'\i}nguez-Fern{\'a}ndez}
  {et~al.}(2024){Dom{\'\i}nguez-Fern{\'a}ndez}, {ZuHone}, {Weinberger},
  {Bellomi}, {Hernquist}, {Nulsen}, \& {Brunetti}}]{paola2024}
{Dom{\'\i}nguez-Fern{\'a}ndez}, P., {ZuHone}, J., {Weinberger}, R., {et~al.}
  2024, arXiv e-prints, arXiv:2406.19681

\bibitem[{Drinkwater {et~al.}(2010)Drinkwater, Jurek, Blake, Woods, Pimbblet,
  Glazebrook, Sharp, Pracy, Brough, Colless, {et~al.}}]{drinkwater2010wigglez}
Drinkwater, M.~J., Jurek, R.~J., Blake, C., {et~al.} 2010, Monthly Notices of
  the Royal Astronomical Society, 401, 1429

\bibitem[{Edwards {et~al.}(2010)Edwards, Fadda, \& Frayer}]{edwards2010first}
Edwards, L.~O., Fadda, D., \& Frayer, D.~T. 2010, The Astrophysical Journal
  Letters, 724, L143

\bibitem[{Fabian(2012)}]{fabian2012observational}
Fabian, A.~C. 2012, Annual Review of Astronomy and Astrophysics, 50, 455

\bibitem[{Falle(1991)}]{falle1991self}
Falle, S. 1991, Monthly Notices of the Royal Astronomical Society, 250, 581

\bibitem[{Fanaroff \& Riley(1974)}]{fanaroff1974morphology}
Fanaroff, B.~L. \& Riley, J.~M. 1974, Monthly Notices of the Royal Astronomical
  Society, 167, 31P

\bibitem[{Fevre {et~al.}(2013)Fevre, Cassata, Cucciati, Garilli, Ilbert, Brun,
  Maccagni, Moreau, Scodeggio, Tresse, {et~al.}}]{fevre2013vimos}
Fevre, O.~L., Cassata, P., Cucciati, O., {et~al.} 2013, arXiv preprint
  arXiv:1307.0545

\bibitem[{Finoguenov {et~al.}(2009)Finoguenov, Connelly, Parker, Wilman,
  Mulchaey, Saglia, Balogh, Bower, \& McGee}]{finoguenov2009roadmap}
Finoguenov, A., Connelly, J., Parker, L., {et~al.} 2009, The Astrophysical
  Journal, 704, 564

\bibitem[{Finoguenov {et~al.}(2007)Finoguenov, Guzzo, Hasinger, Scoville,
  Aussel, B{\"o}hringer, Brusa, Capak, Cappelluti, Comastri,
  {et~al.}}]{finoguenov2007xmm}
Finoguenov, A., Guzzo, L., Hasinger, G., {et~al.} 2007, The Astrophysical
  Journal Supplement Series, 172, 182

\bibitem[{Freeland \& Wilcots(2011)}]{freeland2011intergalactic}
Freeland, E. \& Wilcots, E. 2011, The Astrophysical Journal, 738, 145

\bibitem[{{Galvin} {et~al.}(2020){Galvin}, {Huynh}, {Norris}, {Wang},
  {Hopkins}, {Polsterer}, {Ralph}, {O'Brien}, \& {Heald}}]{Galvin2020}
{Galvin}, T.~J., {Huynh}, M.~T., {Norris}, R.~P., {et~al.} 2020, \mnras, 497,
  2730

\bibitem[{Garilli {et~al.}(2014)Garilli, Guzzo, Scodeggio, Bolzonella, Abbas,
  Adami, Arnouts, Bel, Bottini, Branchini, {et~al.}}]{garilli2014vimos}
Garilli, B., Guzzo, L., Scodeggio, M., {et~al.} 2014, Astronomy \&
  Astrophysics, 562, A23

\bibitem[{Garon {et~al.}(2019)Garon, Rudnick, Wong, Jones, Kim, Andernach,
  Shabala, Kapi{\'n}ska, Norris, De~Gasperin, {et~al.}}]{garon2019radio}
Garon, A.~F., Rudnick, L., Wong, O.~I., {et~al.} 2019, The Astronomical
  Journal, 157, 126

\bibitem[{Gaspari {et~al.}(2011)Gaspari, Brighenti, D'Ercole, \&
  Melioli}]{gaspari2011agn}
Gaspari, M., Brighenti, F., D'Ercole, A., \& Melioli, C. 2011, Monthly Notices
  of the Royal Astronomical Society, 415, 1549

\bibitem[{Girardi {et~al.}(1995)Girardi, Fadda, Giuricin, Mardirossian,
  Mezzetti, \& Biviano}]{girardi1995velocity}
Girardi, M., Fadda, D., Giuricin, G., {et~al.} 1995, arXiv preprint
  astro-ph/9507031

\bibitem[{Golden-Marx {et~al.}(2019)Golden-Marx, Blanton, Paterno-Mahler,
  Brodwin, Ashby, Lemaux, Lubin, Gal, \& Tomczak}]{golden2019high}
Golden-Marx, E., Blanton, E., Paterno-Mahler, R., {et~al.} 2019, The
  Astrophysical Journal, 887, 50

\bibitem[{Golden-Marx {et~al.}(2021)Golden-Marx, Blanton, Paterno-Mahler,
  Brodwin, Ashby, Moravec, Shen, Lemaux, Lubin, Gal, {et~al.}}]{golden2021high}
Golden-Marx, E., Blanton, E., Paterno-Mahler, R., {et~al.} 2021, The
  Astrophysical Journal, 907, 65

\bibitem[{{Golden-Marx} {et~al.}(2023){Golden-Marx}, {Moravec}, {Shen}, {Cai},
  {Blanton}, {Gendron-Marsolais}, {R{\"o}ttgering}, {van Weeren}, {Buiten},
  {Grumitt}, {Golden-Marx}, {Pinjarkar}, \& {Tang}}]{golden2023}
{Golden-Marx}, E., {Moravec}, E., {Shen}, L., {et~al.} 2023, \apj, 956, 87

\bibitem[{{Gordon} {et~al.}(2020){Gordon}, {Boyce}, {O'Dea}, {Rudnick},
  {Andernach}, {Vantyghem}, {Baum}, {Bui}, \& {Dionyssiou}}]{Gordon2020}
{Gordon}, Y.~A., {Boyce}, M.~M., {O'Dea}, C.~P., {et~al.} 2020, Research Notes
  of the American Astronomical Society, 4, 175

\bibitem[{Gozaliasl {et~al.}(2020)Gozaliasl, Finoguenov, Khosroshahi, Laigle,
  Kirkpatrick, Kiiveri, Devriendt, Dubois, \&
  Ahoranta}]{gozaliasl2020kinematic}
Gozaliasl, G., Finoguenov, A., Khosroshahi, H., {et~al.} 2020, Astronomy \&
  Astrophysics, 635, A36

\bibitem[{Gozaliasl {et~al.}(2014)Gozaliasl, Finoguenov, Khosroshahi,
  Mirkazemi, Salvato, Jassur, Erfanianfar, Popesso, Tanaka, Lerchster,
  {et~al.}}]{gozaliasl2014mining}
Gozaliasl, G., Finoguenov, A., Khosroshahi, H., {et~al.} 2014, Astronomy \&
  Astrophysics, 566, A140

\bibitem[{Gozaliasl {et~al.}(2019)Gozaliasl, Finoguenov, Tanaka, Dolag,
  Montanari, Kirkpatrick, Vardoulaki, Khosroshahi, Salvato, Laigle,
  {et~al.}}]{gozaliasl2019chandra}
Gozaliasl, G., Finoguenov, A., Tanaka, M., {et~al.} 2019, Monthly Notices of
  the Royal Astronomical Society, 483, 3545

\bibitem[{G{\"u}rkan {et~al.}(2018)G{\"u}rkan, Hardcastle, Smith, Best, Bourne,
  Calistro-Rivera, Heald, Jarvis, Prandoni, R{\"o}ttgering,
  {et~al.}}]{gurkan2018lofar}
G{\"u}rkan, G., Hardcastle, M.~J., Smith, D.~J., {et~al.} 2018, Monthly Notices
  of the Royal Astronomical Society, 475, 3010

\bibitem[{Hale {et~al.}(2018)Hale, Jarvis, Delvecchio, Hatfield, Novak,
  Smol{\v{c}}i{\'c}, \& Zamorani}]{hale2018clustering}
Hale, C., Jarvis, M., Delvecchio, I., {et~al.} 2018, Monthly Notices of the
  Royal Astronomical Society, 474, 4133

\bibitem[{{Hale} {et~al.}(2024){Hale}, {Heywood}, {Jarvis}, {Whittam}, {Best},
  {An}, {Bowler}, {Harrison}, {Matthews}, {Smith}, {Taylor}, \&
  {Vaccari}}]{Hale2024}
{Hale}, C.~L., {Heywood}, I., {Jarvis}, M.~J., {et~al.} 2024, arXiv e-prints,
  arXiv:2411.04958

\bibitem[{Hardcastle {et~al.}(2019)Hardcastle, Williams, Best, Croston, Duncan,
  R{\"o}ttgering, Sabater, Shimwell, Tasse, Callingham,
  {et~al.}}]{hardcastle2019radio}
Hardcastle, M., Williams, W., Best, P., {et~al.} 2019, Astronomy \&
  Astrophysics, 622, A12

\bibitem[{Hardcastle \& Sakelliou(2004)}]{hardcastle2004jet}
Hardcastle, M.~J. \& Sakelliou, I. 2004, Monthly Notices of the Royal
  Astronomical Society, 349, 560

\bibitem[{Hasinger {et~al.}(2018)Hasinger, Capak, Salvato, Barger, Cowie,
  Faisst, Hemmati, Kakazu, Kartaltepe, Masters, {et~al.}}]{hasinger2018deimos}
Hasinger, G., Capak, P., Salvato, M., {et~al.} 2018, The Astrophysical Journal,
  858, 77

\bibitem[{{Hatfield} {et~al.}(2022){Hatfield}, {Jarvis}, {Adams}, {Bowler},
  {H{\"a}u{\ss}ler}, \& {Duncan}}]{Hatfield2022}
{Hatfield}, P.~W., {Jarvis}, M.~J., {Adams}, N., {et~al.} 2022, \mnras, 513,
  3719

\bibitem[{Hearin {et~al.}(2017)Hearin, Campbell, Tollerud, Behroozi, Diemer,
  Goldbaum, Jennings, Leauthaud, Mao, More, {et~al.}}]{hearin2017forward}
Hearin, A.~P., Campbell, D., Tollerud, E., {et~al.} 2017, The Astronomical
  Journal, 154, 190

\bibitem[{{Helfand} {et~al.}(2015){Helfand}, {White}, \&
  {Becker}}]{Helfand2015}
{Helfand}, D.~J., {White}, R.~L., \& {Becker}, R.~H. 2015, \apj, 801, 26

\bibitem[{Helsdon \& Ponman(2000)}]{helsdon2000intragroup}
Helsdon, S.~F. \& Ponman, T.~J. 2000, Monthly Notices of the Royal Astronomical
  Society, 315, 356

\bibitem[{Hern{\'a}ndez-Lang {et~al.}(2022)Hern{\'a}ndez-Lang, Mohr, Klein,
  Grandis, Melin, Tarr{\'\i}o, Arnaud, Pratt, Abbott, Aguena,
  {et~al.}}]{hernandez2022madpsz}
Hern{\'a}ndez-Lang, D., Mohr, J., Klein, M., {et~al.} 2022, arXiv preprint
  arXiv:2210.04666

\bibitem[{Heywood {et~al.}(2022)Heywood, Jarvis, Hale, Whittam, Bester, Hugo,
  Kenyon, Prescott, Smirnov, Tasse, {et~al.}}]{heywood2022mightee}
Heywood, I., Jarvis, M., Hale, C., {et~al.} 2022, Monthly Notices of the Royal
  Astronomical Society, 509, 2150

\bibitem[{Hintzen(1984)}]{hintzen1984wide}
Hintzen, P. 1984, The Astrophysical Journal Supplement Series, 55, 533

\bibitem[{Hunter(2007)}]{matplotlib}
Hunter, J.~D. 2007, Computing In Science \& Engineering, 9, 90

\bibitem[{{Hurley-Walker} {et~al.}(2017){Hurley-Walker}, {Callingham},
  {Hancock}, {Franzen}, {Hindson}, {Kapi{\'n}ska}, {Morgan}, {Offringa},
  {Wayth}, {Wu}, {Zheng}, {Murphy}, {Bell}, {Dwarakanath}, {For}, {Gaensler},
  {Johnston-Hollitt}, {Lenc}, {Procopio}, {Staveley-Smith}, {Ekers}, {Bowman},
  {Briggs}, {Cappallo}, {Deshpande}, {Greenhill}, {Hazelton}, {Kaplan},
  {Lonsdale}, {McWhirter}, {Mitchell}, {Morales}, {Morgan}, {Oberoi}, {Ord},
  {Prabu}, {Shankar}, {Srivani}, {Subrahmanyan}, {Tingay}, {Webster},
  {Williams}, \& {Williams}}]{Hurley-Walker2017}
{Hurley-Walker}, N., {Callingham}, J.~R., {Hancock}, P.~J., {et~al.} 2017,
  \mnras, 464, 1146

\bibitem[{Ider~Chitham {et~al.}(2020)Ider~Chitham, Comparat, Finoguenov, Clerc,
  Kirkpatrick, Damsted, Kukkola, Nandra, Merloni,
  {et~al.}}]{ider2020cosmological}
Ider~Chitham, J., Comparat, J., Finoguenov, A., {et~al.} 2020, Monthly Notices
  of the Royal Astronomical Society, 499, 4768

\bibitem[{Ilbert {et~al.}(2006)Ilbert, Arnouts, Mccracken, Bolzonella, Bertin,
  Le~F{\`e}vre, Mellier, Zamorani, Pello, Iovino,
  {et~al.}}]{ilbert2006accurate}
Ilbert, O., Arnouts, S., Mccracken, H.~J., {et~al.} 2006, Astronomy \&
  Astrophysics, 457, 841

\bibitem[{Ilbert {et~al.}(2008)Ilbert, Capak, Salvato, Aussel, McCracken,
  Sanders, Scoville, Kartaltepe, Arnouts, Le~Floc'h,
  {et~al.}}]{ilbert2008cosmos}
Ilbert, O., Capak, P., Salvato, M., {et~al.} 2008, The Astrophysical Journal,
  690, 1236

\bibitem[{Ilbert {et~al.}(2013)Ilbert, McCracken, Le~F{\`e}vre, Capak, Dunlop,
  Karim, Renzini, Caputi, Boissier, Arnouts, {et~al.}}]{ilbert2013mass}
Ilbert, O., McCracken, H.~J., Le~F{\`e}vre, O., {et~al.} 2013, Astronomy \&
  Astrophysics, 556, A55

\bibitem[{Jarvis {et~al.}(2017)Jarvis, Taylor, Agudo, Allison, Deane, Frank,
  Gupta, Heywood, Maddox, McAlpine, {et~al.}}]{jarvis2017meerkat}
Jarvis, M.~J., Taylor, A., Agudo, I., {et~al.} 2017, arXiv preprint
  arXiv:1709.01901

\bibitem[{Jonas \& Team(2016)}]{jonas2016meerkat}
Jonas, J. \& Team, M. 2016, MeerKAT Science: On the Pathway to the SKA, 1

\bibitem[{Jones {et~al.}(2009)Jones, Read, Saunders, Colless, Jarrett, Parker,
  Fairall, Mauch, Sadler, Watson, {et~al.}}]{jones20096df}
Jones, D.~H., Read, M.~A., Saunders, W., {et~al.} 2009, Monthly Notices of the
  Royal Astronomical Society, 399, 683

\bibitem[{Jones \& Owen(1979)}]{jones1979hot}
Jones, T. \& Owen, F. 1979, Astrophysical Journal, Part 1, vol. 234, Dec. 15,
  1979, p. 818-824. Research supported by the University of Minnesota., 234,
  818

\bibitem[{Kelson {et~al.}(2014)Kelson, Williams, Dressler, McCarthy, Shectman,
  Mulchaey, Villanueva, Crane, \& Quadri}]{kelson2014carnegie}
Kelson, D.~D., Williams, R.~J., Dressler, A., {et~al.} 2014, The Astrophysical
  Journal, 783, 110

\bibitem[{Klein {et~al.}(2018)Klein, Lisenfeld, \& Verley}]{klein2018radio}
Klein, U., Lisenfeld, U., \& Verley, S. 2018, Astronomy \& Astrophysics, 611,
  A55

\bibitem[{{Kluge} {et~al.}(2024){Kluge}, {Comparat}, {Liu}, {Balzer}, {Bulbul},
  {Ider Chitham}, {Ghirardini}, {Garrel}, {Bahar}, {Artis}, {Bender}, {Clerc},
  {Dwelly}, {Fabricius}, {Grandis}, {Hern{\'a}ndez-Lang}, {Hill}, {Joshi},
  {Lamer}, {Merloni}, {Nandra}, {Pacaud}, {Predehl}, {Ramos-Ceja}, {Reiprich},
  {Salvato}, {Sanders}, {Schrabback}, {Seppi}, {Zelmer}, {Zenteno}, \&
  {Zhang}}]{Kluge2024}
{Kluge}, M., {Comparat}, J., {Liu}, A., {et~al.} 2024, \aap, 688, A210

\bibitem[{Lacy {et~al.}(2020)Lacy, Baum, Chandler, Chatterjee, Clarke, Deustua,
  English, Farnes, Gaensler, Gugliucci, {et~al.}}]{lacy2020karl}
Lacy, M., Baum, S., Chandler, C., {et~al.} 2020, Publications of the
  Astronomical Society of the Pacific, 132, 035001

\bibitem[{Laigle {et~al.}(2016)Laigle, McCracken, Ilbert, Hsieh, Davidzon,
  Capak, Hasinger, Silverman, Pichon, Coupon, {et~al.}}]{laigle2016cosmos2015}
Laigle, C., McCracken, H.~J., Ilbert, O., {et~al.} 2016, The Astrophysical
  Journal Supplement Series, 224, 24

\bibitem[{Leauthaud {et~al.}(2009)Leauthaud, Finoguenov, Kneib, Taylor, Massey,
  Rhodes, Ilbert, Bundy, Tinker, George, {et~al.}}]{leauthaud2009weak}
Leauthaud, A., Finoguenov, A., Kneib, J.-P., {et~al.} 2009, The Astrophysical
  Journal, 709, 97

\bibitem[{Lilly {et~al.}(2009)Lilly, Le~Brun, Maier, Mainieri, Mignoli,
  Scodeggio, Zamorani, Carollo, Contini, Kneib, {et~al.}}]{lilly2009zcosmos}
Lilly, S.~J., Le~Brun, V., Maier, C., {et~al.} 2009, The Astrophysical Journal
  Supplement Series, 184, 218

\bibitem[{Liske {et~al.}(2015)Liske, Baldry, Driver, Tuffs, Alpaslan, Andrae,
  Brough, Cluver, Grootes, Gunawardhana, {et~al.}}]{liske2015galaxy}
Liske, J., Baldry, I.~K., Driver, S.~P., {et~al.} 2015, Monthly Notices of the
  Royal Astronomical Society, 452, 2087

\bibitem[{Loken {et~al.}(2002)Loken, Norman, Nelson, Burns, Bryan, \&
  Motl}]{loken2002universal}
Loken, C., Norman, M.~L., Nelson, E., {et~al.} 2002, The Astrophysical Journal,
  579, 571

\bibitem[{Lubin \& Bahcall(1993)}]{lubin1993relation}
Lubin, L.~M. \& Bahcall, N.~A. 1993, The Astrophysical Journal, 415, L17

\bibitem[{Magliocchetti(2022)}]{magliocchetti2022hosts}
Magliocchetti, M. 2022, The Astronomy and Astrophysics Review, 30, 6

\bibitem[{Mahatma {et~al.}(2023)Mahatma, Basu, Hardcastle, Morabito, \& van
  Weeren}]{mahatma2023low}
Mahatma, V., Basu, A., Hardcastle, M., Morabito, L., \& van Weeren, R. 2023,
  Monthly Notices of the Royal Astronomical Society, 520, 4427

\bibitem[{Malarecki {et~al.}(2015)Malarecki, Jones, Saripalli, Staveley-Smith,
  \& Subrahmanyan}]{malarecki2015giant}
Malarecki, J.~M., Jones, D.~H., Saripalli, L., Staveley-Smith, L., \&
  Subrahmanyan, R. 2015, Monthly Notices of the Royal Astronomical Society,
  449, 955

\bibitem[{Mamon {et~al.}(2013)Mamon, Biviano, \& Bou{\'e}}]{mamon2013mamposst}
Mamon, G.~A., Biviano, A., \& Bou{\'e}, G. 2013, Monthly Notices of the Royal
  Astronomical Society, 429, 3079

\bibitem[{Marshall {et~al.}(2018)Marshall, Shabala, Krause, Pimbblet, Croton,
  \& Owers}]{marshall2018triggering}
Marshall, M.~A., Shabala, S.~S., Krause, M.~G., {et~al.} 2018, Monthly Notices
  of the Royal Astronomical Society, 474, 3615

\bibitem[{{Massey} {et~al.}(2011){Massey}, {Kitching}, \& {Nagai}}]{Massey2011}
{Massey}, R., {Kitching}, T., \& {Nagai}, D. 2011, \mnras, 413, 1709

\bibitem[{Masters {et~al.}(2017)Masters, Stern, Cohen, Capak, Rhodes,
  Castander, \& Paltani}]{masters2017complete}
Masters, D.~C., Stern, D.~K., Cohen, J.~G., {et~al.} 2017, The Astrophysical
  Journal, 841, 111

\bibitem[{Masters {et~al.}(2019)Masters, Stern, Cohen, Capak, Stanford,
  Hernitschek, Galametz, Davidzon, Rhodes, Sanders,
  {et~al.}}]{masters2019complete}
Masters, D.~C., Stern, D.~K., Cohen, J.~G., {et~al.} 2019, The Astrophysical
  Journal, 877, 81

\bibitem[{McLure {et~al.}(2013)McLure, Pearce, Dunlop, Cirasuolo, Curtis-Lake,
  Bruce, Caputi, Almaini, Bonfield, Bradshaw, {et~al.}}]{mclure2013sizes}
McLure, R., Pearce, H., Dunlop, J., {et~al.} 2013, Monthly Notices of the Royal
  Astronomical Society, 428, 1088

\bibitem[{McMullin {et~al.}(2007)McMullin, Waters, Schiebel, Young, \&
  Golap}]{mcmullin2007casa}
McMullin, J.~P., Waters, B., Schiebel, D., Young, W., \& Golap, K. 2007, in
  Astronomical data analysis software and systems XVI, Vol. 376, 127

\bibitem[{Mendygral {et~al.}(2012)Mendygral, Jones, \&
  Dolag}]{mendygral2012mhd}
Mendygral, P., Jones, T., \& Dolag, K. 2012, The Astrophysical Journal, 750,
  166

\bibitem[{Mguda {et~al.}(2015)Mguda, Faltenbacher, Heyden, Gottl{\"o}ber,
  Cress, Vaisanen, \& Yepes}]{mguda2015ram}
Mguda, Z., Faltenbacher, A., Heyden, K. v.~d., {et~al.} 2015, Monthly Notices
  of the Royal Astronomical Society, 446, 3310

\bibitem[{{Miley}(1980)}]{Miley1980}
{Miley}, G. 1980, \araa, 18, 165

\bibitem[{Mingo {et~al.}(2019)Mingo, Croston, Hardcastle, Best, Duncan,
  Morganti, Rottgering, Sabater, Shimwell, Williams,
  {et~al.}}]{mingo2019revisiting}
Mingo, B., Croston, J., Hardcastle, M., {et~al.} 2019, Monthly Notices of the
  Royal Astronomical Society, 488, 2701

\bibitem[{{Mohan} \& {Rafferty}(2015)}]{mohan2015}
{Mohan}, N. \& {Rafferty}, D. 2015, {PyBDSF: Python Blob Detection and Source
  Finder}, Astrophysics Source Code Library, record ascl:1502.007

\bibitem[{Momcheva {et~al.}(2016)Momcheva, Brammer, Van~Dokkum, Skelton,
  Whitaker, Nelson, Fumagalli, Maseda, Leja, Franx, {et~al.}}]{momcheva20163d}
Momcheva, I.~G., Brammer, G.~B., Van~Dokkum, P.~G., {et~al.} 2016, The
  Astrophysical Journal Supplement Series, 225, 27

\bibitem[{Moravec {et~al.}(2019)Moravec, Gonzalez, Stern, Brodwin, Clarke,
  Decker, Eisenhardt, Mo, O’Donnell, Pope, {et~al.}}]{moravec2019massive}
Moravec, E., Gonzalez, A.~H., Stern, D., {et~al.} 2019, The Astrophysical
  Journal, 871, 186

\bibitem[{Moravec {et~al.}(2020)Moravec, Gonzalez, Stern, Clarke, Brodwin,
  Decker, Eisenhardt, Mo, Pope, Stanford, {et~al.}}]{moravec2020massive}
Moravec, E., Gonzalez, A.~H., Stern, D., {et~al.} 2020, The Astrophysical
  Journal, 888, 74

\bibitem[{More {et~al.}(2009)More, Van Den~Bosch, \&
  Cacciato}]{more2009satellite}
More, S., Van Den~Bosch, F.~C., \& Cacciato, M. 2009, Monthly Notices of the
  Royal Astronomical Society, 392, 917

\bibitem[{Morris {et~al.}(2022)Morris, Wilcots, Hooper, \&
  Heinz}]{morris2022does}
Morris, M.~E., Wilcots, E., Hooper, E., \& Heinz, S. 2022, The Astronomical
  Journal, 163, 280

\bibitem[{Muzzin {et~al.}(2013)Muzzin, Marchesini, Stefanon, Franx,
  Milvang-Jensen, Dunlop, Fynbo, Brammer, Labb{\'e}, \&
  Van~Dokkum}]{muzzin2013public}
Muzzin, A., Marchesini, D., Stefanon, M., {et~al.} 2013, The Astrophysical
  Journal Supplement Series, 206, 8

\bibitem[{Nagai {et~al.}(2007)Nagai, Kravtsov, \& Vikhlinin}]{nagai2007effects}
Nagai, D., Kravtsov, A.~V., \& Vikhlinin, A. 2007, The Astrophysical Journal,
  668, 1

\bibitem[{Navarro(1996)}]{navarro1996structure}
Navarro, J.~F. 1996, in Symposium-international astronomical union, Vol. 171,
  Cambridge University Press, 255--258

\bibitem[{Navarro {et~al.}(1995)Navarro, Frenk, \&
  White}]{navarro1995simulations}
Navarro, J.~F., Frenk, C.~S., \& White, S.~D. 1995, Monthly Notices of the
  Royal Astronomical Society, 275, 720

\bibitem[{Navarro {et~al.}(1997)Navarro, Frenk, \&
  White}]{navarro1997universal}
Navarro, J.~F., Frenk, C.~S., \& White, S.~D. 1997, The Astrophysical Journal,
  490, 493

\bibitem[{{Neronov} {et~al.}(2024){Neronov}, {Vazza}, {Brandenburg}, \&
  {Caprini}}]{neronov2024}
{Neronov}, A., {Vazza}, F., {Brandenburg}, A., \& {Caprini}, C. 2024, arXiv
  e-prints, arXiv:2411.01640

\bibitem[{Newman {et~al.}(2013)Newman, Cooper, Davis, Faber, Coil,
  Guhathakurta, Koo, Phillips, Conroy, Dutton, {et~al.}}]{newman2013deep2}
Newman, J.~A., Cooper, M.~C., Davis, M., {et~al.} 2013, The Astrophysical
  Journal Supplement Series, 208, 5

\bibitem[{Nishizawa {et~al.}(2020)Nishizawa, Hsieh, Tanaka, \&
  Takata}]{nishizawa2020photometric}
Nishizawa, A.~J., Hsieh, B.-C., Tanaka, M., \& Takata, T. 2020, arXiv preprint
  arXiv:2003.01511

\bibitem[{O'Dea \& Owen(1985)}]{o1985global}
O'Dea, C.~P. \& Owen, F. 1985, The Astronomical Journal, 90, 954

\bibitem[{O'Donoghue {et~al.}(1993)O'Donoghue, Eilek, \& Owen}]{o1993flow}
O'Donoghue, A.~A., Eilek, J.~A., \& Owen, F.~N. 1993, Astrophysical Journal,
  Part 1 (ISSN 0004-637X), vol. 408, no. 2, p. 428-445., 408, 428

\bibitem[{Owen \& Rudnick(1976)}]{owen1976radio}
Owen, F.~N. \& Rudnick, L. 1976, Astrophysical Journal, Lett., Vol. 205, p.
  L1-L4, 205, L1

\bibitem[{O’Dea \& Baum(2023)}]{o2023wide}
O’Dea, C.~P. \& Baum, S.~A. 2023, Galaxies, 11, 67

\bibitem[{{Padovani} {et~al.}(2017){Padovani}, {Alexander}, {Assef}, {De
  Marco}, {Giommi}, {Hickox}, {Richards}, {Smol{\v{c}}i{\'c}},
  {Hatziminaoglou}, {Mainieri}, \& {Salvato}}]{Padovani2017}
{Padovani}, P., {Alexander}, D.~M., {Assef}, R.~J., {et~al.} 2017, \aapr, 25, 2

\bibitem[{Pentericci {et~al.}(2018)Pentericci, McLure, Garilli, Cucciati,
  Franzetti, Iovino, Amorin, Bolzonella, Bongiorno, Carnall,
  {et~al.}}]{pentericci2018vandels}
Pentericci, L., McLure, R., Garilli, B., {et~al.} 2018, Astronomy \&
  Astrophysics, 616, A174

\bibitem[{P\'erez \& Granger(2007)}]{ipython}
P\'erez, F. \& Granger, B.~E. 2007, Computing in Science and Engineering, 9, 21

\bibitem[{Perley {et~al.}(1979)Perley, Willis, \& Scott}]{perley1979structure}
Perley, R., Willis, A., \& Scott, J. 1979, Nature, 281, 437

\bibitem[{Pinjarkar {et~al.}(2023)Pinjarkar, Hardcastle, Harwood, Lal,
  Hatfield, Jarvis, Randriamanakoto, \& Whittam}]{SiddhantRLAGN}
Pinjarkar, S., Hardcastle, M.~J., Harwood, J.~J., {et~al.} 2023, Monthly
  Notices of the Royal Astronomical Society, 523, 620

\bibitem[{{Polsterer} {et~al.}(2019){Polsterer}, {Gieseke}, \&
  {Doser}}]{Polsterer2019}
{Polsterer}, K.~L., {Gieseke}, F., \& {Doser}, B. 2019, {PINK: Parallelized
  rotation and flipping INvariant Kohonen maps}, Astrophysics Source Code
  Library, record ascl:1910.001

\bibitem[{Ponman {et~al.}(1999)Ponman, Cannon, \& Navarro}]{ponman1999thermal}
Ponman, T.~J., Cannon, D.~B., \& Navarro, J.~F. 1999, Nature, 397, 135

\bibitem[{{Popping} \& {Braun}(2007)}]{popping2007}
{Popping}, A. \& {Braun}, R. 2007, \nar, 51, 24

\bibitem[{Prestage \& Peacock(1988)}]{prestage1988cluster}
Prestage, R.~M. \& Peacock, J.~A. 1988, Monthly Notices of the Royal
  Astronomical Society, 230, 131

\bibitem[{Rykoff {et~al.}(2014)Rykoff, Rozo, Busha, Cunha, Finoguenov, Evrard,
  Hao, Koester, Leauthaud, Nord, {et~al.}}]{rykoff2014redmapper}
Rykoff, E., Rozo, E., Busha, M., {et~al.} 2014, The Astrophysical Journal, 785,
  104

\bibitem[{Sakelliou {et~al.}(1996)Sakelliou, Merrifield, \&
  McHardy}]{sakelliou1996bent}
Sakelliou, I., Merrifield, M., \& McHardy, I. 1996, Monthly Notices of the
  Royal Astronomical Society, 283, 673

\bibitem[{Sanderson {et~al.}(2003)Sanderson, Ponman, Finoguenov, Lloyd-Davies,
  \& Markevitch}]{sanderson2003birmingham}
Sanderson, A.~J., Ponman, T., Finoguenov, A., Lloyd-Davies, E., \& Markevitch,
  M. 2003, Monthly Notices of the Royal Astronomical Society, 340, 989

\bibitem[{Saro {et~al.}(2013)Saro, Mohr, Bazin, \& Dolag}]{saro2013toward}
Saro, A., Mohr, J.~J., Bazin, G., \& Dolag, K. 2013, The Astrophysical Journal,
  772, 47

\bibitem[{Sawant {et~al.}(2022)Sawant, Kosak, Li, Avachat, Perlman, \&
  Mitra}]{sawant2022jetcurry}
Sawant, S.~M., Kosak, K., Li, K., {et~al.} 2022, Astronomy and Computing, 41,
  100653

\bibitem[{Schinnerer {et~al.}(2010)Schinnerer, Sargent, Bondi,
  Smol{\v{c}}i{\'c}, Datta, Carilli, Bertoldi, Blain, Ciliegi, Koekemoer,
  {et~al.}}]{schinnerer2010vla}
Schinnerer, E., Sargent, M., Bondi, M., {et~al.} 2010, The Astrophysical
  Journal Supplement Series, 188, 384

\bibitem[{Schinnerer {et~al.}(2007)Schinnerer, Smol{\v{c}}i{\'c}, Carilli,
  Bondi, Ciliegi, Jahnke, Scoville, Aussel, Bertoldi, Blain,
  {et~al.}}]{schinnerer2007vla}
Schinnerer, E., Smol{\v{c}}i{\'c}, V., Carilli, C.~L., {et~al.} 2007, The
  Astrophysical Journal Supplement Series, 172, 46

\bibitem[{Scoville {et~al.}(2013)Scoville, Arnouts, Aussel, Benson, Bongiorno,
  Bundy, Calvo, Capak, Carollo, Civano, {et~al.}}]{scoville2013evolution}
Scoville, N., Arnouts, S., Aussel, H., {et~al.} 2013, The Astrophysical Journal
  Supplement Series, 206, 3

\bibitem[{Sejake {et~al.}(2023)Sejake, White, Heywood, Thorat, Bester,
  Makhathini, \& Fanaroff}]{sejake2023meerkat}
Sejake, P.~K., White, S.~V., Heywood, I., {et~al.} 2023, Monthly Notices of the
  Royal Astronomical Society, 518, 4290

\bibitem[{Shimwell {et~al.}(2017)Shimwell, R{\"o}ttgering, Best, Williams,
  Dijkema, de~Gasperin, Hardcastle, Heald, Hoang, Horneffer,
  {et~al.}}]{shimwell2017lofar}
Shimwell, T., R{\"o}ttgering, H., Best, P.~N., {et~al.} 2017, Astronomy \&
  Astrophysics, 598, A104

\bibitem[{Shimwell {et~al.}(2019)Shimwell, Tasse, Hardcastle, Mechev, Williams,
  Best, R{\"o}ttgering, Callingham, Dijkema, De~Gasperin,
  {et~al.}}]{shimwell2019lofar}
Shimwell, T., Tasse, C., Hardcastle, M., {et~al.} 2019, Astronomy \&
  Astrophysics, 622, A1

\bibitem[{{Shimwell} {et~al.}(2022){Shimwell}, {Hardcastle}, {Tasse}, {Best},
  {R{\"o}ttgering}, {Williams}, {Botteon}, {Drabent}, {Mechev}, {Shulevski},
  {van Weeren}, {Bester}, {Br{\"u}ggen}, {Brunetti}, {Callingham}, {Chy{\.z}y},
  {Conway}, {Dijkema}, {Duncan}, {de Gasperin}, {Hale}, {Haverkorn}, {Hugo},
  {Jackson}, {Mevius}, {Miley}, {Morabito}, {Morganti}, {Offringa}, {Oonk},
  {Rafferty}, {Sabater}, {Smith}, {Schwarz}, {Smirnov}, {O'Sullivan},
  {Vedantham}, {White}, {Albert}, {Alegre}, {Asabere}, {Bacon}, {Bonafede},
  {Bonnassieux}, {Brienza}, {Bilicki}, {Bonato}, {Calistro Rivera}, {Cassano},
  {Cochrane}, {Croston}, {Cuciti}, {Dallacasa}, {Danezi}, {Dettmar}, {Di
  Gennaro}, {Edler}, {En{\ss}lin}, {Emig}, {Franzen}, {Garc{\'\i}a-Vergara},
  {Grange}, {G{\"u}rkan}, {Hajduk}, {Heald}, {Heesen}, {Hoang}, {Hoeft},
  {Horellou}, {Iacobelli}, {Jamrozy}, {Jeli{\'c}}, {Kondapally}, {Kukreti},
  {Kunert-Bajraszewska}, {Magliocchetti}, {Mahatma}, {Ma{\l}ek}, {Mandal},
  {Massaro}, {Meyer-Zhao}, {Mingo}, {Mostert}, {Nair}, {Nakoneczny},
  {Nikiel-Wroczy{\'n}ski}, {Orr{\'u}}, {Pajdosz-{\'S}mierciak}, {Pasini},
  {Prandoni}, {van Piggelen}, {Rajpurohit}, {Retana-Montenegro}, {Riseley},
  {Rowlinson}, {Saxena}, {Schrijvers}, {Sweijen}, {Siewert}, {Timmerman},
  {Vaccari}, {Vink}, {West}, {Wo{\l}owska}, {Zhang}, \& {Zheng}}]{Shimwell2022}
{Shimwell}, T.~W., {Hardcastle}, M.~J., {Tasse}, C., {et~al.} 2022, \aap, 659,
  A1

\bibitem[{Silverman {et~al.}(2015)Silverman, Kashino, Sanders, Kartaltepe,
  Arimoto, Renzini, Rodighiero, Daddi, Zahid, Nagao,
  {et~al.}}]{silverman2015fmos}
Silverman, J.~D., Kashino, D., Sanders, D., {et~al.} 2015, The Astrophysical
  Journal Supplement Series, 220, 12

\bibitem[{Silverstein {et~al.}(2017)Silverstein, Anderson, \&
  Bregman}]{silverstein2017increased}
Silverstein, E.~M., Anderson, M.~E., \& Bregman, J.~N. 2017, The Astronomical
  Journal, 155, 14

\bibitem[{Skelton {et~al.}(2014)Skelton, Whitaker, Momcheva, Brammer,
  Van~Dokkum, Labb{\'e}, Franx, Van Der~Wel, Bezanson, Da~Cunha,
  {et~al.}}]{skelton20143d}
Skelton, R.~E., Whitaker, K.~E., Momcheva, I.~G., {et~al.} 2014, The
  Astrophysical Journal Supplement Series, 214, 24

\bibitem[{Smol{\v{c}}i{\'c} {et~al.}(2011)Smol{\v{c}}i{\'c}, Finoguenov,
  Zamorani, Schinnerer, Tanaka, Giodini, \& Scoville}]{smolvcic2011occupation}
Smol{\v{c}}i{\'c}, V., Finoguenov, A., Zamorani, G., {et~al.} 2011, Monthly
  Notices of the Royal Astronomical Society: Letters, 416, L31

\bibitem[{Smol{\v{c}}i{\'c} {et~al.}(2018)Smol{\v{c}}i{\'c}, Intema,
  {\v{S}}laus, Raychaudhury, Novak, Horellou, Chiappetti, Delhaize, Birkinshaw,
  Bondi, {et~al.}}]{smolvcic2018xxl}
Smol{\v{c}}i{\'c}, V., Intema, H., {\v{S}}laus, B., {et~al.} 2018, Astronomy \&
  Astrophysics, 620, A14

\bibitem[{Smol{\v{c}}i{\'c} {et~al.}(2017{\natexlab{a}})Smol{\v{c}}i{\'c},
  Novak, Bondi, Ciliegi, Mooley, Schinnerer, Zamorani, Navarrete, Bourke,
  Karim, {et~al.}}]{smolvcic2017vlaSource}
Smol{\v{c}}i{\'c}, V., Novak, M., Bondi, M., {et~al.} 2017{\natexlab{a}},
  Astronomy \& Astrophysics, 602, A1

\bibitem[{Smol{\v{c}}i{\'c} {et~al.}(2017{\natexlab{b}})Smol{\v{c}}i{\'c},
  Novak, Delvecchio, Ceraj, Bondi, Delhaize, Marchesi, Murphy, Schinnerer,
  Vardoulaki, {et~al.}}]{smolvcic2017vla}
Smol{\v{c}}i{\'c}, V., Novak, M., Delvecchio, I., {et~al.} 2017{\natexlab{b}},
  Astronomy \& Astrophysics, 602, A6

\bibitem[{Smol{\v{c}}i{\'c} {et~al.}(2007)Smol{\v{c}}i{\'c}, Schinnerer,
  Finoguenov, Sakelliou, Carilli, Botzler, Brusa, Scoville, Ajiki, Capak,
  {et~al.}}]{smolvcic2007wide}
Smol{\v{c}}i{\'c}, V., Schinnerer, E., Finoguenov, A., {et~al.} 2007, The
  Astrophysical Journal Supplement Series, 172, 295

\bibitem[{{Springel} \& {Farrar}(2007)}]{Springel2007}
{Springel}, V. \& {Farrar}, G.~R. 2007, \mnras, 380, 911

\bibitem[{Straatman {et~al.}(2018)Straatman, van~der Wel, Bezanson, Pacifici,
  Gallazzi, Wu, Noeske, Bari{\v{s}}i{\'c}, Bell, Brammer,
  {et~al.}}]{straatman2018large}
Straatman, C.~M., van~der Wel, A., Bezanson, R., {et~al.} 2018, The
  Astrophysical Journal Supplement Series, 239, 27

\bibitem[{{Sweijen} {et~al.}(2022){Sweijen}, {van Weeren}, {R{\"o}ttgering},
  {Morabito}, {Jackson}, {Offringa}, {van der Tol}, {Veenboer}, {Oonk}, {Best},
  {Bondi}, {Shimwell}, {Tasse}, \& {Thomson}}]{Sweijen2022}
{Sweijen}, F., {van Weeren}, R.~J., {R{\"o}ttgering}, H.~J.~A., {et~al.} 2022,
  Nature Astronomy, 6, 350

\bibitem[{Tasse {et~al.}(2007)Tasse, R{\"o}ttgering, Best, Cohen, Pierre, \&
  Wilman}]{tasse2007gmrt}
Tasse, C., R{\"o}ttgering, H., Best, P., {et~al.} 2007, Astronomy \&
  Astrophysics, 471, 1105

\bibitem[{Taylor {et~al.}(1990)Taylor, Perley, Inoue, Kato, Tabara, \&
  Aizu}]{taylor1990vla}
Taylor, G., Perley, R., Inoue, M., {et~al.} 1990, Astrophysical Journal, Part 1
  (ISSN 0004-637X), vol. 360, Sept. 1, 1990, p. 41-54., 360, 41

\bibitem[{Taylor(2011)}]{taylor2011topcat}
Taylor, M. 2011, Astrophysics Source Code Library, 1

\bibitem[{Turner \& Shabala(2015)}]{turner2015energetics}
Turner, R.~J. \& Shabala, S.~S. 2015, The Astrophysical Journal, 806, 59

\bibitem[{Vardoulaki {et~al.}(2021{\natexlab{a}})Vardoulaki, Andrade,
  Delvecchio, Smol{\v{c}}i{\'c}, Schinnerer, Sargent, Gozaliasl, Finoguenov,
  Bondi, Zamorani, {et~al.}}]{vardoulaki_3GHz_FR}
Vardoulaki, E., Andrade, E.~J., Delvecchio, I., {et~al.} 2021{\natexlab{a}},
  Astronomy \& Astrophysics, 648, A102

\bibitem[{Vardoulaki {et~al.}(2019)Vardoulaki, Andrade, Karim, Novak, Leslie,
  Tisani{\'c}, Smol{\v{c}}i{\'c}, Schinnerer, Sargent, Bondi,
  {et~al.}}]{vardoulaki2019closer}
Vardoulaki, E., Andrade, E.~J., Karim, A., {et~al.} 2019, Astronomy \&
  Astrophysics, 627, A142

\bibitem[{Vardoulaki {et~al.}(2023)Vardoulaki, Gozaliasl, Finoguenov, Novak, \&
  Khosroshahi}]{vardoulaki2023evolution}
Vardoulaki, E., Gozaliasl, G., Finoguenov, A., Novak, M., \& Khosroshahi, H.~G.
  2023, The evolution of the radio luminosity function of group galaxies in
  COSMOS

\bibitem[{Vardoulaki {et~al.}(2021{\natexlab{b}})Vardoulaki, Vazza,
  Jim{\'e}nez-Andrade, Gozaliasl, Finoguenov, \& Wittor}]{vardoulaki2021bent}
Vardoulaki, E., Vazza, F., Jim{\'e}nez-Andrade, E.~F., {et~al.}
  2021{\natexlab{b}}, Galaxies, 9, 93

\bibitem[{Vazza {et~al.}(2021)Vazza, Wittor, Brunetti, \&
  Br{\"u}ggen}]{vazza2021}
Vazza, F., Wittor, D., Brunetti, G., \& Br{\"u}ggen, M. 2021, arXiv preprint
  arXiv:2102.04193

\bibitem[{Vazza {et~al.}(2023)Vazza, Wittor, Di~Federico, Br{\"u}ggen, Brienza,
  Brunetti, Brighenti, Pasini, {et~al.}}]{vazza2023life}
Vazza, F., Wittor, D., Di~Federico, L., {et~al.} 2023, Astron. Astrophys, 669,
  A50

\bibitem[{{Virtanen} {et~al.}(2020){Virtanen}, {Gommers}, {Oliphant},
  {Haberland}, {Reddy}, {Cournapeau}, {Burovski}, {Peterson}, {Weckesser},
  {Bright}, {van der Walt}, {Brett}, {Wilson}, {Jarrod Millman}, {Mayorov},
  {Nelson}, {Jones}, {Kern}, {Larson}, {Carey}, {Polat}, {Feng}, {Moore}, {Vand
  erPlas}, {Laxalde}, {Perktold}, {Cimrman}, {Henriksen}, {Quintero}, {Harris},
  {Archibald}, {Ribeiro}, {Pedregosa}, {van Mulbregt}, \&
  {Contributors}}]{scipy}
{Virtanen}, P., {Gommers}, R., {Oliphant}, T.~E., {et~al.} 2020, Nature
  Methods, 17, 261

\bibitem[{Weaver {et~al.}(2022)Weaver, Kauffmann, Ilbert, McCracken, Moneti,
  Toft, Brammer, Shuntov, Davidzon, Hsieh, {et~al.}}]{weaver2022cosmos2020}
Weaver, J.~R., Kauffmann, O., Ilbert, O., {et~al.} 2022, The Astrophysical
  Journal Supplement Series, 258, 11

\bibitem[{Wetzel {et~al.}(2014)Wetzel, Tinker, Conroy, \&
  Bosch}]{wetzel2014galaxy}
Wetzel, A.~R., Tinker, J.~L., Conroy, C., \& Bosch, F. C. v.~d. 2014, Monthly
  Notices of the Royal Astronomical Society, 439, 2687

\bibitem[{White {et~al.}(2020{\natexlab{a}})White, Franzen, Riseley, Wong,
  Kapi{\'n}ska, Hurley-Walker, Callingham, Thorat, Wu, Hancock,
  {et~al.}}]{white2020gleamI}
White, S.~V., Franzen, T.~M., Riseley, C.~J., {et~al.} 2020{\natexlab{a}},
  arXiv preprint arXiv:2004.13125

\bibitem[{White {et~al.}(2020{\natexlab{b}})White, Franzen, Riseley, Wong,
  Kapi{\'n}ska, Hurley-Walker, Callingham, Thorat, Wu, Hancock,
  {et~al.}}]{white2020gleamII}
White, S.~V., Franzen, T.~M., Riseley, C.~J., {et~al.} 2020{\natexlab{b}},
  arXiv preprint arXiv:2004.13025

\bibitem[{White {et~al.}(2015)White, Jarvis, H{\"a}u{\ss}ler, \&
  Maddox}]{white2015radio}
White, S.~V., Jarvis, M.~J., H{\"a}u{\ss}ler, B., \& Maddox, N. 2015, Monthly
  Notices of the Royal Astronomical Society, 448, 2665

\bibitem[{White {et~al.}(2017)White, Jarvis, Kalfountzou, Hardcastle, Verma,
  Cao~Orjales, \& Stevens}]{white2017evidence}
White, S.~V., Jarvis, M.~J., Kalfountzou, E., {et~al.} 2017, Monthly Notices of
  the Royal Astronomical Society, 468, 217

\bibitem[{Whittam {et~al.}(2022)Whittam, Jarvis, Hale, Prescott, Morabito,
  Heywood, Adams, Afonso, An, Ao, {et~al.}}]{whittam2022mightee}
Whittam, I., Jarvis, M., Hale, C., {et~al.} 2022, Monthly Notices of the Royal
  Astronomical Society, 516, 245

\bibitem[{Wing \& Blanton(2011)}]{wing2011galaxy}
Wing, J.~D. \& Blanton, E.~L. 2011, The Astronomical Journal, 141, 88

\bibitem[{Wright {et~al.}(2010)Wright, Eisenhardt, Mainzer, Ressler, Cutri,
  Jarrett, Kirkpatrick, Padgett, McMillan, Skrutskie,
  {et~al.}}]{wright2010wide}
Wright, E.~L., Eisenhardt, P.~R., Mainzer, A.~K., {et~al.} 2010, The
  Astronomical Journal, 140, 1868

\bibitem[{{Zhao} {et~al.}(2024){Zhao}, {Xu}, {Liu}, {Zhang}, {Ji}, {Chang},
  {Hu}, {Werner}, {Zhang}, {Cui}, \& {Wu}}]{zhao2024}
{Zhao}, Y., {Xu}, H., {Liu}, A., {et~al.} 2024, arXiv e-prints,
  arXiv:2410.06836

\end{thebibliography}

\begin{appendix}

\section{Very Bent Sources}
\label{peculiar Sources}

\subsection{Very Bent Sources in X-ray Galaxy Groups}
\label{sec:verybentxgroups}

\begin{table}
\begin{center}
\caption{Properties of the very bent sources located in X-ray galaxy groups.}
\label{tab:verybentinXray}
\scalebox{0.64}{
\renewcommand{\arraystretch}{1.5}
\begin{tabular}[t]{l c c c c c}
\hline
\textbf{Source} & BA (deg.) &  log$_{10}$(M$_{200}$/M$_{\odot}$) & $kT$ (keV) & $r/r_{200}$ & $P_{\rm ICM}$ (keV cm$^{-3}$)\\
\hline
XMM-LSS Source 1 & 23 & 14.34 & 2.95 & 0.15 & $4.21\times10^{-3}$\\
XMM-LSS Source 200 & 16 & 14.29 & 2.77 & 0.56 & $2.13\times10^{-4}$\\
\hline
COSMOS Source 247 & 52 & 13.6 & 0.9 & 0.07 & $1.79\times10^{-3}$\\
COSMOS Source 252 & 45 & 13.34 & 0.66 & 0.08 & $1.11\times10^{-3}$\\
\hline
\end{tabular}
}
\end{center}

\end{table}

In Table \ref{tab:verybentinXray}, we compare the X-ray galaxy group properties for the four very bent sources discussed above. 
We find that the two NATs in XMM-LSS are in different X-ray galaxy group environments compared to the two WATs in COSMOS:
While the two NATs in XMM-LSS are located in comparatively more massive groups (M$_{200}\approx10^{14.3}$M$_{\odot}$) at higher temperatures ($kT\approx\SI{2.8}{keV}$) and are part of the \emph{inner region} of groups ($0.1<r/r_{200}<1$), the two WATs in COSMOS are found in comparatively less massive groups (M$_{200}\approx10^{13.5}$M$_{\odot}$) at lower temperatures ($kT<\SI{1}{keV}$) and are in the \emph{core region} of their groups ($r/r_{200}<0.1$). As we mentioned in Section \ref{Discussion}, the X-ray galaxy groups in COSMOS probe lower halo masses and group temperatures compared to XMM-LSS. This most likely plays a role for the question on why we do not observe sources with group properties like Source 1 and Source 200 from the XMM-LSS sample in COSMOS. Figure \ref{fig:kTM200Histo} shows that the very bent sources in groups are at the lower end of the group temperatures and halo masses for the bent sources in COSMOS groups, where there is little to no overlap to mass and temperature ranges of the XMM-LSS groups. In other words, the WATs in COSMOS and the NATs in XMM-LSS probe different parameter space and the reasons for the bent jets could differ, where in the NATs the rapid infall can be the cause of severe bending of the jets.

This may explain why we find different group environments for the very bent sources between the COSMOS and XMM-LSS sample, but not why we find very bent sources under such different environmental conditions. Under the assumption that the bending for the two NATs in XMM-LSS is driven by the ram pressure exerted on the jets as the galaxy moves through the ICM \citep{o1985global}, the environments we find for Source 1 and Source 200 from the XMM-LSS sample are well suited to explain the bending, since high temperatures statistically correspond higher velocities of the galaxies that move through the ICM \citep{girardi1995velocity}. Out of the four very bent sources in our samples in galaxy groups, Source 200 in XMM-LSS is the furthest away from the X-ray group centre ($r/r_{200}=0.56$, $\sim$\SI{600}{kpc}). To induce the jet bending the ICM pressure should be higher, which would indicate that the source could be closer to the centre than calculated here. But if one assumes a flattening of the gas density \citep[e.g.][]{ponman1999thermal} and thus of the ICM profile, in the mass range we are probing, this would not play a significant role. Since the ram pressure exerted on a galaxy by the ICM scales with the square of the galaxy's velocity and only linearly with the ICM density ($P_{\rm ram} = \rho_{\rm ICM}\,v^2_{\rm gal}$), this could explain the observed bending. This further supports the NAT classification of Source 200, as NATs are typically moving with high velocities through the ICM and are found at larger distances from the core region of groups and clusters \citep{owen1976radio}.

The two very bent sources in COSMOS are WATs, which are also believed to be predominantly shaped by ram pressure we observe for NATs \citep[e.g.][]{smolvcic2007wide,o2023wide}. WATs are usually found near the group or cluster centre \citep{hardcastle2004jet}, as they tend to be the dominant galaxy of the group. This is consistent with our results for distance to the group centre, as seen in Table \ref{tab:verybentinXray}. One would expect WATs not to be located in cool cores of galaxy groups \citep{o1993flow}, but the low group temperatures we observe for Source 247 and Source 252 are not necessarily indicative of the group's core temperature, because we only have access to the mean group temperatures obtained from scaling relations. Indeed, more disturbed AGN are found in cool cores \citep{o2023wide}. \cite{smolvcic2007wide} found that Source 252 is located in a merging group environment, where the velocities needed to explain the observed bending due to ram pressure (and buoyancy forces, dominating at the jet-tail transition) are induced by the merging event of 3 galaxy groups that will result in a massive galaxy cluster. In such a dynamical scenario, the measured temperature of the group environment does not reflect the final state after the merger, while the velocities already do. While \cite{smolvcic2007wide} gives concrete evidence for this scenario for Source 252, we do not find merger candidates for Source 247 from the galaxy groups in COSMOS. Since this group is located near the edge of the X-ray coverage in COSMOS, it could be possible that Source 247 is also part of a group merger event not probed by the current X-ray data coverage. In the following section, we estimate the expected temperature one would expect from the jet bending as opposed to the mean group temperature we looked at so far.

\section{Sample Properties}
\label{Appendix}

 \newpage
 \begin{sidewaystable*}
 \caption{Properties of bent AGN in XMM-LSS within X-ray galaxy groups (Table is available in the online version)}
 \label{tab:xmmlsssample}
 \centering
 \scalebox{0.9}{
 \begin{tabular}[t]{c c c c c c c c c c c c c c c c}
 \hline\hline
 &\multicolumn{2}{c}{Radio} & \multicolumn{2}{c}{Host}&\multicolumn{4}{c}{Redshift}& LAS & Size &\multicolumn{2}{c}{bending angle}&Flux &$log_{10}(L_{1.4GHz})$ & Tags\\
 Object ID & RA & DEC & RA & DEC & z & z$_{l68}$ & z$_{u68}$ & z$_{ref}$ &  &  & BA$_{flux}$ & BA$_{edge}$ &  & \\
 & \multicolumn{4}{c}{(deg., J2000)} & & & & & (") & (kpc) & (deg.) & (deg.) & (mJy) & (W/Hz)\\
 (1) & (2) & (3) & (4) & (5) & (6) & (7) & (8) & (9) & (10) & (11)& (12) & (13) & (14) & (15) & (16)\\
 \hline\hline

 1 & 33.88255 & -4.68081 & 33.88022 & -4.68296 & 0.348$^{S}$ & - & - & PRIMUS-DR1 & 68 & 337 & 67 & 23 & 14.6 & 24.71&Jets?, NAT? \\
 5 & 34.98366 & -5.46760 & 34.98318 & -5.46831 & 0.278$^{S}$ & - & - & SDSS-DR13 & 108 & 458 & 163 & 108 & 53.432 & 25.05&Jets, bent, S-shaped \\
 12 & 36.11746 & -4.83127 & 36.11798 & -4.8309 & 0.494$^{S}$ & - & - & VIPERS & 41 & 254 & 170 & 170 & 13.192 & 25.01&Jets, not-bent \\
 32 & 34.60201 & -5.41025 & 34.59796 & -5.41684 & 0.648$^{S}$ & - & - & PRIMUS-DR1 & 85 & 589 & 121 & 122 & 11.248 & 25.21&Jets, bent, WAT \\
 35 & 34.94020 & -4.89374 & 34.93856 & -4.89252 & 0.333$^{S}$ & - & - & SDSS-DR15 & 45 & 218 & 76 & 102 & 13.953 & 24.64&Jets? NAT? \\
 49 & 35.77514 & -4.20915 & 35.77554 & -4.20929 & 0.631$^{S}$ & - & - & PRIMUS-DR1 & 45 & 311 & 161 & 175 & 2.989 & 24.61&Jets, S-shaped \\
 57 & 34.49016 & -5.46774 & 34.48867 & -5.46556 & 0.7$^{P}$ & 0.64 & 0.769 & HSC-SSP-PDR2 & 70 & 506 & 151 & 138 & 7.016 & 25.09&Jets, bent \\
 60 & 34.40662 & -5.22500 & 34.40484 & -5.22493 & 0.646$^{S}$ & - & - & SDSS-DR15 & 33 & 232 & - & 159 & 3.133 & 24.66&Jets? one-sided? \\
 84 & 37.61886 & -4.61858 & 37.62289 & -4.62078 & 0.292$^{S}$ & - & - & SDSS-DR16 & 75 & 329 & 168 & 172 & 9.565 & 24.35&Jets, bent, blended \\
 146 & 36.33764 & -3.79950 & 36.33812 & -3.79928 & 0.28$^{P}$ & 0.24 & 0.318 & HSC-SSP-PDR2 & 41 & 176 & 142 & 144 & 1.935 & 23.62&Jets, bent, blended \\
 161 & 34.87760 & -4.07180 & 34.87757 & -4.07146 & 0.601$^{S}$ & - & - & SDSS-DR15 & 82 & 550 & 171 & 168 & 2.765 & 24.53&Jets, not-bent \\
 176 & 33.73470 & -3.79616 & 33.73487 & -3.79625 & 0.14$^{S}$ & - & - & SDSS-DR15 & 68 & 169 & 153 & 140 & 2.579 & 23.09&Jets, bent, blended \\
 178 & 34.01987 & -4.23261 & 34.01978 & -4.23244 & 0.154$^{S}$ & - & - & SDSS-DR15 & 228 & 608 & 160 & 174 & 25.329 & 24.17&Jets, bent, blended? \\
 179 & 33.67119 & -4.07914 & 33.67097 & -4.07891 & 0.448$^{S}$ & - & - & PRIMUS-DR1 & 72 & 414 & 133 & 133 & 16.996 & 25.02&Jets, bent, WAT?, blended \\
 197 & 33.87294 & -5.54909 & 33.87288 & -5.54858 & 0.29$^{S}$ & - & - & SDSS-DR15 & 48 & 209 & 164 & 177 & 32.385 & 24.88&Jets, not-bent, blended? \\
 200 & 34.08918 & -5.99792 & 34.08913 & -5.99772 & 0.397$^{S}$ & - & - & GAMA-DR3 & 88 & 471 & 89 & 16 & 2.191 & 24.01&Jets, bent, NAT \\
 303 & 33.73724 & -4.49900 & 33.73882 & -4.49991 & 0.136$^{S}$ & - & - & SDSS-DR15 & 56 & 137 & 118 & 107 & 8.337 & 23.57&Jets, bent \\

 \hline

 \end{tabular}
 }
 \tablefoot{
 \textbf{Column 1:} The \SI{1.28}{GHz} radio ID assigned to the source. \textbf{Columns 2\&3:} Right Ascension and Declination of the radio position in degrees. \textbf{Columns 4\&5:} Right Ascension and Declination of the host position in degrees. \textbf{Column 5:} Redshift of the radio source. The superscript 'S' and 'P' are assigned for spectroscopic and photometric redshifts, respectively. \textbf{Columns 6\&7:} The lower and upper bounds of the redshift at $68\%$ confidence level, respectively. \textbf{Column 9:} The survey from which the redshift value is obtained from. \textbf{Column 10:} The largest angular size (LAS) of the radio source. \textbf{Column 11:} The projected linear size of the radio source, calculated from the largest angular size and the redshift. \textbf{Column 12\&13:} The bending angle, measured from the peak flux positions of the jets/lobes (\textbf{Column 12}) and from the edges of the jets/lobes (\textbf{Column 13)}, respectively. \textbf{Column 14:} Flux density of the radio source in mJy. \textbf{Column 15:} Radio Luminosity at \SI{1.4}{GHz} in W Hz$^{-1}$, calculated using the median frequency at 1.28 GHz and a typical radio spectral index of $a=0.7$. \textbf{Column 16:} Tags that describe the radio morphology. The assigned tags are chosen from jets/no-jets, one-sided, bent/not-bent, X/S/Z-shaped, WAT, NAT, blended, peculiar and giant radio galaxy (GRG), for sources that are at or greater than \SI{1}{Mpc} in linear size. A '?' is added to the tag if the assessment of the tag is deemed uncertain.
 }
 \end{sidewaystable*}

 \clearpage

 \begin{sidewaystable*}
 \caption{Properties of bent AGN in COSMOS within X-ray galaxy groups (Table is available in the online version)}
 \label{tab:cosmossample}
 \centering
 \scalebox{0.9}{
 \begin{tabular}[t]{c c c c c c c c c c c c c c c c}
 \hline\hline
 &\multicolumn{2}{c}{Radio} & \multicolumn{2}{c}{Host}&\multicolumn{4}{c}{Redshift}& LAS & Size &\multicolumn{2}{c}{bending angle}&Flux &$log_{10}(L_{1.4GHz})$ & Tags\\
 Object ID & RA & DEC & RA & DEC & z & z$_{l68}$ & z$_{u68}$ & z$_{ref}$ &  &  & BA$_{flux}$ & BA$_{edge}$ &  & \\
 & \multicolumn{4}{c}{(deg., J2000)} & & & & & (") & (kpc) & (deg.) & (deg.) & (mJy) & (W/Hz)\\
 (1) & (2) & (3) & (4) & (5) & (6) & (7) & (8) & (9) & (10) & (11)& (12) & (13) & (14) & (15) & (16)\\
 \hline\hline
 38 & 150.23038 & 2.49483 & 150.23055 & 2.49502 & 0.378$^{S}$ & - & - & PRIMUS-DR1 & 35 & 183 & - & 154 & 0.442 & 23.27&Jets, not-bent, blended? \\
 92 & 150.57515 & 1.93226 & 150.5735 & 1.93386 & 0.31$^{S}$ & - & - & DEIMOS10k & 31 & 141 & - & 125 & 0.407 & 23.04&Jets, bent, blended \\
 178 & 149.99838 & 2.76914 & 149.99847 & 2.76906 & 0.166$^{S}$ & - & - & SDSS-DR15 & 806 & 2289 & 150 & 152 & 31.1 & 24.32&Jets, bent, GRG \\
 182 & 150.19832 & 1.98653 & 150.19832 & 1.98653 & 0.44$^{S}$ & - & - & PRIMUS-DR1 & 33 & 189 & 175 & 176 & 21.0 & 25.1&Jets, not-bent, blended \\
 192 & 149.75746 & 2.89340 & 149.75722 & 2.89339 & 0.351$^{S}$ & - & - & SDSS-DR15 & 59 & 296 & 177 & 158 & 5.44 & 24.29&Jets, bent, Z-shaped \\
 210 & 149.59712 & 2.44124 & 149.59711 & 2.44123 & 1.168$^{S}$ & - & - & zCOSMOS & 94 & 776 & 171 & 172 & 121.0 & 26.84&Jets, not-bent \\
 213 & 150.62456 & 2.54032 & 150.62459 & 2.54031 & 0.432$^{S}$ & - & - & VLT/FORS2 & 71 & 403 & 174 & 171 & 6.66 & 24.58&Jets, bent \\
 218 & 150.20662 & 1.82325 & 150.20663 & 1.82326 & 0.53$^{S}$ & - & - & SDSS-DR15 & 52 & 330 & 113 & 130 & 12.4 & 25.05&Jets, bent, S-shaped \\
  220 & 149.50874 & 2.26135 & 149.50874 & 2.26138 & 0.943$^{S}$ & - & - & IMACS & 31 & 250 & 142 & 139 & 2.5 & 24.94&Jet, one-sided, blended \\
 221 & 150.11786 & 2.68427 & 150.11782 & 2.68428 & 0.348$^{S}$ & - & - & PRIMUS-DR1 & 106 & 525 & 114 & 101 & 76.9 & 25.43&Jets, bent, blended \\
 225 & 149.60008 & 2.82116 & 149.60008 & 2.82116 & 0.345$^{S}$ & - & - & SDSS-DR15 & 108 & 531 & 164 & 120 & 57.4 & 25.29&Jets, bent, S-shaped, blended \\
 238 & 150.17995 & 1.76885 & 150.17999 & 1.76884 & 0.346$^{S}$ & - & - & SDSS-DR9 & 48 & 236 & 175 & 174 & 89.2 & 25.49&Jets,bent, peculiar \\
 241 & 150.11766 & 1.58572 & 150.11763 & 1.58571 & 0.84$^{S}$ & - & - & DEIMOS10k & 70 & 537 & 174 & 167 & 27.7 & 25.87&Jets, bent \\
 245 & 150.09076 & 1.99999 & 150.09076 & 2.0 & 0.219$^{S}$ & - & - & SDSS-DR15 & 167 & 592 & 157 & 160 & 4.84 & 23.78&Jets, bent, WAT? \\
 247 & 149.95424 & 2.92122 & 149.95422 & 2.92122 & 0.126$^{S}$ & - & - & SDSS-DR15 & 57 & 130 & 46 & 52 & 2.83 & 23.03&Jets, bent, NAT \\
 250 & 150.07710 & 2.54897 & 150.07709 & 2.54894 & 0.889$^{S}$ & - & - & zCOSMOS & 42 & 326 & 167 & 165 & 2.8 & 24.93&Jets, not-bent, blended \\
 252 & 150.11432 & 2.35647 & 150.11434 & 2.35647 & 0.22$^{S}$ & - & - & 3D-HST & 96 & 341 & 73 & 45 & 12.0 & 24.18&Jets, bent, WAT \\
 257 & 150.27806 & 1.55567 & 150.27807 & 1.55558 & 0.358$^{P}$ & 0.348 & 0.373 & COSMOS2020 & 25 & 127 & 170 & 163 & 2.4 & 23.95&Jets, not-bent \\
 258 & 149.94295 & 2.60061 & 149.94297 & 2.60063 & 0.344$^{S}$ & - & - & zCOSMOS & 26 & 130 & - & 156 & 18.6 & 24.8&Jets, not-bent, blended? \\

 \hline

 \end{tabular}
 }

 \tablefoot{
 \textbf{Column 1:} The \SI{1.28}{GHz} radio ID assigned to the source. \textbf{Columns 2\&3:} Right Ascension and Declination of the radio position in degrees. \textbf{Columns 4\&5:} Right Ascension and Declination of the host position in degrees. \textbf{Column 5:} Redshift of the radio source. The superscript 'S' and 'P' are assigned for spectroscopic and photometric redshifts, respectively. \textbf{Columns 6\&7:} The lower and upper bounds of the redshift at $68\%$ confidence level, respectively. \textbf{Column 9:} The survey from which the redshift value is obtained from. \textbf{Column 10:} The largest angular size (LAS) of the radio source. \textbf{Column 11:} The projected linear size of the radio source, calculated from the largest angular size and the redshift. \textbf{Column 12\&13:} The bending angle, measured from the peak flux positions of the jets/lobes (\textbf{Column 12}) and from the edges of the jets/lobes (\textbf{Column 13)}, respectively. \textbf{Column 14:} Flux density of the radio source in mJy. \textbf{Column 15:} Radio Luminosity at \SI{1.4}{GHz} in W Hz$^{-1}$, calculated using the median frequency at 1.28 GHz and a typical radio spectral index of $a=0.7$. \textbf{Column 16:} Tags that describe the radio morphology. The assigned tags are chosen from jets/no-jets, one-sided, bent/not-bent, X/S/Z-shaped, WAT, NAT, blended, peculiar and giant radio galaxy (GRG), for sources that are at or greater than \SI{1}{Mpc} in linear size. A '?' is added to the tag if the assessment of the tag is deemed uncertain.
 }

 \end{sidewaystable*}

 \end{appendix}

 \end{document}